\documentclass{article}

\usepackage{arxiv}

\usepackage[utf8]{inputenc} 
\usepackage[T1]{fontenc}    
\usepackage{hyperref}       
\usepackage{url}            
\usepackage{booktabs}       
\usepackage{amsfonts}       
\usepackage{nicefrac}       
\usepackage{microtype}      
\usepackage{lipsum}
\usepackage{graphicx}
\graphicspath{ {./images/} }
\usepackage{epstopdf}
\ifpdf
  \DeclareGraphicsExtensions{.eps,.pdf,.png,.jpg}
\else
  \DeclareGraphicsExtensions{.eps}
\fi

\usepackage{algorithm}
\usepackage{algpseudocode}
\usepackage{threeparttable}

\usepackage{caption}
\usepackage{subcaption}
\usepackage{mathtools}
\usepackage{amssymb}
\usepackage{multirow}
\usepackage{siunitx}

\usepackage[shortlabels]{enumitem}
\setlist[enumerate]{leftmargin=.5in}
\setlist[itemize]{leftmargin=.5in}

\newenvironment{codefont}{\fontfamily{lmtt}\selectfont}{\par}
\DeclareTextFontCommand{\codetext}{\codefont}

\DeclareMathOperator{\atantwo}{atan2}

\title{Predict future sale}

\title{Intrepid MCMC: Metropolis-Hastings with Exploration}

\author{
 Promit Chakroborty \\
  Department of Civil and Systems Engineering\\
  Johns Hopkins University\\
  Baltimore, MD 21218 \\
  \texttt{pchakro1@jhu.edu} \\
   \And
 Michael D. Shields \\
  Department of Civil and Systems Engineering\\
  Johns Hopkins University\\
  Baltimore, MD 21218 \\
  \texttt{michael.shields@jhu.edu} \\
}

\begin{document}

\newtheorem{proof}{Proof}
\newtheorem{theorem}{Theorem}
\newtheorem{lemma}{Lemma}
\newtheorem{corollary}{Corollary}
\newtheorem{proposition}{Proposition}
\newtheorem{definition}{Definition}
\newtheorem{remark}{Remark}
\newtheorem{stipulation}{Stipulation}

\maketitle

\begin{abstract}
In engineering examples, one often encounters the need to sample from unnormalized distributions with complex shapes that may also be implicitly defined through a physical or numerical simulation model, making it computationally expensive to evaluate the associated density function. For such cases, MCMC has proven to be an invaluable tool. Random-walk Metropolis Methods (also known as Metropolis-Hastings (MH)), in particular, are highly popular for their simplicity, flexibility, and ease of implementation. However, most MH algorithms suffer from significant limitations when attempting to sample from distributions with multiple modes (particularly disconnected ones). In this paper, we present Intrepid MCMC - a novel MH scheme that utilizes a simple coordinate transformation to significantly improve the mode-finding ability and convergence rate to the target distribution of random-walk Markov chains while retaining most of the simplicity of the vanilla MH paradigm. Through multiple examples, we showcase the improvement in the performance of Intrepid MCMC over vanilla MH for a wide variety of target distribution shapes. We also provide an analysis of the mixing behavior of the Intrepid Markov chain, as well as the efficiency of our algorithm for increasing dimensions. A thorough discussion is presented on the practical implementation of the Intrepid MCMC algorithm. Finally, its utility is highlighted through a Bayesian parameter inference problem for a two-degree-of-freedom oscillator under free vibration.
\end{abstract}


\section{Introduction}
\label{section:introduction}

The vast majority of uncertainty quantification problems in engineering -- forward and inverse problems alike -- cannot be solved analytically. Therefore, one resorts to some manner of Monte Carlo estimation, which requires drawing samples from the distribution of interest. To ensure accurate and efficient estimation, the drawn set of samples must be of `good quality', i.e., being statistically representative, or ideally independent, leading to unbiased statistical estimates. 
Furthermore, the distribution of interest is usually tied to a physical system, which requires either experimental or numerical simulation to characterize. This makes it impossible
to define the distribution analytically (through its mass/density function or cumulative distribution function). To overcome these challenges, it is necessary to have robust and efficient techniques to draw samples from implicitly defined distributions that are easy to use, are widely applicable for general distributions (including those with multiple modes and disjoint support), and produce high-quality statistical samples. This sampling problem is typically addressed using Markov Chain Monte Carlo (MCMC) methods. 
In this paper, we present a versatile random-walk MCMC technique with enhanced exploration capability, built on the ubiquitous Metropolis-Hastings Algorithm~\cite{MetropolisMCMC,UnderstandingMH}, that is designed to sample from distributions with complex shapes, particularly multimodal distributions or those having disjoint support, with significantly accelerated convergence and improved statistical properties compared to existing random-walk methods.

\subsection{Problem Setup}
\label{section:problem_description}

Consider the random vector $ \mathbf{x} = \begin{bmatrix} x_1 & x_2 & \dots & x_d \end{bmatrix}^T \in  \Omega \subseteq \mathbb{R}^d$ having probability distribution 
$ \pi (\mathbf{x}) $ (referred to herein as the \textit{target distribution}) that is implicitly defined through a particular engineering task, i.e. $ \pi (\mathbf{x}) $ cannot be defined analytically. 
It is difficult to draw samples of $\mathbf{x}$ because conventional techniques require the density/distribution function to be known explicitly.

Let us assume that $ \pi (\mathbf{x}) $ can be constructed as the product of a simpler underlying distribution $ p (\mathbf{x}) $ (which we will call the \textit{parent distribution}) and some transformation function $ T (\mathbf{x}) $, i.e.
\begin{equation}
\label{eqn:target_structure}
    \pi (\mathbf{x}) \propto T (\mathbf{x}) p (\mathbf{x})
\end{equation}
Numerous engineering applications result in distributions of this form. Two broad classes of problems that fall under this setup are:

\begin{itemize}
    \item \textbf{Bayesian Inference}, which deals with updating the unknown uncertainties in $\mathbf{x}$ based on observations of some related quantity of interest of the system. In Bayesian inference, the parent distribution is called the \textit{prior distribution}, and is often an assumed distribution over $\mathbf{x}$ based on existing knowledge, beliefs, or assumptions.
    The distribution is then updated into the target distribution, called the \textit{posterior distribution}, based on observations $ \mathcal{D} $ made on the system through the transformation $ T(\mathbf{x}) = \mathcal{L} (\mathcal{D} | \mathbf{x}) $, which is called the \textit{Likelihood function}. The likelihood function encodes the probability of making the observations $ \mathcal{D} $ given that the random vector takes the value $ \mathbf{x} $. Bayesian inference finds a wide variety of engineering applications, especially in model calibration~\cite{beck1998updating,asaadi2017computationalbayesian, patsialis2020bayesian}, model selection~\cite{diazdelao2017bayesian, zhang2018effect} and general inverse problems~\cite{beck2010bayesian, betz2018bayesian, kim2021bayesianpipeline,guo2022bayesian}.
    \item \textbf{Reliability Analysis}, which deals with estimating the probability of (typically rare) failure events.
    Reliability analysis is a forward problem in which the parent distribution is the known distribution of the model inputs. Meanwhile, the target distribution is the conditional distribution of the inputs given the failure event $ F \subset \Omega $, which depends on the physics of the system. The transformation function $ T(\mathbf{x}) = I_F (\mathbf{x}) $ indicates failure by taking nonzero values only when $ \mathbf{x} \in F $. Engineering applications of reliability analysis are ubiquitous~\cite{pradlwarter2005realistic, goller2013reliability, dhulipala2022reliability} and developing efficient estimators of small failure probabilities is a major challenge in engineering analysis~\cite{papadopoulos2012accelerated,papaioannou2015mcmc, papaioannou2016sis, moustapha2022active}. 
\end{itemize}

The formulation in Eq.~\eqref{eqn:target_structure} is useful because $ p (\mathbf{x}) $ is usually some known, simple distribution that can be directly sampled from, and only the function $ T (\mathbf{x}) $ represents the physical system and requires experiments or simulations to evaluate. This allows for the use of techniques that first generate a sample set from $ p (\mathbf{x}) $, then modify it to follow $ \pi (\mathbf{x}) $. However, even this becomes challenging when the shape of the parent and target distributions and/or their important regions are significantly different. In such cases, simple sampling techniques such as Rejection Sampling (using $ p (\mathbf{x}) $ as the proposal distribution), Importance Sampling (using $ p (\mathbf{x}) $ as the Importance Sampling Distribution) result in a significant waste of computational effort. This is because many samples generated from $ p (\mathbf{x}) $ will not be suitable candidates for a sample set from $ \pi (\mathbf{x}) $, and judging the suitability requires an evaluation of $ T (\mathbf{x}) $ for each sample, which is typically expensive computationally.

\section{Markov Chain Monte Carlo (MCMC)}
\label{section:MCMC}

Markov chain Monte Carlo (MCMC) is a class of highly versatile methods that are used to generate samples from arbitrary distributions. The general idea is to construct a discrete-time Markov chain on a continuous state space, such that the invariant distribution of the Markov chain exists and is equal to the target distribution $ \pi(\mathbf{x}) $. An added benefit of MCMC is that samples can be generated from the target distribution even if it is \textit{unnormalized}, i.e., only known up to a multiplicative constant as is the case in Eq.~\eqref{eqn:target_structure}. 
Thus, MCMC techniques are useful to sample in this setting, although MCMC methods, in general, do not require the multiplicative decomposition presented in Eq~\eqref{eqn:target_structure}.

MCMC theory can be formalized as follows (using notation from~\cite{UnderstandingMH}). For a given target distribution $ \pi(\mathbf{x}) $, a transition kernel $ K(\mathbf{x}, A) $ is constructed that represents the probability of the chain moving from state $\mathbf{x}$ to a new state in the set $A$ in a given step. This kernel is the conditional distribution function of the Markov chain starting at point $ \mathbf{x} \in \Omega $ and moving to a point in the set $ A \subset \Omega $. The kernel must be constructed such that its invariant density is $ \pi (\mathbf{x}) $, i.e., when iterated $ n $ times, and as $ n \to \infty $, the kernel 
converges to $ \pi(\mathbf{x}) $. That is, for the Markov chain $ \left\{ \mathbf{x}_{t} : t \geq 0 \right\} $ starting at some arbitrary point $ \mathbf{x}_0 \in \Omega $ 
and propagated using the transition kernel $ K(\mathbf{x}, A) $, $ \mathbf{x}_{t} \sim \pi(\mathbf{x}) $ $ \forall t > b $, for sufficiently large $ b $ (called the \textit{burn-in} period).

Often, the transition kernel is constructed to satisfy \textit{time-reversibility} or \textit{detailed balance}, defined as follows. Let there be a function $ k(\mathbf{x}, \mathbf{y}) $ (for $ \mathbf{x}, \mathbf{y} \in \Omega $) -- referred to as the \textit{transition kernel generating function} -- such that the transition kernel can be written as
\begin{equation}
    \label{eqn:transition_kernel_expansion}
    K(\mathbf{x}, d\mathbf{y}) = k(\mathbf{x}, \mathbf{y})d\mathbf{y} + r(\mathbf{x}) \delta_x (d\mathbf{y})
\end{equation}
where $ \delta_x (d\mathbf{y}) = 1 $ if $ \mathbf{x} \in d \mathbf{y} $ and $ 0 $ otherwise, $ r(\mathbf{x}) = 1 - \int_{\Omega} k(\mathbf{x}, \mathbf{y})d\mathbf{y} $ is the probability that the chain does not move from $ \mathbf{x} $, and $ k(\mathbf{x}, \mathbf{x}) = 0 $. Then, $ k(\mathbf{x}, \mathbf{y}) $ satisfies time-reversibility if
\begin{equation}
    \label{eqn:time_reversibility_definition}
    \pi(\mathbf{x}) k(\mathbf{x}, \mathbf{y}) = \pi(\mathbf{y}) k(\mathbf{y}, \mathbf{x}),
\end{equation}
which is a sufficient condition for the resultant transition kernel $ K(\mathbf{x}, d\mathbf{y}) $ to have $ \pi (\mathbf{x}) $ as its invariant distribution~\cite{tierney1994, UnderstandingMH}.

\subsection{Random-Walk Metropolis and its Extensions}
\label{section:MH_description}

The random-walk Metropolis, or Metropolis-Hastings (MH) algorithm~\cite{MetropolisMCMC, Hastings_MH, UnderstandingMH}, is one of the most popular and widely used MCMC techniques due to its simplicity and ease of implementation. In the MH algorithm, the transition kernel generating function is formed from a proposal density function $ q (\mathbf{y} | \mathbf{x}) $ and an acceptance probability $ \alpha (\mathbf{x}, \mathbf{y}) $ as
\begin{equation}
    \label{eqn:MH_transition_kernel_generating_function}
    k_{\text{MH}} (\mathbf{x}, \mathbf{y}) = q (\mathbf{y} | \mathbf{x}) \alpha (\mathbf{x}, \mathbf{y}) \; , \qquad \mathbf{x} \neq \mathbf{y}
\end{equation}
and the transition kernel itself is constructed as per Eq.~\eqref{eqn:transition_kernel_expansion}. The proposal density function encodes the probability of proposing $ \mathbf{y} $ as the subsequent state of the Markov chain, given that the chain is currently in state $ \mathbf{x} $, and must be chosen such that the resultant Markov chain satisfies the mild regularity conditions of \textit{irreducibility} and \textit{aperiodicity}. The acceptance ratio $ \alpha (\mathbf{x}, \mathbf{y}) $ is the probability that the proposed state $ \mathbf{y} $ is accepted as the next state of the Markov chain and guarantees time reversibility for $ k_{\text{MH}} (\mathbf{x}, \mathbf{y}) $, and takes the following form
\begin{equation}
    \label{eqn:acceptnace_ratio_MH}
    \alpha (\mathbf{x}, \mathbf{y}) = \begin{cases}
        \min \left[ \frac{\pi(\mathbf{y}) q (\mathbf{x} | \mathbf{y})}{\pi(\mathbf{x}) q (\mathbf{y} | \mathbf{x})} , 1 \right] & \text{if } \pi(\mathbf{x}) q (\mathbf{y} | \mathbf{x}) > 0 \\
        1 & \text{if } \pi(\mathbf{x}) q (\mathbf{y} | \mathbf{x}) = 0
    \end{cases}
\end{equation}

With these choices, the MH algorithm can be summarized in the following steps:
\begin{enumerate}
    \item Let the $ t $-th state of the chain $ \mathbf{x}_t = \mathbf{x} $.
    \item Randomly sample $ \mathbf{y} $ from $ q (\mathbf{y} | \mathbf{x}) $ and propose $ \mathbf{y} $ as the next state of the chain.
    \item Compute the acceptance probability $ \alpha (\mathbf{x}, \mathbf{y}) $.
    \item Set the $ (t+1) $-th state $ \mathbf{x}_{t+1} = \mathbf{y} $ with probability $ \alpha (\mathbf{x}, \mathbf{y}) $, otherwise $ \mathbf{x}_{t+1} = \mathbf{x} $.
\end{enumerate}

The quality of an MH algorithm depends on the appropriate choice of $ q (\mathbf{y} | \mathbf{x}) $, which influences the acceptance rate of proposed states (governing the number of unique samples in the chain) and the correlations between the successive states (governing how quickly the chain can explore the space). These two factors are usually at odds; a small correlation between states implies a large step size at each iteration, which can lead to a high rejection rate. Meanwhile, proposals that take small steps tend to increase the acceptance rate but significantly increase correlation and hamper exploration. The proposal distribution $ q (\mathbf{y} | \mathbf{x}) $ must be chosen to strike a balance between these factors. 

A common choice for $ q (\mathbf{y} | \mathbf{x}) $ is the $ d $-dimensional Gaussian distribution $ \mathcal{N}_d \left( \mathbf{x}, \mathbf{\Sigma} \right) $, as its symmetric nature also simplifies Eq.~\eqref{eqn:acceptnace_ratio_MH}. The covariance $ \mathbf{\Sigma} $ is typically chosen such that some, usually heuristic, optimal acceptance rate is achieved. Guidelines on selecting $ \mathbf{\Sigma} $ to achieve optimal acceptance rates under different conditions have been proposed in the literature~\cite{GelmanCarlinBayesianAnalysis, RobertsRosenthal2001}. Schemes have also been proposed to adaptively tune the proposal distribution to improve sampling efficiency, such as selecting $ \mathbf{\Sigma} $ based on the previous samples in the chain~\cite{AdaptiveMetropolisHaario1999, AdaptiveMetropolisHaario2001}. Another strategy to improve acceptance rates is delayed rejection~\cite{TierneyMira1999, mira2001metropolisDelayedRejection, green2001delayed}, whereby a new candidate is proposed from a modified proposal distribution when the initial candidate for the $ (t+1) $-th step is rejected, instead of the $ t $-th state being immediately replicated. Further, combining delayed rejection with adaptively tuned proposals (i.e. Delayed Rejection Adaptive Metropolis -- DRAM~\cite{haario2006dram}) has been shown to improve the performance of MH.

\begin{figure}[!htbp]
\centering
\begin{subfigure}{.45\textwidth}
  \centering
  \includegraphics[width=\linewidth]{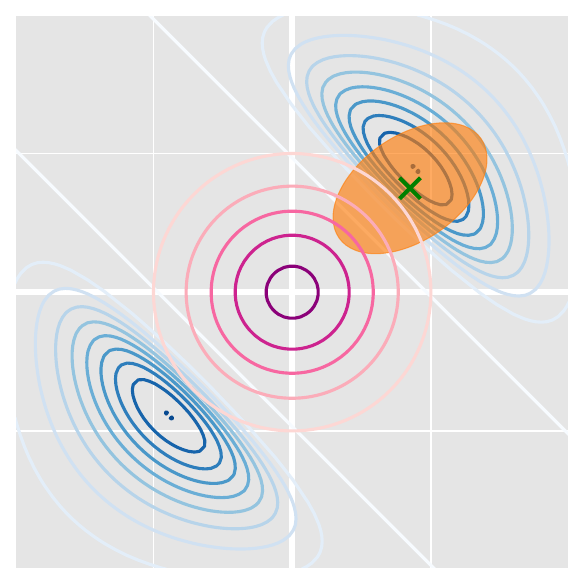}
  \caption{}
  \label{fig:exploitation_proposal}
\end{subfigure}%
\begin{subfigure}{.45\textwidth}
  \centering
  \includegraphics[width=\linewidth]{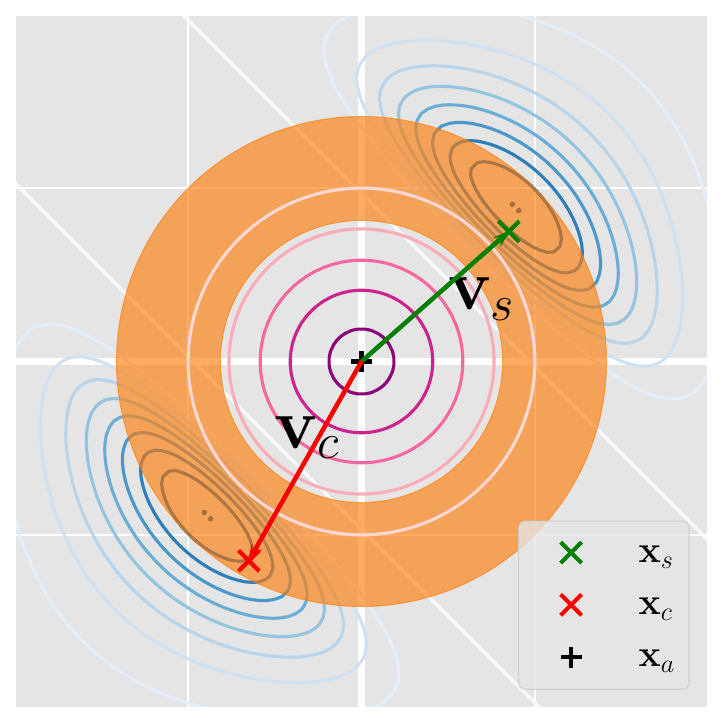}
  \caption{}
  \label{fig:exploration_proposal}
\end{subfigure}
\caption{Visualization of (a) a standard Metropolis-Hastings proposal, and (b) the Intrepid proposal. Both figures show the target distribution (which is bimodal with a valley of zero density between the modes) in blue contours and the parent distribution in red-purple contours. The green cross represents the current state of the Markov chain, and the orange-shaded region represents the area containing $ 99 \% $ of the proposal density. The difference in proposal shapes highlights why a standard MH proposal struggles to sample from multimodal distributions and how the Intrepid proposal overcomes this issue. Additionally, the anchor point, proposed candidate, seed vector $ \mathbf{v}_s $, and candidate vector $ \mathbf{v}_c $ for Intrepid MCMC, as defined in Section~\ref{section:Intrepid_proposal_theory} are shown in plot (b).}
\label{fig:proposal_visualization}
\end{figure}

Most applications of the MH algorithm use a proposal distribution that is unimodal (or uniform) and centered at the current state of the Markov chain. Unfortunately, this choice has a significant drawback for multimodal targets. Although theoretically such proposals often have non-zero density everywhere in $ \Omega $ (e.g., Gaussian proposal), in practice the proposed states nearly always lie within a bounded region centered at the current state. 
For example, consider the target distribution presented in Figure~\ref{fig:exploitation_proposal}, 
which is bimodal with a region of identically zero density separating the modes. Consider that the current state of the Markov chain is shown using the green cross and the shaded orange ellipse represents the region containing $ 99\% $ of the proposal density. 
Due to the large valley of zero density between the modes and the size of the $ 99 $-th quantile of the proposal distribution, it is easy to see why this chain can practically almost never explore the bottom mode. One may argue that the proposal density should be made larger, but if the bottom mode has never been discovered, such insights are not obvious as the proposal scale is appropriate for the mode that has been observed and even adaptive methods such as DRAM will fail to find the second mode.  

Clearly, vanilla MH methods are not well-suited to discovering new modes whose locations are unknown a priori. One approach to mitigate this issue is to use an ensemble of chains that communicate with each other. This is done by propagating a collection of $ m $ points, such that the resultant Markov chain $ \left\{ \mathbf{X}_t : t > 0 \right\} $ has as its stationary distribution $ \Pi (\mathbf{X}) = \left[ \pi (\mathbf{x}) \right]^m $
, where $ \mathbf{X}_t = \begin{bmatrix} \mathbf{x}_{1, t} & \dots & \mathbf{x}_{m, t} \end{bmatrix} $ and $ \mathbf{x}_{i, t} \in \Omega $ $ \forall i = 1, \dots, m $ and $ t > 0 $. Such chains can be constructed using algorithms from a class of ensemble methods
, such as the Affine Invariant samplers~\cite{goodman2010AffineInvariantMCMC} and those based on differential evolution algorithms (i.e. DE-MC~\cite{terBraak2006demc, terBraakVrugtDEMCSnooker2008} or DREAM~\cite{vrugt2009DREAM, vrugt2016MatlabDREAM}). However, even these methods do not have an explicit discovery or exploration capability because they adapt new samples based on existing observed states. Consequently, the proposed candidate for $ \mathbf{x}_{i, (t+1)} $ usually lies in the vicinity of an existing sample $ \mathbf{x}_{j, t} $. These algorithms are capable of jumping between modes, 
but new samples usually fall within a mode that's already been found by a different sample within the ensemble $ \mathbf{X}_t $. 
Thus, while these ensemble Markov chains cover more of the parameter space $ \Omega $ than a single chain, much of their exploratory ability 
depends on the dispersion of the samples that form the initial ensemble. If the initial ensemble misses an important mode, the ensemble MCMC is likely to miss that mode.

Some attention has been devoted to constructing algorithms that explicitly try to locate previously undiscovered modes. The Repelling-Attracting Metropolis (RAM) algorithm~\cite{TakMeng2018RAM} attempts to do so by first moving ``downhill'' on the target distribution before making an ``uphill'' move, with the idea that the downhill move may potentially break free of the current mode (or ``region of attraction''), such that the uphill step moves toward different mode. However, RAM requires a symmetric proposal and generation of multiple candidate points (therefore multiple, possibly expensive, evaluations of $ \pi (\mathbf{x}) $) for each step of the chain. The Adaptive Gaussian Mixture MH (AGM-MH) method~\cite{LuengoMartino2013} employs a mixture of Gaussians as the proposal distribution, with the parameters of the mixture tuned adaptively as the chain propagates. While potentially improving the acceptance rate, the AGM-MH is highly sensitive to initial conditions, requires the tuning of multiple parameters, and, again, does not have explicit exploration steps. 

Another popular concept that has spawned several algorithms is \textit{Tempering}~\cite{gilks1996ImprovingMCMC, geyerParallelTempering, miasojedow2013AdaptiveParallelTempering, marinari1992SimulatedTempering}, also called Replica Exchange~\cite{SHARMA2023103448}, where samples are generated from auxiliary distributions proportional to $ \left( \pi (\mathbf{x}) \right) ^\beta $ for stages of increasing $ \beta \in \left( 0, 1 \right) $. These auxiliary distributions are flatter/wider than the true target, making it easier for a Markov chain to explore and move between modes. Unfortunately, requiring samples from the auxiliary distributions implies that significant computational resources are expended on samples that are not representative of the target distribution. Additionally, tempering fails when the valley between modes has identically zero density. It has also been noted that tempering algorithms lose efficiency if the modes have different covariance structures~\cite{Woodard2009TemperingMixingConditions}. A final class of approaches first run an optimization algorithm to locate the modes of the target distribution, and then tune different proposals for each of the discovered modes~\cite{andricioaei2001SmartDartingMonteCarlo, sminchisescu2007GeneralizedDartingMC, ahn2013DistributedAdaptiveDartingMC, Pompe2020AdaptiveMultimodalMCMC}. In this case, no usable samples from $ \pi (\mathbf{x}) $ are generated during the optimization procedure, and since a multimodal distribution by definition has a (potentially very) non-convex density function, implementing a proper optimization procedure poses a whole new set of challenges.

\subsection{A Summary of the Proposed Method}
\label{section:proposed_method_summary}

In this paper, we showcase how a simple modification to the MH algorithm can markedly accelerate exploration of the parameter space $ \Omega $, and significantly improve its ability to locate previously undiscovered modes of the target distribution without prior knowledge. The foundational idea of the proposed method -- \textit{Intrepid MCMC} -- is to inject a fraction of highly exploratory steps into a traditional MH algorithm by using a mixture-type transition kernel, formed as the weighted sum of the so-called Intrepid proposal density and a traditional MH proposal density. The Intrepid proposal is specially constructed to rapidly sweep the parameter space by leveraging the probability structure of the parent distribution $ p(\mathbf{x}) $, thus incorporating additional known information about the target distribution $ \pi (\mathbf{x}) $ that is not harnessed by most MH methods. However, we further show that Intrepid MCMC can be used even in cases where a decomposition of the form of Eq.~\eqref{eqn:target_structure} does not exist (see Section~\ref{section:general_applicability}).

Notably, the popularity and versatility of MH is directly attributable to its simplicity. We therefore focus on retaining this simplicity in Intrepid MCMC. Our goal is to highlight how a straightforward implementation of the Intrepid proposal can lead to significant improvements in sampling efficiency for complex-shaped and multimodal distributions. Many of the more sophisticated ideas mentioned above -- such as adaptive proposal tuning, delayed rejection, and ensembling of multiple chains -- can be employed to further improve Intrepid MCMC without changing the basic concept presented here. We leave the exploration of such topics as future work.

Section~\ref{section:Intrepid_MCMC} discusses the intuition upon which Intrepid MCMC rests and goes on to present the theoretical details involved in the construction of the Intrepid proposal, including guarantees of irreducibility, aperiodicity, and reversibility of the resultant Markov chain. The Intrepid MCMC algorithm is also explicitly provided to guide implementation, along with some practical considerations for various use cases. A detailed study of the convergence behavior of Intrepid MCMC for a miscellany of multimodal target distributions, $ \pi (\mathbf{x}) $, is presented in Section~\ref{section:numerical_results}. Insights into the correlation structure of the Intrepid Markov chain and its scalability with increasing dimensions are also provided. The analysis is capped with an engineering example highlighting the method's applicability to real-world problems.

\section{Intrepid MCMC}
\label{section:Intrepid_MCMC}

The aim of Intrepid MCMC is to split the tasks of locally exploring a known mode and globally exploring to discover new ones between two distinct proposals, working in tandem to converge to the target distribution. A traditional MH proposal is employed exclusively to explore (exploit) already located modes. The new Intrepid proposal then focuses purely on locating new modes. This Intrepid proposal is designed to efficiently search the parameter space, $ \Omega $, making as few assumptions about the structure of the target density as possible.

Since Intrepid MCMC is a variant of MH, we utilize the equations and nomenclature defined in Section~\ref{section:MCMC}. We denote the transition kernel for Intrepid MCMC as $ \mathcal{K} (\mathbf{x}, d\mathbf{y}) $, 
where
\begin{equation}
\label{eqn:intrepid_transition_kernel}
    \mathcal{K} (\mathbf{x}, d\mathbf{y}) = \begin{cases}
        \mathcal{K}_L (\mathbf{x}, d\mathbf{y}) & \text{w.p. } 1 - \beta \\
        \mathcal{K}_I (\mathbf{x}, d\mathbf{y}) & \text{w.p. } \beta
    \end{cases}
\end{equation}

Eq.~\eqref{eqn:intrepid_transition_kernel} forms a mixture type kernel, as defined by Tierney~\cite{tierney1994}. It is formed from a \textit{locally exploitative transition kernel} $ \mathcal{K}_L (\mathbf{x}, d\mathbf{y}) $ and a \textit{globally explorative transition kernel} $ \mathcal{K}_I (\mathbf{x}, d\mathbf{y}) $, with $ \beta \in \left( 0, 1 \right) $ being the selection probability of the globally explorative kernel, referred to as the exploration ratio. Mixture kernels result in irreducible and aperiodic Markov chains as long as any of the individual kernels individually produce irreducible and aperiodic Markov chains~\cite{tierney1994}.

Following from Eq.~\eqref{eqn:transition_kernel_expansion}, we can write
\begin{align}
    \mathcal{K}_L (\mathbf{x}, d\mathbf{y}) &= \kappa_L (\mathbf{x}, \mathbf{y})d\mathbf{y} + r_L(\mathbf{x}) \delta_x (d\mathbf{y}) \label{eqn:CMH_kernel_expansion} \\
    \mathcal{K}_I (\mathbf{x}, d\mathbf{y}) &= \kappa_I (\mathbf{x}, \mathbf{y})d\mathbf{y} + r_I(\mathbf{x}) \delta_x (d\mathbf{y}) \label{eqn:intrepid_kernel_expansion}
\end{align}
where $ \kappa_L (\mathbf{x}, \mathbf{y}) $ and $ \kappa_I (\mathbf{x}, \mathbf{y}) $ are the transition kernel generating functions for $ \mathcal{K}_L (\mathbf{x}, d\mathbf{y}) $ and $ \mathcal{K}_I (\mathbf{x}, d\mathbf{y}) $, respectively, and $ r_L(\mathbf{x}) $, $ r_I(\mathbf{x}) $, and $ \delta_x (d\mathbf{y}) $ have definitions similar to those presented in Section~\ref{section:MCMC}. 

For $ \kappa_L (\mathbf{x}, \mathbf{y}) $, we recommend the \textit{Component-wise Metropolis-Hastings} approach, first introduced by Au and Beck~\cite{AuBeck2001SuS} as \textit{Modified Metropolis Hastings}. It is a single-chain variant of MH designed to reduce the number of replicated states in the Markov chain. 
The method uses different symmetric proposal densities (usually Gaussians) $ q_{L_i} ( y_i | x_i) $ for each dimension $ i \in \left\{ 1, \dots, d \right\} $, successively accepting or rejecting proposed values for each component of the candidate state. 
Component-wise Metropolis-Hastings is known to generate irreducible and aperiodic chains for appropriate proposal distributions; thus, Intrepid MCMC will satisfy these conditions as well, regardless of whether or not $ \mathcal{K}_I (\mathbf{x}, d\mathbf{y}) $ on its own results in irreducible and aperiodic Markov chains.

The construction of $ \kappa_I (\mathbf{x}, \mathbf{y}) $ follows the MH structure as well, with
\begin{equation}
\label{eqn:intrepid_kernel_generating_function}
    \kappa_I (\mathbf{x}, \mathbf{y}) = q_I (\mathbf{y} | \mathbf{x}) \alpha_I (\mathbf{x}, \mathbf{y})
\end{equation}
where $ q_I (\mathbf{y} | \mathbf{x}) $ is the Intrepid proposal, and $ \alpha_I (\mathbf{x}, \mathbf{y}) $ is the Intrepid acceptance probability. This ensures that the invariant distribution of the kernel $ \mathcal{K}_I (\mathbf{x}, d\mathbf{y}) $ is $ \pi (\mathbf{x}) $. The form of $ q_I (\mathbf{y} | \mathbf{x}) $ 
is motivated by the following observations that are readily apparent from the structure of $ \pi (\mathbf{x}) $ stipulated in Eq.~\eqref{eqn:target_structure}:
\begin{enumerate}
    \item If points $ \mathbf{x}_1, \mathbf{x}_2 \in \Omega $ have the same value of the transformation function, i.e., $ T (\mathbf{x}_1) = T (\mathbf{x}_2) $ and  $ \pi (\mathbf{x}_1) >  \pi (\mathbf{x}_2) $, then $ p (\mathbf{x}_1) > p (\mathbf{x}_2) $.
    \item If points $ \mathbf{x}_1, \mathbf{x}_2 \in \Omega $ have the same value of the parent density function, i.e., $ p (\mathbf{x}_1) = p (\mathbf{x}_2) $ and  $ \pi (\mathbf{x}_1) >  \pi (\mathbf{x}_2) $ then $ T (\mathbf{x}_1) > T (\mathbf{x}_2) $. 
     
\end{enumerate}
Assuming no knowledge of the behavior of $ T (\mathbf{x}) $ and that it is expensive to evaluate, we draw the following conclusions. Point 1 implies that MCMC sampling should adequately explore regions of higher density in the parent distribution $p(\mathbf{x})$ to ensure that $\pi(\mathbf{x})$ is sufficiently explored. Classical MH algorithms are typically reasonably good at sampling from higher density regions before moving to lower ones if the parent distribution is smooth, unimodal, and does not have very large gradients (as under the condition of point 1, $ \pi (\mathbf{x}) $ behaves the same as $ p (\mathbf{x}) $).
Point 2 implies that, given a current state $\mathbf{x}_t$ with $\pi(\mathbf{x}_t)=T(\mathbf{x}_t)p(\mathbf{x}_t)$, it is desirable to explore along approximately constant contours of $p(\mathbf{x}_t)$ to identify new regions, which may be far away in parameter space, but have similar probability $p(\mathbf{x})$ and potentially high values of $T(\mathbf{x})$. Conventional MH algorithms do not perform well in this regard because they involve potentially large proposal steps that must be guided by the parent distribution (existing MH methods are not). This is the goal of the Intrepid proposal step. For comparison with classical MH proposals, the 99\% probability range for a proposal density of this type is illustrated in Figure~~\ref{fig:exploration_proposal}. 



Algorithm~\ref{Algo:Intrepid_MCMC} demonstrates how the proposed mixture kernel $ \mathcal{K} (\mathbf{x}, d\mathbf{y}) $ from Eq.~\eqref{eqn:intrepid_transition_kernel} can be implemented in a general setting to run Intrepid MCMC for a chain of length $ N $, starting at the point $ \mathbf{x}_0 \in \Omega $ using notation as defined in Table~\ref{tab:general_notation}.

\begin{algorithm}[!ht]
\caption{Intrepid MCMC}
\label{Algo:Intrepid_MCMC}
\begin{codefont}
\begin{algorithmic}[1]
\Require $ \mathbf{x}_0 $, $ N $, $ \beta $
\State Create array SAMPLES
\State SAMPLES[1] $ \gets \mathbf{x}_0 $
\For{j = 2:$ N $}
\State $ \mathbf{x}_s \gets $ SAMPLES[j-1]
\State Draw $ u \sim \mathcal{U} (0, 1) $
\If{$ u \leq \beta $}
\State Generate $ \mathbf{x}_n $ from $ \mathbf{x}_s $ using $\kappa_I (\cdot)$, see Algorithm~\ref{Algo:Intrepid_Proposal}
\Else
\State Generate $ \mathbf{x}_n $ from $ \mathbf{x}_s $ using $\kappa_L (\cdot)$, i.e., using Component-wise MH.
\EndIf
\State SAMPLES[j] $ \gets \mathbf{x}_n $
\EndFor
\end{algorithmic}
\end{codefont}
\end{algorithm}

\begin{table}[!ht]
    \centering
    \begin{tabular}{c|c}
        Symbol & Description \\
        \hline \hline
        $ \mathbf{x}_s $ & Current state of the Markov chain \\
        $ \mathbf{x}_n $ & Next state of the Markov chain \\
        $ \mathbf{x}_c $ & Candidate proposed for next state of the Markov chain \\
        $ \mathbf{x}_a $ & Anchor point (see Section~\ref{section:Intrepid_proposal_theory}) \\
        $ r_i $ & Euclidean distance of $ \mathbf{x}_i $ from $ \mathbf{x}_a $ (i.e., $ \lVert \mathbf{x}_i - \mathbf{x}_a \rVert_2 $) \\
        $ \boldsymbol{\theta}_i $ & Unit direction vector pointing from $ \mathbf{x}_a $ towards $ \mathbf{x}_i $ (i.e., $ \left( \mathbf{x}_i - \mathbf{x}_a \right)/r_i $) \\
        $ \theta_{i, j} $ & $ j $-th angular coordinate of $ \boldsymbol{\theta}_i $ represented in $ d $-dimensional hyperspherical coordinates with origin at $ \mathbf{x}_a $ \\
        \hline
    \end{tabular}
    \caption{Some general notation for implementation of Intrepid MCMC. Each point $ \mathbf{x}_{(\cdot)} \in \Omega $.}
    \label{tab:general_notation}
\end{table}


\subsection{Theoretical Construction of the Intrepid Proposal}
\label{section:Intrepid_proposal_theory}

In order for the Intrepid proposal, $ q_I (\mathbf{y} | \mathbf{x}) $, to freely explore along the contours of $ p (\mathbf{x}) $, we first define a hyperspherical coordinate system.
The new coordinate system mentioned above is centered on a selected \textit{anchor point}, $ \mathbf{x}_a \in \Omega $, which is chosen a priori based on the geometry of $ p (\mathbf{x}) $, and is held constant for the duration of the propagation of the Intrepid Markov chain. 
More details on the selection of the anchor point are presented later in this section.

    
In this hyperspherical coordinate system, the coordinates 
for any point $ \mathbf{x} \neq \mathbf{x}_a $ can be expressed through the vector $ \mathbf{v} = \mathbf{x} - \mathbf{x}_a = \begin{bmatrix} r & \theta_1 & \dots & \theta_{(d-1)} \end{bmatrix} $ (notated in shorthand as $ \mathbf{v} = \left( r, \boldsymbol{\theta} \right) $, $ \boldsymbol{\theta} = \begin{bmatrix} \theta_{1} & \dots & \theta_{(d-1)} \end{bmatrix} $), where
\begin{equation}
\label{eqn:hyperspherical_coordinate_system_definition}
    \begin{aligned}
        r &= \lVert \mathbf{x} - \mathbf{x}_a \rVert_2 = \sqrt{\sum_{i=1}^d \left( x_i - x_{a, i} \right)^2} \\
        \theta_j &= \atantwo \left( \sqrt{\sum_{i=j+1}^d \left( x_i - x_{a, i} \right)^2}, \left( x_j - x_{a, j} \right) \right) \; , \qquad \forall j \in \left\{1, \dots, d-1 \right\}
    \end{aligned}
\end{equation}
Note that $ \theta_1, \dots, \theta_{(d-2)} \in \left[ 0, \pi \right] $, $ \theta_{(d-1)} \in \left[ 0, 2 \pi \right) $, and $ r \in \left[ 0, \infty \right) $. The coordinates of the current state of the Markov chain are then determined as $ r_s = \lVert \mathbf{x}_s - \mathbf{x}_a \rVert_2 $ and $ \boldsymbol{\theta}_s = \left( \mathbf{x}_s - \mathbf{x}_a \right)/r_s $, and represented by the vector $ \mathbf{v}_s = \left( r_s, \boldsymbol{\theta}_s \right) $. 
    
To generate a proposed/candidate state, we first generate its angular coordinates 
$ \boldsymbol{\theta}_c = \begin{bmatrix} \theta_{c, 1} & \dots & \theta_{c, (d-1)} \end{bmatrix} $
by adding a random perturbation angle $ \phi_j $ to each of the angular coordinates of the current state $ \theta_{s, j} $. This $ \phi_j $ is randomly sampled from an \textit{angular proposal distribution} $ \mathfrak{q}_{j} (\phi | \theta_{s, j}) $ $ \forall j \in \left\{ 1, \dots, (d-1) \right\} $. That is,
    \begin{equation}
        \begin{aligned}
            \theta_{c, j} &= \theta_{s, j} + \phi_j      \\
            \phi_j &\sim \mathfrak{q}_{j} (\phi | \theta_{s, j})    
        \end{aligned}
        \label{eqn:candidate_direction_component_angle}
    \end{equation}
Section~\ref{section:Intrepid_proposal_selection} discusses how to choose these angular proposal distributions in practice.
    
Next, we generate the radial magnitude of the candidate $ r_c $ as
    \begin{equation}
    \label{eqn:intrepid_candidate_radius}
        r_c = \begin{cases}
            R_{0, c} \left( \gamma R_{s, 0} \left( r_s \right) \right) & \text{if } R_{ \left[ \cdot \right] } (\cdot) \text{ exists} \\
            \gamma r_s & \text{otherwise}
        \end{cases}
    \end{equation}
    This involves the following quantities that require some explanation:
    \begin{enumerate}
        \item The \textit{Radial Transformation Function} (RTF), $ R_{1, 2}(r_1) $, is a function whose output is the radial magnitude $ r_2 $ such that the points $ \left( \mathbf{x}_a + \mathbf{v}_2 \right) $ and $ \left( \mathbf{x}_a + \mathbf{v}_1 \right) $ lie on the same contour of $ p (\mathbf{x}) $ (i.e. $p\left( \mathbf{x}_a + \mathbf{v}_2 \right) = p\left( \mathbf{x}_a + \mathbf{v}_1 \right)$). Note that, for simplicity in notation, the subscript $_{1,2}$ indexes the dependence on $ \boldsymbol{\theta}_1, \boldsymbol{\theta}_2$. In other words, the function $ R_{1, 2}(r_1) $ determines the radial coordinate $ r_2 $ for the vector $ \mathbf{v}_2 =\left( r_2, \boldsymbol{\theta}_2 \right) $ such that the points  $ \mathbf{x}_a + \mathbf{v}_1 $ and $ \mathbf{x}_a + \mathbf{v}_2 $ have equal probability
        according to the parent distribution  $ p (\mathbf{x}) $.
        To be of practical use, the RTF must be continuous and differentiable, having derivative $R_{1, 2}'(r) = \left.\dfrac{dR_{1,2}}{dr}\right|_{\boldsymbol{\theta}_1}$. The conditions for the existence and the construction of the RTF are discussed in detail in Appendix~\ref{appendix:PRE_and_RTF}, with some useful cases discussed later in this section.
        
        \item The \textit{radial perturbation factor}, $ \gamma $, is a random variable drawn from a \textit{radial proposal distribution} $ \mathfrak{q}_r (\gamma | \boldsymbol{\theta}_c) $, i.e., $ \gamma \sim \mathfrak{q}_r (\gamma | \boldsymbol{\theta}_c) $, and defines a random perturbation from the proposal distribution contour. Section~\ref{section:Intrepid_proposal_selection} discusses how to choose this radial proposal distribution in practice.

        \item The \textit{reference direction}, $ \boldsymbol{\theta}_0 $, is an angle vector used as an intermediate step in the candidate generation procedure. The radial perturbation happens along the reference direction to ensure the Markov chain can make reverse jumps even with the (usually) highly nonlinear radial transformation. This preserves reversibility when the RTF exists by allowing the radial proposal distribution to be independent of the candidate direction $ \boldsymbol{\theta}_c $, i.e., $ \mathfrak{q}_r (\gamma | \boldsymbol{\theta}_c) \equiv \mathfrak{q}_r (\gamma) $, as discussed in Appendix~\ref{appendix:intrepid_proposal_and_acceptance_derivation}. The reference direction is selected a priori and kept constant throughout the propagation of the Markov chain, but it is used only when the RTF exists. 
        If the RTF does not exist, the radial perturbation happens along the candidate direction. Additionally, some of the most common classes of parent distributions have RTFs of much simpler forms, which makes the reference direction unnecessary in practice (see Section~\ref{section:Anchor_and_RTF}). For all the examples in Section~\ref{section:numerical_results}, the reference direction was not necessary to implement due to the use of Gaussian parent distributions.
    \end{enumerate}
        
Together the radial and angular coordinates of the candidate define the vector $ \mathbf{v}_c =\left( r_c, \boldsymbol{\theta}_c \right) $. Finally, the candidate point is given by \footnote{Note a small abuse of notation in Eq.~\eqref{eqn:intrepid_candidate} and previous expressions. The vector $ \mathbf{v}_c$, 
which is defined in hyperspherical coordinates, must be converted back into Cartesian coordinates before being added to $ \mathbf{x}_a $. Similar notation is used in the entirety of this manuscript in the interest of conciseness; the coordinate system representation of a point can be inferred from context.}
        \begin{equation}
            \label{eqn:intrepid_candidate}
            \mathbf{x}_c = \mathbf{x}_a + \mathbf{v}_c 
        \end{equation}

Based on the steps above, the full form of the Intrepid proposal $ q_I (\mathbf{y} | \mathbf{x}) $ can be written in terms of current state $ \mathbf{x}_s $ and proposed candidate $ \mathbf{x}_c $ (or equivalently their hyperspherical representations $ \mathbf{v}_s $ and proposed candidate $ \mathbf{v}_c $) as follows
\begin{equation}
    \label{eqn:intrepid_proposal_full_form}
    q_I (\mathbf{x}_c | \mathbf{x}_s) = q_I (\mathbf{v}_c | \mathbf{v}_s) = \begin{dcases}
        \frac{\mathfrak{q}_r (\gamma) \left[ \prod_{j=1}^{d-1} \mathfrak{q}_{j} (\phi_j | \theta_{s, j}) \right]}{\left[ \left\{ r_c \right\}^{d-1} R_{0, c}' \left( R_{c, 0} \left( r_c \right) \right) R_{s, 0} \left( r_s \right) \right] \left[ \prod_{j=1}^{d-1} \sin^{\left( d-j-1 \right)} \left( \theta_{c, j} \right) \right]} & \text{if } R_{ \left[ \cdot \right] } (\cdot) \text{ exists} \\
        \frac{ \mathfrak{q}_r (\gamma | \boldsymbol{\theta}_c) \left[ \prod_{j=1}^{d-1} \mathfrak{q}_{j} (\phi_j | \theta_{s, j}) \right]}{ \gamma ^{(d-1)} r_s^d \left[ \prod_{j=1}^{d-1} \sin^{\left( d-j-1 \right)} \left( \theta_{c, j} \right) \right]} & \text{otherwise}
    \end{dcases}
\end{equation}
Eq.~\eqref{eqn:intrepid_proposal_full_form} is derived in Appendix~\ref{appendix:intrepid_proposal_and_acceptance_derivation}. Pseudo-code associated with the above procedure is presented in Algorithm~\ref{Algo:Intrepid_Proposal}.

\begin{algorithm}[!ht]
\caption{Exploration using the Intrepid Proposal}
\label{Algo:Intrepid_Proposal}
\begin{codefont}
\begin{algorithmic}[1]
\Require $ \mathbf{x}_s $, $ \mathbf{x}_a $, $ \mathfrak{q}_{j} (\phi | \theta_{s, j}) $ $ \forall j \in \left\{ 1, \dots, (d-1) \right\} $, $ \mathfrak{q}_r (\gamma | \boldsymbol{\theta}) $ $ \forall \boldsymbol{\theta} $
\Statex \Comment{See definitions in Section~\ref{section:Intrepid_proposal_theory}}
\Require The RTF $ R_{1, 2} (r_1) $ for any two directions $ \boldsymbol{\theta}_1, \boldsymbol{\theta}_2 $ if it exists.
\State Compute $ r_s = \lVert \mathbf{x}_s - \mathbf{x}_a \rVert_2 $
\State Compute $ \boldsymbol{\theta}_s = \left( \mathbf{x}_s - \mathbf{x}_a \right) / r_s $
\For{j = 1:$ (d - 1) $}
\State Draw $ \phi_j \sim \mathfrak{q}_{j} (\phi | \theta_{s, j}) $     
\State Compute $ \theta_{c, j} = \theta_{s, j} + \phi_j $     
\EndFor
\State Construct $ \boldsymbol{\theta}_c = \begin{bmatrix} \theta_{c, 1} & \dots & \theta_{c, (d-1)} \end{bmatrix} $
\State Draw $ \gamma \sim \mathfrak{q}_r (\gamma | \boldsymbol{\theta}_c) $ or $ \gamma \sim \mathfrak{q}_r (\gamma) $ as appropriate
\If{The RTF exists}
\State Compute $ r_c = R_{0, c} \left( \gamma R_{s, 0} \left( r_s \right) \right) $
\Else
\State Compute $ r_c = \gamma r_s $
\EndIf
\State Define $ \mathbf{v}_c = \left( r_c, \boldsymbol{\theta}_c \right) $
\State $ \mathbf{x}_c = \mathbf{x}_a + \mathbf{v}_c $
\State Compute $ \alpha_I (\mathbf{x}_s, \mathbf{x}_c) $ \Comment{From Eq.~\eqref{eqn:Intrepid_acceptance_rate}}
\State Draw $ u \sim \mathcal{U} (0, 1) $
\If{$ u \leq \alpha_I (\mathbf{x}_s, \mathbf{x}_c) $}
\State $ \mathbf{x}_n = \mathbf{x}_c $
\Else
\State $ \mathbf{x}_n = \mathbf{x}_s $
\EndIf
\State Return $ \mathbf{x}_n $
\end{algorithmic}
\end{codefont}
\end{algorithm}

\subsection{Selection of Anchor Point and Radial Transformation Function}
\label{section:Anchor_and_RTF}

Although the Intrepid MCMC methodology is valid for any arbitrary choice of the anchor point, careful selection of $ \mathbf{x}_a $ can make its implementation significantly easier. The anchor point should be chosen such that 
the behavior of the parent distribution $ p (\mathbf{x}) $ along different radial directions emanating from $ \mathbf{x}_a $ is as similar as possible. Usually, a good choice for the anchor point is some notion of centrality of $ p ( \mathbf{x} ) $, such as the mean/median/mode, depending on the form of $ p ( \mathbf{x} ) $. In ideal cases, all equal probability contours of the parent distribution $ p (\mathbf{x}) $ are convex and enclose $ \mathbf{x}_a $, ensuring the existence of the RTF (which depends on the anchor point, see Appendix~\ref{appendix:PRE_and_RTF}). To maintain conciseness and clarity of presentation, in this section we only discuss anchor point selection and RTF construction for the following three cases of practical interest. Further details, including mathematically rigorous constructions and existence criteria for the RTF, are presented in Appendix~\ref{appendix:PRE_and_RTF}.

\begin{enumerate}
    \item \textbf{\textit{Radially Symmetric Distributions}}  -- e.g., Gaussian distribution -- have a density function that only depends on the distance of any point from a fixed point that we select as the anchor point, i.e., $ p (\mathbf{x}) \equiv p \left( \lVert \mathbf{x} - \mathbf{x}_a \rVert_2 \right) $ (e.g., for the Gaussian distribution the mean is chosen as the anchor point). 
    This results in 
    $ R_{1, 2} (r) = r $ ($ R'_{1, 2} (r) = 1 $) for any two directions $ \boldsymbol{\theta}_1 $ and $ \boldsymbol{\theta}_2 $ (i.e., the RTF is the identity function).
    
    \item \textbf{\textit{Uniform Distributions}} have constant density $ p(\mathbf{x}) = p_0 $ $ \forall \mathbf{x} \in \Omega $. 
    Regardless of the choice of $ \mathbf{x}_a $, The RTF is given by
    \begin{equation}
    \label{eqn:RTF_for_uniform}
        R_{1, 2} (r) = r \left( \frac{\lambda_2}{\lambda_1} \right)
    \end{equation}
    where
    \begin{equation}
    \label{eqn:uniform_parent_radial_extent}
        \lambda_i = \max_r \left( r \; | \; p \left( \mathbf{x}_a + \mathbf{v}_i \right) > 0 \right) \quad \text{given } \mathbf{v}_i = (r, \boldsymbol{\theta}_i)
    \end{equation}
    For practical purposes, we recommend selecting 
    the mean of the distribution as $ \mathbf{x}_a $.
    
    \item \textbf{\textit{Unimodal Distribution with Convex Contours.}} -- e.g., Gumbel distribution -- have density functions $ p (\mathbf{x}) $ that are monotonically decreasing in every radial direction from the mode (they have no plateaus). Every equal probability contour, therefore, forms a convex shape containing the mode. Theorem~\ref{them:RTF_definition_contour_based} in Appendix~\ref{appendix:PRE_and_RTF} proves that an RTF exists when $ \mathbf{x}_a $ is chosen as the mode. 
    In this case,
    \begin{equation}
    \label{eqn:RTF_for_unimodal_convex}
        R_{1, 2} (r) = \Psi_{\boldsymbol{\theta}_2}^{-1} \left( \Psi_{\boldsymbol{\theta}_1} (r) \right)
    \end{equation}
    where $ \Psi_{\boldsymbol{\theta}} (r) $ is the \textit{unnormalized radial conditional} of $ p (\mathbf{x}) $ \footnote{Note that $ \Psi_{\boldsymbol{\theta}}^{-1} \left( r \right) $ exists, as $ \Psi_{\boldsymbol{\theta}} \left( r \right) $ is continuous and monotonically decreasing $ \forall \boldsymbol{\theta} $ in this case.}, defined as
    \begin{equation}
        \label{eqn:radial_conditional}
        \Psi_{\boldsymbol{\theta}} \left( r \right) = p \left( \mathbf{x}_a + \mathbf{v} \right) \quad \text{given } \mathbf{v} = (r, \boldsymbol{\theta})
    \end{equation}
\end{enumerate}
For other classes of parent distributions (e.g. multimodal distributions), the RTF often doesn't exist. In some cases, even if an RTF exists, it might be too cumbersome to implement. In such cases, the same goal of sampling approximately equal probability contours can be achieved by carefully varying the radial proposal distribution $ \mathfrak{q}_r (\gamma | \boldsymbol{\theta}) $ with the direction $ \boldsymbol{\theta} $. 
However, the development of such proposals is beyond the scope of this work. Similarly, when no natural choice for the anchor point exists, it can be drawn at random from the parent distribution, i.e., $ \mathbf{x}_a \sim p (\mathbf{x}) $, but we again leave the exploration of this strategy for future work.


\subsection{Theoretical Construction of the Intrepid Acceptance Ratio}
\label{section:Intrepid_acceptance_ratio_theory}

Building from the classical MH acceptance probability in Eq.~\eqref{eqn:acceptnace_ratio_MH}, we can write $ \alpha_I (\mathbf{x}, \mathbf{y}) $ in terms of current state $ \mathbf{x}_s $ and proposed candidate $ \mathbf{x}_c $ as
\begin{equation}
\label{eqn:Intrepid_acceptance_rate}
    \alpha_I (\mathbf{x}_s, \mathbf{x}_c) = \begin{cases}
        \min \left[ \rho_I (\mathbf{x}_s, \mathbf{x}_c) , 1 \right] & \text{if } \pi(\mathbf{x}_s) q_I (\mathbf{x}_c | \mathbf{x}_s) > 0 \\
        1 & \text{if } \pi(\mathbf{x}_s) q_I (\mathbf{x}_c | \mathbf{x}_s) = 0 
    \end{cases}
\end{equation}
where, substituting in Eq.~\eqref{eqn:intrepid_proposal_full_form}, we write the ratio $ \rho_I (\mathbf{x}_s, \mathbf{x}_c) $ as
\begin{align}
    \rho_I (\mathbf{x}_s, \mathbf{x}_c) &= \frac{\pi(\mathbf{x}_c) q_I (\mathbf{x}_s | \mathbf{x}_c)}{\pi(\mathbf{x}_s) q_I (\mathbf{x}_c | \mathbf{x}_s)} \label{eqn:rho_original_defn} \\
    &= \Gamma \frac{\pi(\mathbf{x}_c)}{\pi(\mathbf{x}_s)} \left[ \prod_{j=1}^{d-1} \frac{\mathfrak{q}_{j} (- \phi_j | \theta_{c, j})}{\mathfrak{q}_{j} (\phi_j | \theta_{s, j})} \right] \left[ \prod_{j=1}^{d-1} \frac{\sin^{\left( d-j-1 \right)} \left( \theta_{c, j} \right)}{\sin^{\left( d-j-1 \right)} \left( \theta_{s, j} \right)} \right] \label{eqn:rho_intrepid_defn} \\
    \text{where} \quad \Gamma &= \begin{dcases*}
        \frac{1}{\gamma} \frac{\mathfrak{q}_r (\nicefrac{1}{\gamma})}{\mathfrak{q}_r (\gamma)} \left[ \frac{r_c}{r_s} \right]^{(d-1)} \left[ \frac{R_{0, c}' \left( R_{c, 0} \left( r_c \right) \right)}{R_{0, s}' \left( R_{s, 0} \left( r_s \right) \right)} \right] & If RTF exists \\
        \gamma^{(d-2)} \frac{\mathfrak{q}_r \left( \left( \nicefrac{1}{\gamma} \right) | \boldsymbol{\theta}_s \right)}{\mathfrak{q}_r (\gamma | \boldsymbol{\theta}_c)} & If RTF does not exist
    \end{dcases*} \label{eqn:Gamma_defn_for_rho}
\end{align}
The derivations of Eqs.~\eqref{eqn:rho_intrepid_defn} and~\eqref{eqn:Gamma_defn_for_rho} are provided in Appendix~\ref{appendix:intrepid_proposal_and_acceptance_derivation}. A special case occurs for radially symmetric parent distributions (as described in the previous section), where $ R_{1, 2} (r) = r $. In this case, $ \Gamma $ simplifies to
\begin{equation}
\label{eqn:gamma_for_radially_symmetric_parent}
    \Gamma = \gamma^{(d-2)} \frac{\mathfrak{q}_r (\nicefrac{1}{\gamma})}{\mathfrak{q}_r (\gamma)}
\end{equation}
which is the same as when the RTF is not used, except that the radial proposal distribution $ \mathfrak{q}_r (\gamma) $ does not need to depend on the direction $ \boldsymbol{\theta}_c $ while still providing the ability to propose candidates near the contour $ \mathbf{x}_s $ lies on. A similar simplification happens when the parent distribution is uniform, which results in
\begin{equation}
    \label{eqn:gamma_for_uniform_parent}
    \Gamma = \gamma^{(d-1)} \frac{\mathfrak{q}_r (\nicefrac{1}{\gamma})}{\mathfrak{q}_r (\gamma)} \left[ \frac{\lambda_c}{\lambda_s} \right]^{d}
\end{equation}

\subsection{A Practical Guide to Intrepid Proposal Selection}
\label{section:Intrepid_proposal_selection}

Selecting a good proposal distribution is an essential step in any MH algorithm. In this section, we present symmetries in $ \mathfrak{q}_j (\phi | \theta_j) $ and $ \mathfrak{q}_r (\gamma) $ that simplify the form of $ \alpha_I (\mathbf{x}_s, \mathbf{x}_c) $, and suggest distributions that follow those symmetries. Strategies can then be explored for optimal tuning of these distributions for different choices of their form to maximize exploration efficiency, but we leave this for future work.

    \textbf{\textit{Symmetry of Angular Proposal Distribution.}} If the angular proposal has the following symmetry
    \begin{equation}
        \label{eqn:angular_distribution_symmetry}
        \mathfrak{q}_j \left( - \phi | \theta_{c, j} \right) = \mathfrak{q}_j \left( \phi | \theta_{s, j} \right) \quad \forall j \in \left\{ 1, \dots, d-1 \right\},
    \end{equation}
    the ratio $ \rho_I (\mathbf{x}_s, \mathbf{x}_c) $ from Eq.~\eqref{eqn:rho_original_defn} simplifies to
    \begin{equation}
    \label{eqn:rho_for_angular_proposal_symmetry}
        \rho_I (\mathbf{x}_s, \mathbf{x}_c) = \Gamma \frac{\pi(\mathbf{x}_c)}{\pi(\mathbf{x}_s)} \left[ \prod_{j=1}^{d-1} \frac{\sin^{\left( d-j-1 \right)} \left( \theta_{c, j} \right)}{\sin^{\left( d-j-1 \right)} \left( \theta_{s, j} \right)} \right]
    \end{equation}
    with $ \Gamma $ as defined in Eq.~\eqref{eqn:Gamma_defn_for_rho}.
    
    The following choices possess this symmetry:~\footnote{$ \text{TruncNorm} \left( 0, \sigma, \left[ a, b \right] \right) $ is a truncated normal distribution, with density function proportional to $ \mathcal{N} (0, \sigma) $ within the interval $ [a, b] $ and $ 0 $ elsewhere. An alternative to the truncated normal as defined here is the cicular normal, or von Mises distribution.}
    \begin{equation}
        \begin{aligned}
            \label{eqn:angular_proposal_symmetric_choices}
                \mathfrak{q}_j \left( \phi | \theta_j \right) &= \begin{dcases}
                    \text{Uniform} \left( - \theta_j , \pi - \theta_j \right) \\
                    \text{TruncNorm} \left( 0, \sigma_j, \left[ - \theta_j , \pi - \theta_j \right] \right)
                \end{dcases} \quad  \forall j \in \left\{ 1, \dots, d-2 \right\} \\
                \mathfrak{q}_{(d-1)} \left( \phi | \theta_{(d-1)} \right) &= \begin{dcases}
                    \text{Uniform} \left( - \theta_{(d-1)} , 2 \pi - \theta_{(d-1)} \right) \\
                    \text{TruncNorm} \left( 0, \sigma_{(d-1)}, \left[ - \theta_{(d-1)} , 2 \pi - \theta_{(d-1)} \right] \right)
                \end{dcases}
        \end{aligned}
    \end{equation}
    Anecdotally, values of $ \sigma_{(d-1)} = \pi $ and $ \sigma_j = \pi/2,  \forall j \in \left\{ 1, \dots, d-2 \right\} $ have been observed to perform well.
    
    \textbf{\textit{Symmetry of Radial Proposal Distribution.}} For any value of $ k $, the relationship of the form
    \begin{equation}
        \label{eqn:radial_distribution_symmetry}
        \mathfrak{q}_r (\gamma) = \gamma^k \mathfrak{q}_r (\nicefrac{1}{\gamma})
    \end{equation}
    simplifies $ \rho_I (\mathbf{x}_s, \mathbf{x}_c) $ by simplifying $ \Gamma $ as follows
    \begin{equation}
    \label{eqn:Gamma_for_radial_proposal_symmetry}
        \Gamma = \begin{dcases}
        \gamma^{(d-k-2)} & \text{If } R_{1, 2} (r) = r \\
        \gamma^{(d-k-1)} \left[ \frac{\lambda_c}{\lambda_s} \right]^{d} & \text{If } R_{1, 2} (r) = r \left( \nicefrac{\lambda_2}{\lambda_1} \right) \quad \text{($ \lambda_i $ as defined in Eq.~\eqref{eqn:uniform_parent_radial_extent}} \\
        \frac{1}{\gamma^{(k+1)}} \left[ \frac{r_c}{r_s} \right]^{(d-1)} \left[ \frac{R_{0, c}' \left( R_{c, 0} \left( r_c \right) \right)}{R_{0, s}' \left( R_{s, 0} \left( r_s \right) \right)} \right] & \text{If RTF has some other form}
        \end{dcases}
    \end{equation}
    Leveraging a similar symmetry for cases when the RTF doesn't exist is more complicated, as the radial proposal distribution can depend on the direction. 
    
    A distribution of the following form possesses this symmetry,
    \begin{equation}
        \label{eqn:radial_proposal_symmetric_choices}
        \mathfrak{q}_r (\gamma) = \begin{cases}
            \frac{\left( k + 2 \right) \sqrt{\gamma_0}^{(k+2)}}{2 \left( \gamma_0^{(k+2)} - 1 \right)} \gamma^{\nicefrac{k}{2}} & \gamma \in \left[ \frac{1}{\gamma_0}, \gamma_0 \right] , \; \gamma_0 > 1 \\
            0 & \text{otherwise}
        \end{cases}
    \end{equation}
    and, in particular, we recommend
    \begin{equation}
    \label{eqn:radial_proposal_uniform}
        \mathfrak{q}_r (\gamma) = \text{Uniform} \left( \nicefrac{1}{\gamma_0} , \gamma_0 \right) , \quad \gamma_0 > 1
    \end{equation}
    with $ \gamma_0 = 2 $ showing promising results.

\subsection{Applicability to more General Problems}
\label{section:general_applicability}

In cases where $ \pi (\mathbf{x}) $ has no natural decomposition of the form presented in Eq.~\eqref{eqn:target_structure}, a simple trick can make Intrepid MCMC viable. We are free to arbitrarily select any parent distribution $ p (\mathbf{x}) $ in this case, and define $ T (\mathbf{x}) $ as
\begin{equation}
    \label{eqn:T_constructed}
    T (\mathbf{x}) = \frac{\pi (\mathbf{x})}{p (\mathbf{x})}
\end{equation}

Naturally, the performance of Intrepid MCMC will depend strongly on the chosen parent distribution. We must be careful to select $ p (\mathbf{x}) $ that encourages Intrepid MCMC to both explore and exploit the geometry of $ \pi (\mathbf{x}) $ to the greatest extent. Although we won't explore this further here, our goal here is to highlight that Intrepid MCMC can be applied even in cases that do not strictly follow the structure in Eq.~\eqref{eqn:target_structure}. 


\section{Numerical Results}
\label{section:numerical_results}

In this section, we study the convergence and mixing behavior of Intrepid MCMC through a variety of examples. Component-wise Metropolis-Hastings (CMH) is used 
as the benchmark to compare against the proposed algorithm's performance.\footnote{This is equivalent to comparing to the case where $ \beta $ (as in Section~\ref{section:Intrepid_MCMC}) is zero.} Note that we intentionally do not compare against ensemble/multi-chain MCMC methods because the intention of this work is to demonstrate that exploration can be achieved within a single random-walk Intrepid Markov chain, while it cannot be achieved with conventional random-walk MH. Extension of the proposed method to ensembles of chains remains a topic for future work. Section~\ref{section:distribution_shape_results} analyzes Intrepid MCMC's ability to sample efficiently from target distributions with non-convex shapes and/or multiple disconnected modes. We then explore its robustness with increasing dimensionality in Section~\ref{section:dimension_results}. Next, we present some insights into the mixing behavior of Intrepid MCMC in Section~\ref{section:intrepid_chain_behavior}. Finally, in Section~\ref{section:bayesian_example}, we apply Intrepid MCMC to a Bayesian parameter inference problem on a 2-degree-of-freedom oscillator under free vibration.

Results in this section are presented using violin plots. Scott's rule is used for bandwidth selection, and the violins are cut off at the measured extremes in the data.
The plot widths are normalized to the number of data points used to construct the violins; thus, the plots are exaggerated to highlight the shape so that multiple peaks in the data can be spotted and, consequently, do not depict density values. This presentation of the results is intended to convey general information about the trends and `groupings' in the measured data.

\subsection{Analytical Multimodal Distributions}
\label{section:distribution_shape_results}

\begin{figure}[!ht]
\centering
\begin{subfigure}{.32\textwidth}
  \centering
  \includegraphics[width=\linewidth]{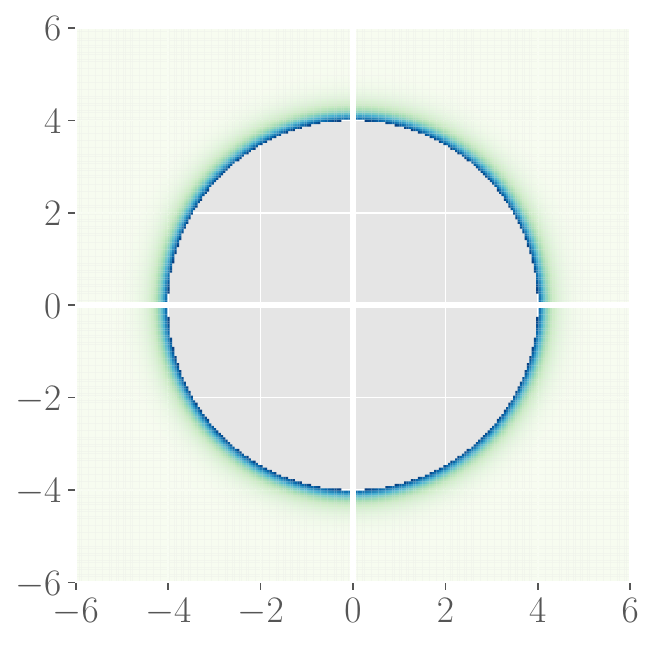}
  \caption{}
  \label{fig:gauss_ring_target}
\end{subfigure}%
\begin{subfigure}{.32\textwidth}
  \centering
  \includegraphics[width=\linewidth]{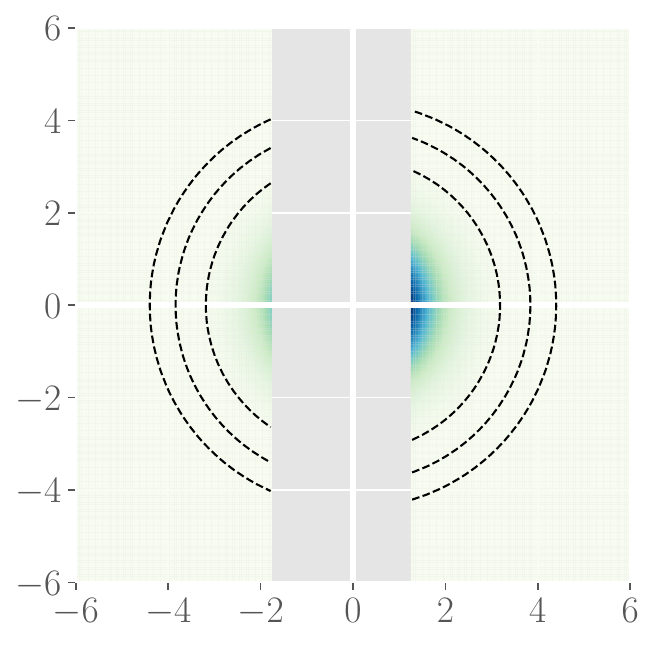}
  \caption{}
  \label{fig:gauss_planes_target}
\end{subfigure}%
\begin{subfigure}{.32\textwidth}
  \centering
  \includegraphics[width=\linewidth]{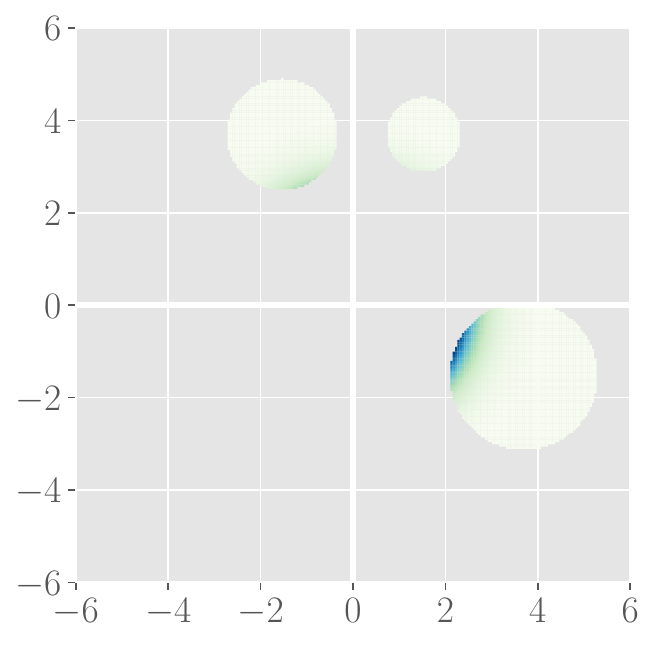}
  \caption{}
  \label{fig:gauss_circles_target}
\end{subfigure}
\begin{subfigure}{.32\textwidth}
  \centering
  \includegraphics[width=\linewidth]{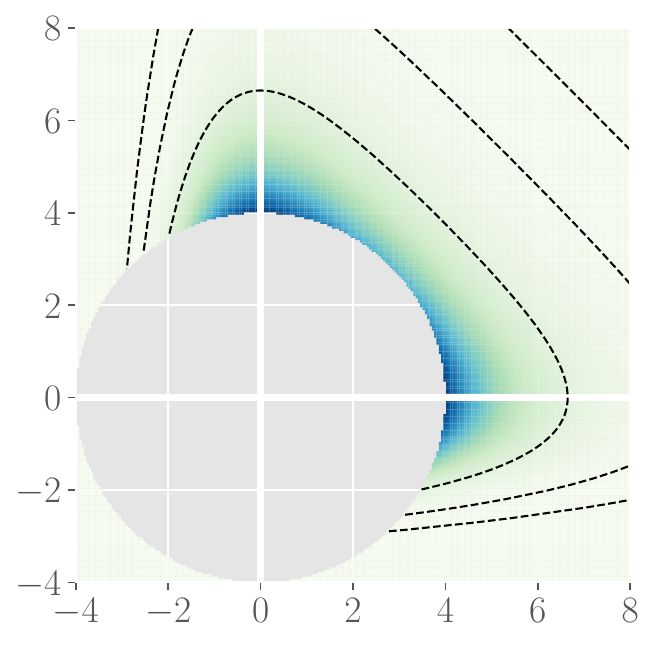}
  \caption{}
  \label{fig:gumbel_ring_target}
\end{subfigure}%
\begin{subfigure}{.32\textwidth}
  \centering
  \includegraphics[width=\linewidth]{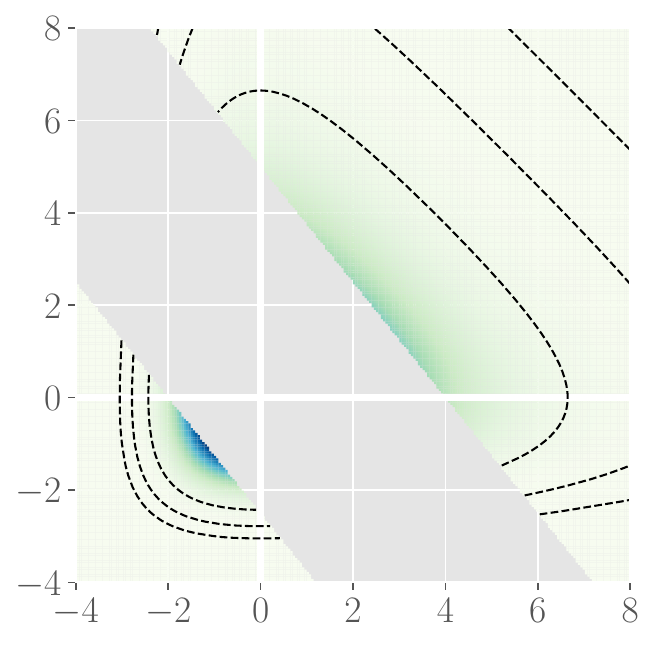}
  \caption{}
  \label{fig:gumbel_planes_target}
\end{subfigure}%
\begin{subfigure}{.32\textwidth}
  \centering
  \includegraphics[width=\linewidth]{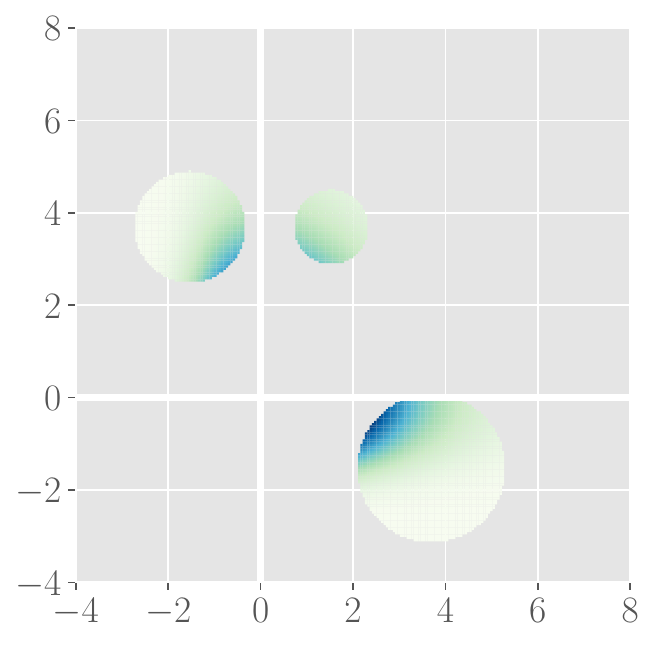}
  \caption{}
  \label{fig:gumbel_circles_target}
\end{subfigure}
\begin{subfigure}{.32\textwidth}
  \centering
  \includegraphics[width=\linewidth]{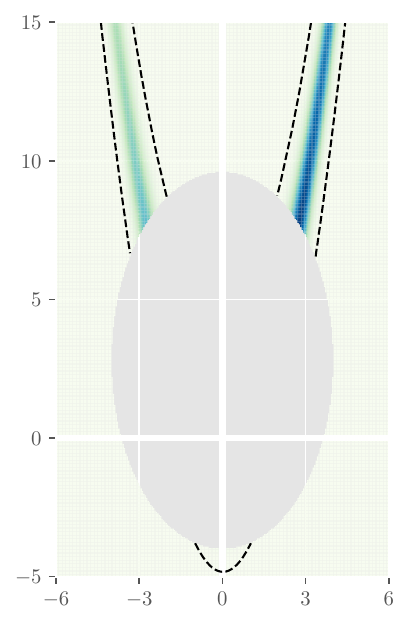}
  \caption{}
  \label{fig:rosenbrock_ring_target}
\end{subfigure}%
\begin{subfigure}{.32\textwidth}
  \centering
  \includegraphics[width=\linewidth]{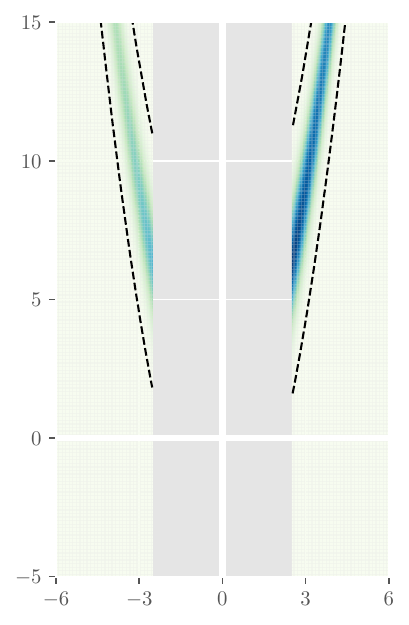}
  \caption{}
  \label{fig:rosenbrock_planes_target}
\end{subfigure}%
\begin{subfigure}{.32\textwidth}
  \centering
  \includegraphics[width=\linewidth]{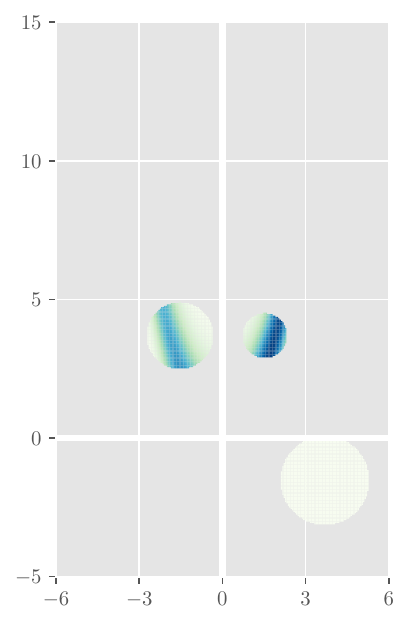}
  \caption{}
  \label{fig:rosenbrock_circles_target}
\end{subfigure}
\caption{Target distributions used in Section~\ref{section:distribution_shape_results}. (a) Case 1 (Gauss-Ring), (b) Case 2 (Gauss-Planes), (c) Case 3 (Gauss-Circles), (d) Case 4 (Gumbel-Ring), (e) Case 5 (Gumbel-Planes), (f) Case 6 (Gumbel-Circles), (g) Case 7 (Rosenbrock-Ring), (h) Case 8 (Rosenbrock-Planes), and (i) Case 9 (Rosenbrock-Circles).  Unshaded regions have identically zero target density.}
\label{fig:target_visualization_for_different_shapes}
\end{figure}

A total of nine two-dimensional target distributions with varying non-convex or multimodal shapes are used to compare the performance of Intrepid MCMC against CMH. Each target distribution takes the form of a transformation function multiplied by a parent density function, per Eq.~\eqref{eqn:target_structure}. We define six indicator functions, $I_j(\mathbf{x})$ in Table~\ref{tab:indicator_definitions} and three density functions $f_i(\mathbf{x})$ in Table~\ref{tab:density_definitions}. In all cases, $ \mathbf{x} = \begin{bmatrix} x_1 & x_2 \end{bmatrix}^T $. The total set of nine examples is then presented in Table~\ref{tab:distribution_shape_definitions}, which shows the target $ \pi (\mathbf{x}) $ and the assumed $ p (\mathbf{x}) $ and $ T (\mathbf{x}) $ for each case. 
Contour maps for all nine target distributions are plotted in Figure~\ref{fig:target_visualization_for_different_shapes}.

\begin{table}[!ht]
    \centering
    \begin{tabular}{l|c|c}
        Indicator Function & Symbol & Iverson Bracket Definition \\
        \hline \hline
        Gauss-Planes & \( I_1 (\mathbf{x}) \) & \( \displaystyle \left[ \min \left[ \left( 1.25 - x_1 \right), \left( 1.75 + x_1 \right) \right] \leq 0 \right] \) \\
        Gumbel-Planes & \( I_2 (\mathbf{x}) \) & \( \displaystyle \left[ \min \left[ \left( 4 - 0.8 x_2 - x_1 \right), \left( 2 + 0.8 x_2 + x_1 \right) \right] \leq 0 \right] \) \\
        Rosenbrock-Planes & \( I_3 (\mathbf{x}) \) & \( \displaystyle \left[ \min \left[ \left( 2.5 - x_1 \right), \left( 2.5 + x_1 \right) \right] \leq 0 \right] \) \\
        Ring  & \( I_4 (\mathbf{x}) \) & \( \displaystyle \left[ 4 - \lVert \mathbf{x} \rVert_2 \leq 0 \right] \) \\
        Rosenbrock-Ring & \( I_5 (\mathbf{x}) \) & \( \displaystyle \left[ 16 - x_1^2 - \left( \nicefrac{(x_2 - 2.8)}{1.7} \right)^2 \leq 0 \right] \) \\
        Circles & \( I_6 (\mathbf{x}) \) & \( \displaystyle \left[ \min_{i \in \left\{ 1, 2, 3 \right\}} \left[ \left(x_1 - 4 \cos \left( \vartheta_i \right) \right)^2 + \left(x_2 - 4 \sin \left( \vartheta_i \right) \right)^2 - R_i^2 \right] \leq 0 \right] \) \\
         & & ( \( \displaystyle \vartheta_1 = \nicefrac{3 \pi}{8} \), \( \displaystyle \vartheta_2 = \nicefrac{5 \pi}{8} \), \( \displaystyle \vartheta_3 = \nicefrac{15 \pi}{8} \), \( \displaystyle R_1 = 0.8 \), \( \displaystyle R_2 = 1.2 \), \( \displaystyle R_3 = 1.6 \) ) \\
        \hline
    \end{tabular}
    \caption{Indicator functions used in Section~\ref{section:distribution_shape_results}. An indicator function takes a value of $ 1 $ wherever the condition within the Iverson bracket is true and $ 0 $ everywhere else.}
    \label{tab:indicator_definitions}
\end{table}

\begin{table}[!ht]
    \centering
    \begin{tabular}{c|c|c}
        Name & Symbol & Definition \\
        \hline \hline
        Gaussian & \( f_1 (\mathbf{x}) \) & \( \displaystyle \exp{ \left[ - \frac{1}{2} \left( x_1^2 + x_2^2 \right) \right]} \) \\
        Gumbel & \( f_2 (\mathbf{x}) \) & \( \displaystyle \exp{ \left[ - \left( x_1 + x_2 + e^{-x_1} + e^{-x_2} \right) \right]} \) \\
        Rosenbrock & \( f_3 (\mathbf{x}) \) & \( \displaystyle \exp{ \left[ - \frac{1}{20} \left( \left( 1 - x_1 \right)^2 + 5 \left(x_2 - x_1^2\right)^2 \right) \right]} \) \\
        \hline
    \end{tabular}
    \caption{Definitions for the density functions used in Section~\ref{section:distribution_shape_results}.}
    \label{tab:density_definitions}
\end{table}

\begin{table}[!ht]
    \centering
    \begin{tabular}{c|c|c|c}
        Case & $ \pi (\mathbf{x}) $ & $ p (\mathbf{x}) $ & $ T (\mathbf{x}) $ \\
        \hline \hline
        Case 1 (\textit{Gauss-Ring}) & $ I_4 (\mathbf{x}) f_1 (\mathbf{x}) $ & $ f_1 (\mathbf{x}) $ & $ I_4 (\mathbf{x}) $ \\
        Case 2 (\textit{Gauss-Planes}) & $ I_1 (\mathbf{x}) f_1 (\mathbf{x}) $ & $ f_1 (\mathbf{x}) $ & $ I_1 (\mathbf{x}) $ \\
        Case 3 (\textit{Gauss-Circles}) & $ I_6 (\mathbf{x}) f_1 (\mathbf{x}) $ & $ f_1 (\mathbf{x}) $ & $ I_6 (\mathbf{x}) $ \\
        Case 4 (\textit{Gumbel-Ring}) & $ I_4 (\mathbf{x}) f_2 (\mathbf{x}) $ & $ f_1 (\mathbf{x}) $ & $ I_4 (\mathbf{x}) \dfrac{f_2 (\mathbf{x})}{f_1 (\mathbf{x})} $ \\
        Case 5 (\textit{Gumbel-Planes}) & $ I_2 (\mathbf{x}) f_2 (\mathbf{x}) $ & $ f_1 (\mathbf{x}) $ & $ I_2 (\mathbf{x}) \dfrac{f_2 (\mathbf{x})}{f_1 (\mathbf{x})} $ \\
        Case 6 (\textit{Gumbel-Circles}) & $ I_6 (\mathbf{x}) f_2 (\mathbf{x}) $ & $ f_1 (\mathbf{x}) $ & $ I_6 (\mathbf{x}) \dfrac{f_2 (\mathbf{x})}{f_1 (\mathbf{x})} $ \\
        Case 7 (\textit{Rosenbrock-Ring}) & $ I_5 (\mathbf{x}) f_3 (\mathbf{x}) $ & $ f_1 (\mathbf{x}) $ & $ I_5 (\mathbf{x}) \dfrac{f_3 (\mathbf{x})}{f_1 (\mathbf{x})} $ \\
        Case 8 (\textit{Rosenbrock-Planes}) & $ I_3 (\mathbf{x}) f_3 (\mathbf{x}) $ & $ f_1 (\mathbf{x}) $ & $ I_3 (\mathbf{x}) \dfrac{f_3 (\mathbf{x})}{f_1 (\mathbf{x})} $ \\
        Case 9 (\textit{Rosenbrock-Circles}) & $ I_6 (\mathbf{x}) f_3 (\mathbf{x}) $ & $ f_1 (\mathbf{x}) $ & $ I_6 (\mathbf{x}) \dfrac{f_3 (\mathbf{x})}{f_1 (\mathbf{x})} $ \\
        \hline
    \end{tabular}
    \caption{Target distributions used in Section~\ref{section:distribution_shape_results}}
    \label{tab:distribution_shape_definitions}
\end{table}


First, 50 million IID samples were drawn from each distribution using rejection sampling and the empirical distributions were estimated to represent the target distributions for all convergence computations. Then, for all nine cases, Intrepid MCMC was run with exploration probability $ \beta = \left\{ 0, 0.01, 0.05, 0.1, 0.3, 0.5, 1.0 \right\} $. 
Note that $ \beta = 0 $ corresponds to CMH. For each target and each $ \beta $, 100 independent Markov chains were propagated from randomly selected locations to  
various lengths after discarding 10,000 samples as burn-in. 
To measure convergence of the Markov chain, the Total Variation Distance (TVD) was computed between the empirical distribution of the IID samples and the empirical distribution from each Markov chain. We also provide a standardized measure of error in the estimated mean by calculating the $ l^2 $-norm of the difference between the Markov chain sample mean and the true mean, normalized by 
the square root of the trace of the target covariance.  
Finally, the acceptance rates of the Markov chains were also evaluated. In all cases, the following proposal distributions were used~\footnote{These proposals are tuned for their respective tasks. The intrepid proposal has a wide spread to facilitate exploration, while the CMH proposal is focused towards populating already discovered modes only.}; for the Intrepid kernel $ \mathfrak{q}_1 \left( \phi | \theta_{s, 1} \right) \equiv \textit{Uniform} \left( - \theta_{s,1} , 2 \pi - \theta_{s,1} \right) $ and $ \mathfrak{q}_r \left( \gamma \right) \equiv \textit{Uniform} \left( 0.5, 2.0 \right) $, and for the CMH kernel $ q_{L_i} \left( z | x_{s, i} \right) \equiv \mathcal{N} \left( x_{s, i}, 1 \right) $, $ i = 1, 2 $. Since $ f_1 (\mathbf{x}) $ -- which is radially symmetric --  was used as the parent distribution for all nine cases, the RTF is the identity function.

\begin{figure}[!ht]
\centering
\begin{subfigure}{.33\textwidth}
  \centering
  \includegraphics[width=\linewidth]{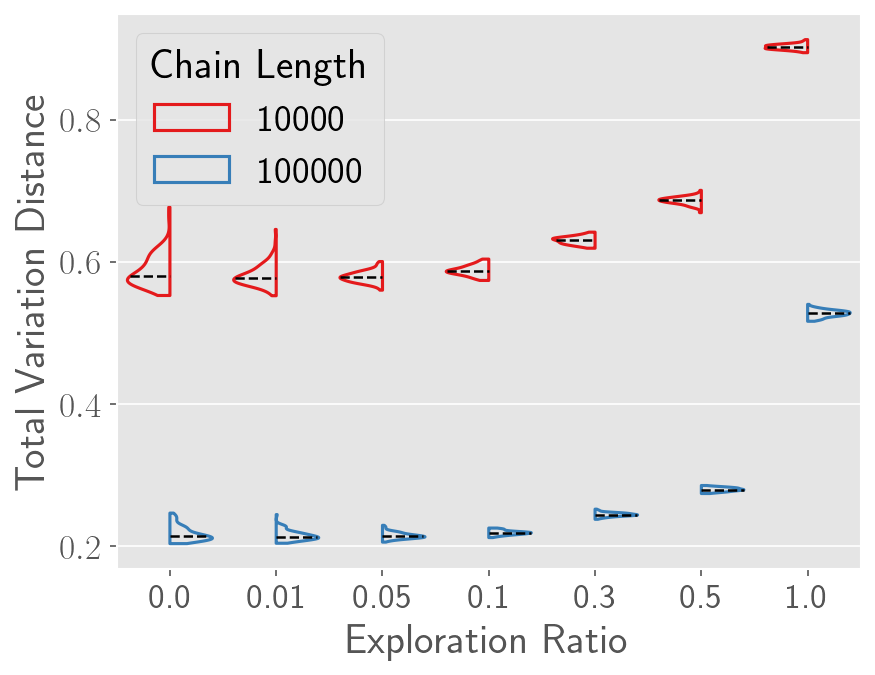}
  \caption{}
  \label{fig:gauss_ring_tvd}
\end{subfigure}%
\begin{subfigure}{.33\textwidth}
  \centering
  \includegraphics[width=\linewidth]{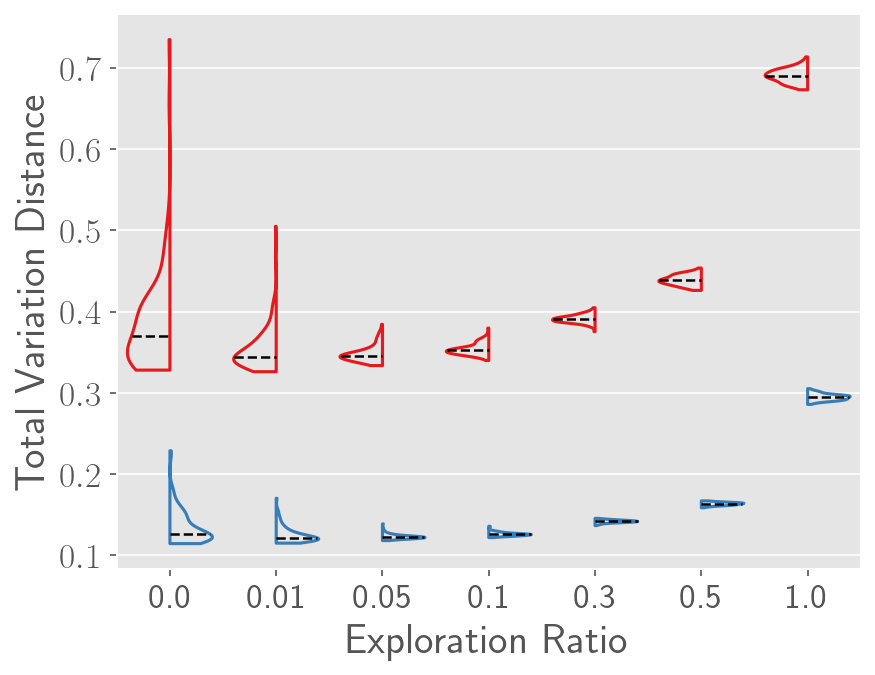}
  \caption{}
  \label{fig:gauss_planes_tvd}
\end{subfigure}%
\begin{subfigure}{.33\textwidth}
  \centering
  \includegraphics[width=\linewidth]{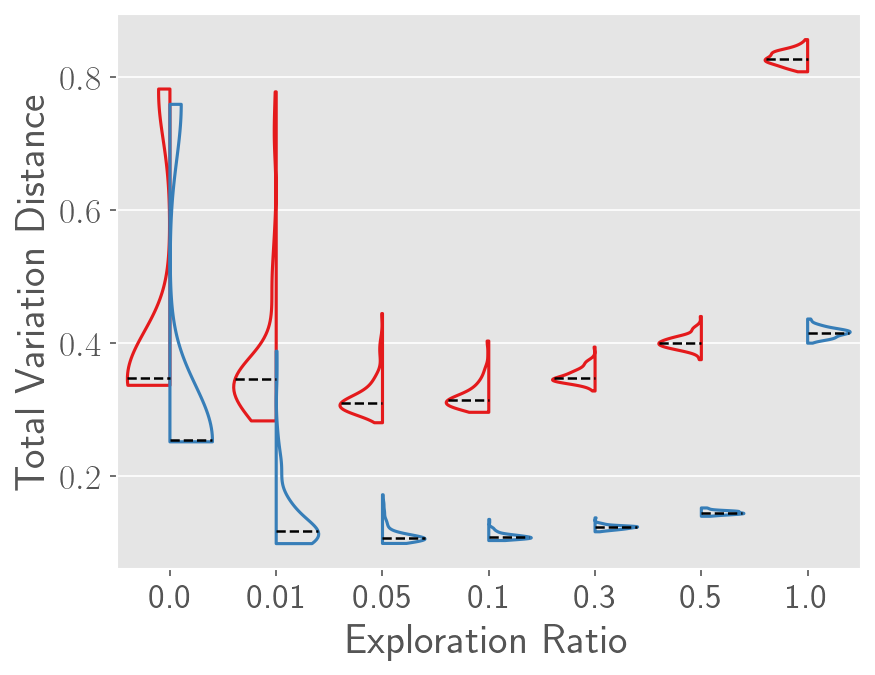}
  \caption{}
  \label{fig:gauss_circles_tvd}
\end{subfigure}
\begin{subfigure}{.33\textwidth}
  \centering
  \includegraphics[width=\linewidth]{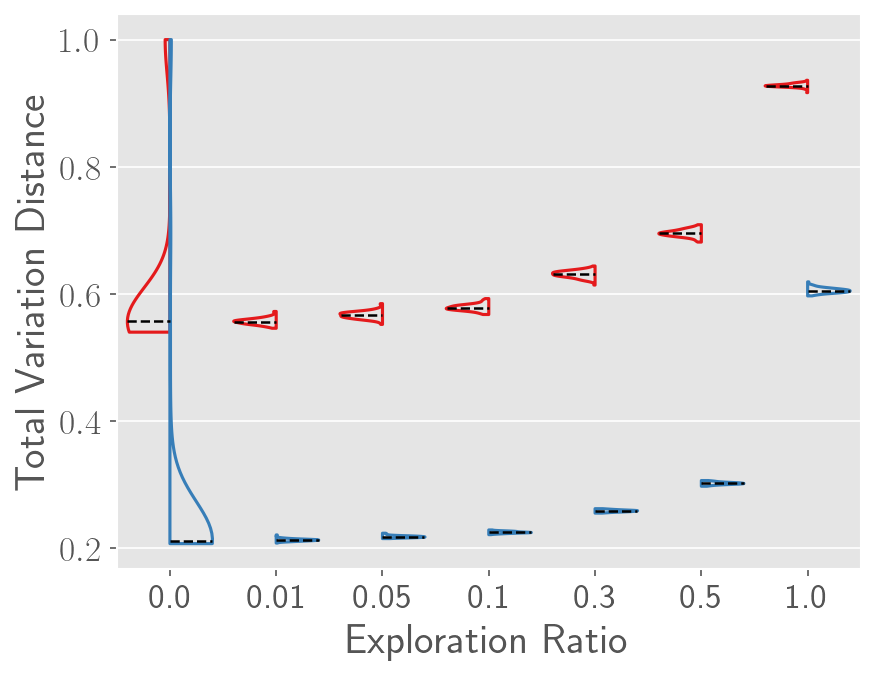}
  \caption{}
  \label{fig:gumbel_ring_tvd}
\end{subfigure}%
\begin{subfigure}{.33\textwidth}
  \centering
  \includegraphics[width=\linewidth]{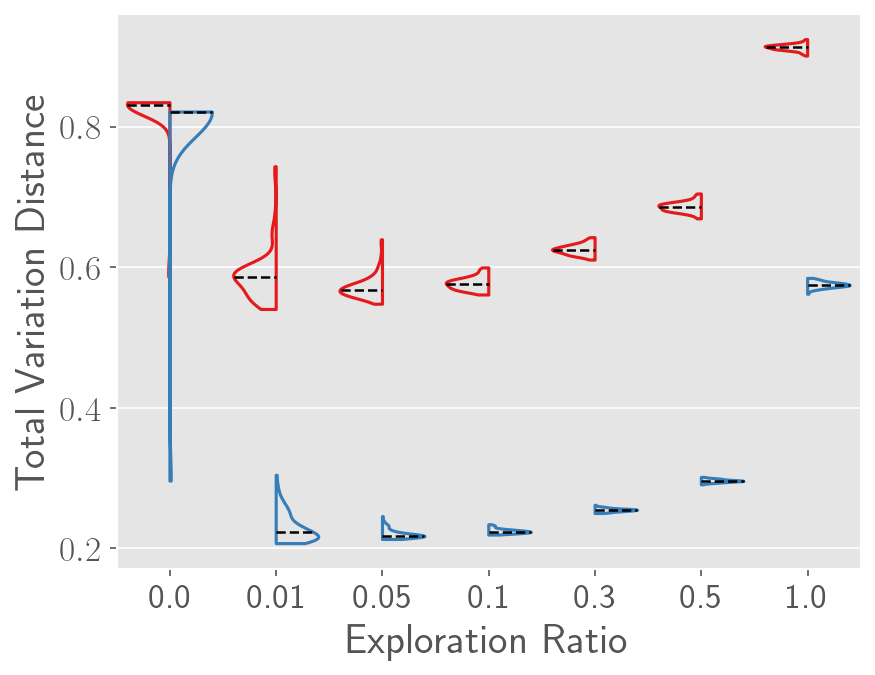}
  \caption{}
  \label{fig:gumbel_planes_tvd}
\end{subfigure}%
\begin{subfigure}{.33\textwidth}
  \centering
  \includegraphics[width=\linewidth]{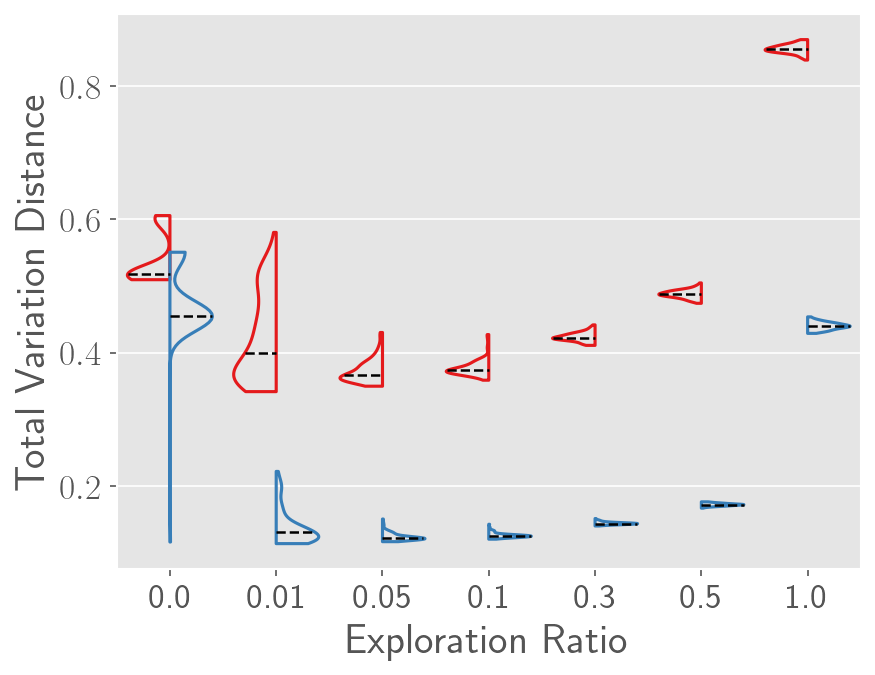}
  \caption{}
  \label{fig:gumbel_circles_tvd}
\end{subfigure}
\begin{subfigure}{.33\textwidth}
  \centering
  \includegraphics[width=\linewidth]{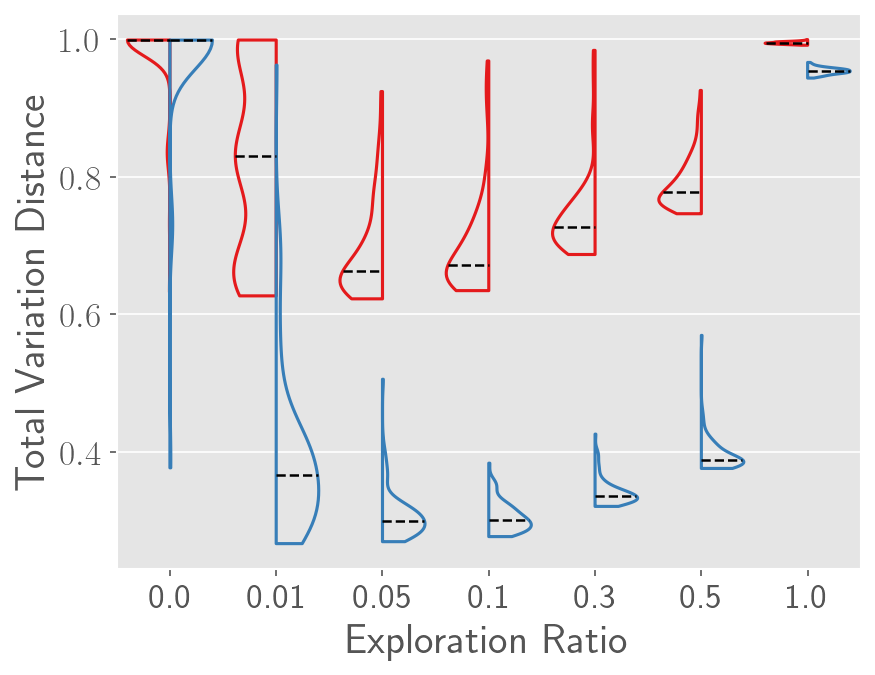}
  \caption{}
  \label{fig:rosenbrock_ring_tvd}
\end{subfigure}%
\begin{subfigure}{.33\textwidth}
  \centering
  \includegraphics[width=\linewidth]{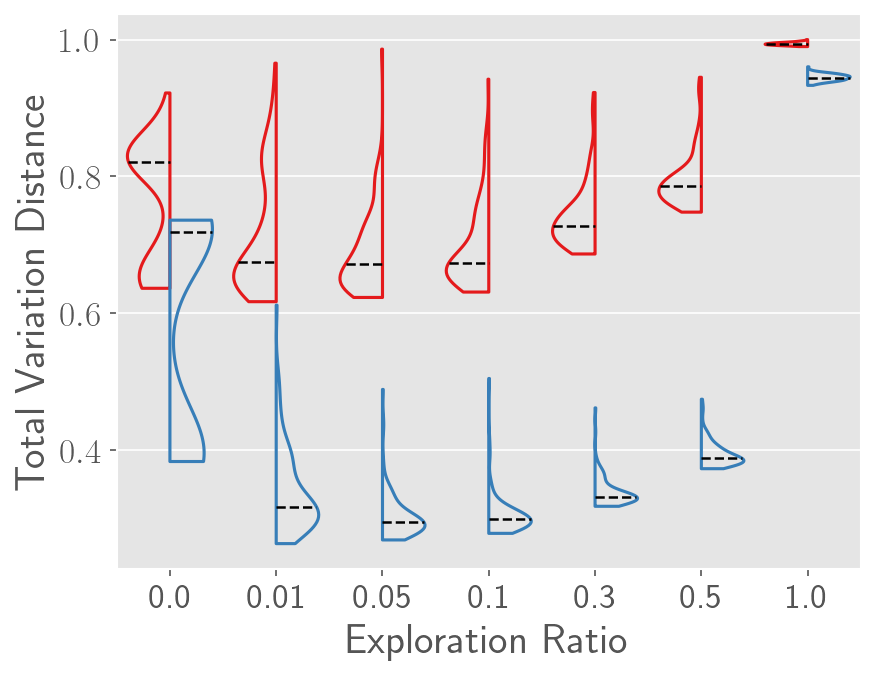}
  \caption{}
  \label{fig:rosenbrock_planes_tvd}
\end{subfigure}%
\begin{subfigure}{.33\textwidth}
  \centering
  \includegraphics[width=\linewidth]{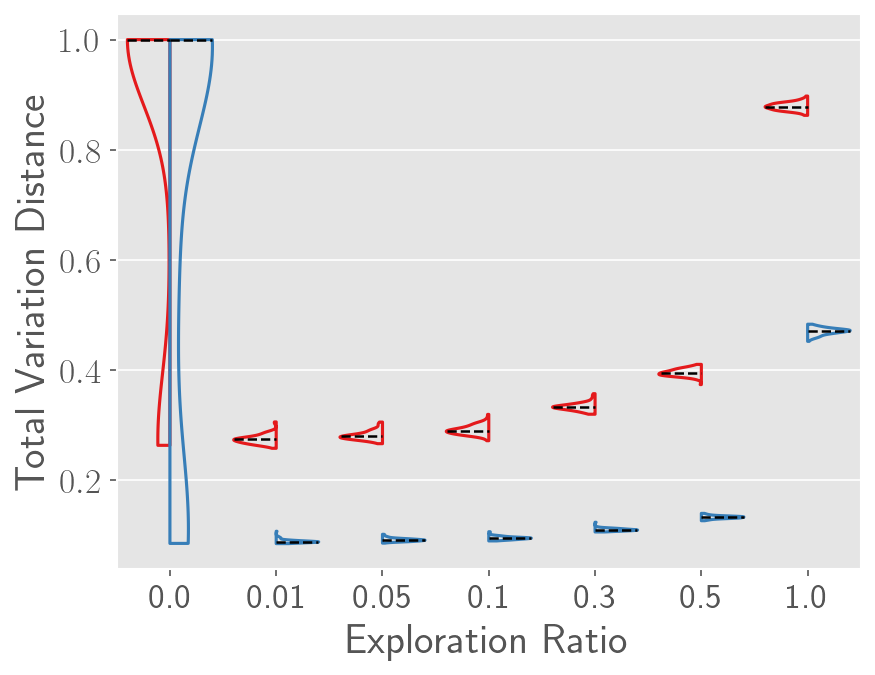}
  \caption{}
  \label{fig:rosenbrock_circles_tvd}
\end{subfigure}
\caption{Violin plots of total variation distance from 100 repeated trials for Intrepid MCMC with varying exploration ratios for the nine target distributions from Section~\ref{section:distribution_shape_results}: (a) Case 1 (Gauss-Ring), (b) Case 2 (Gauss-Planes), (c) Case 3 (Gauss-Circles), (d) Case 4 (Gumbel-Ring), (e) Case 5 (Gumbel-Planes), (f) Case 6 (Gumbel-Circles), (g) Case 7 (Rosenbrock-Ring), (h) Case 8 (Rosenbrock-Planes), and (i) Case 9 (Rosenbrock-Circles). Results for two different chain lengths are included, as shown in the legend provided in (a), which is common for all subfigures.}
\label{fig:tvd_exploration}
\end{figure}

\begin{figure}[!ht]
\centering
\includegraphics[width=\linewidth]{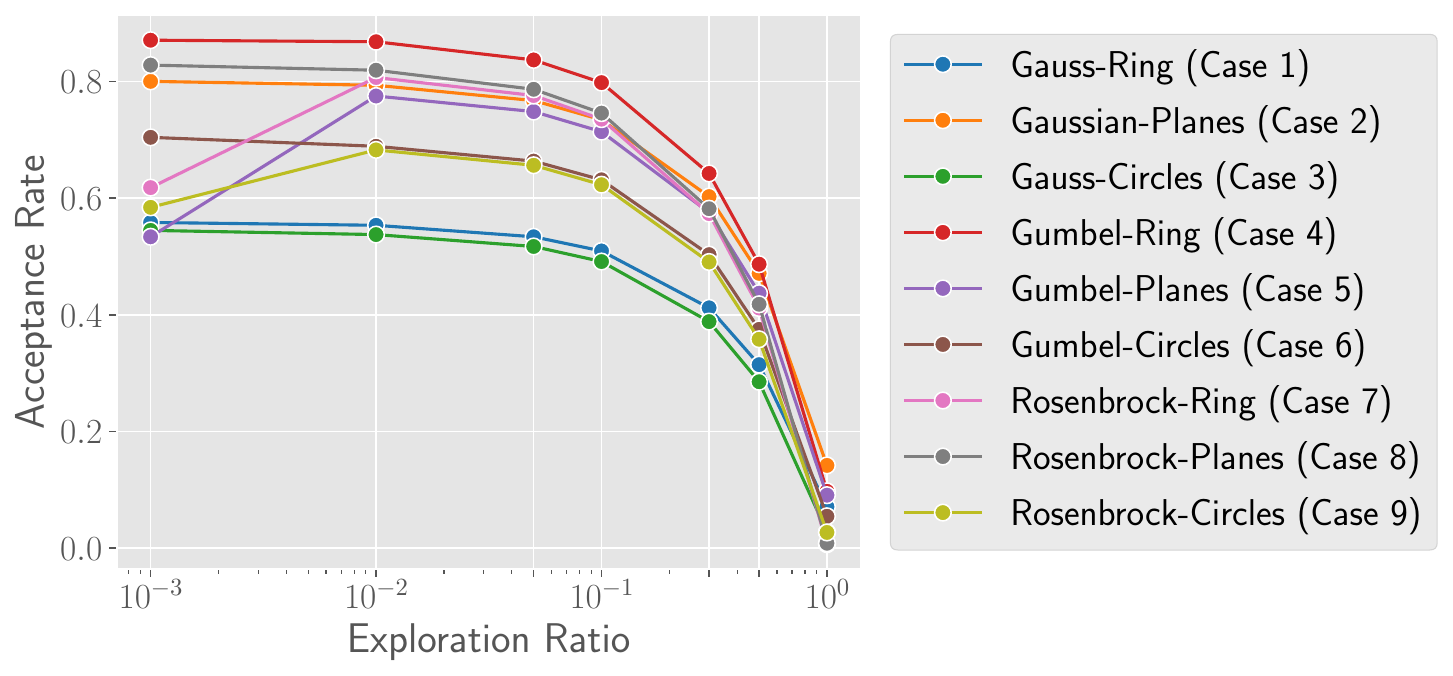}
\caption{Acceptance rate for varying exploration ratios for the nine distributions considered in Section~\ref{section:distribution_shape_results}.}
\label{fig:acceptance_rate_exploration}
\end{figure}

\begin{figure}[!ht]
\centering
\begin{subfigure}{.33\textwidth}
  \centering
  \includegraphics[width=\linewidth]{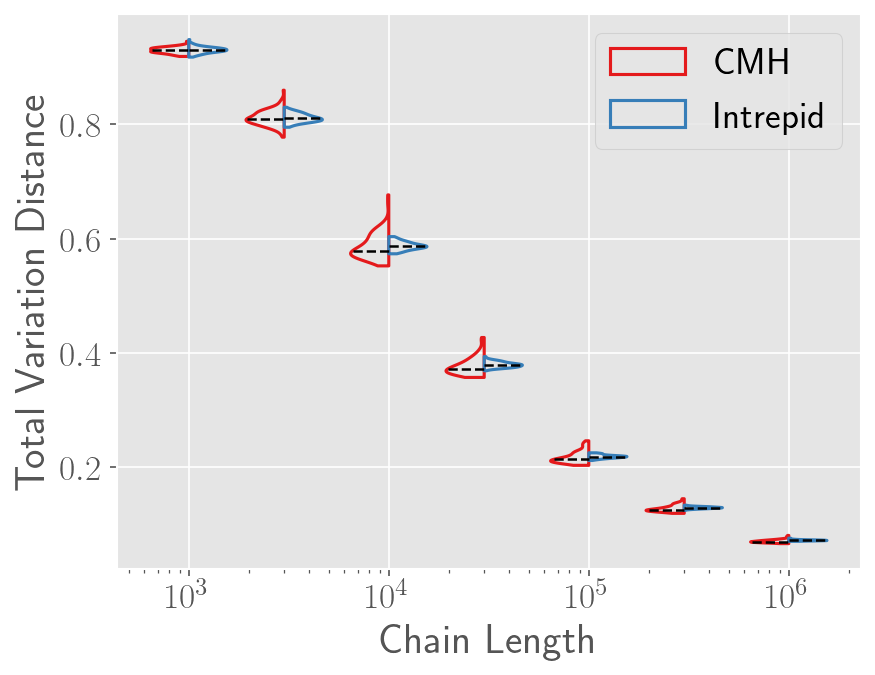}
  \caption{}
  \label{fig:gauss_ring_tvd_cmh_intrepid}
\end{subfigure}%
\begin{subfigure}{.33\textwidth}
  \centering
  \includegraphics[width=\linewidth]{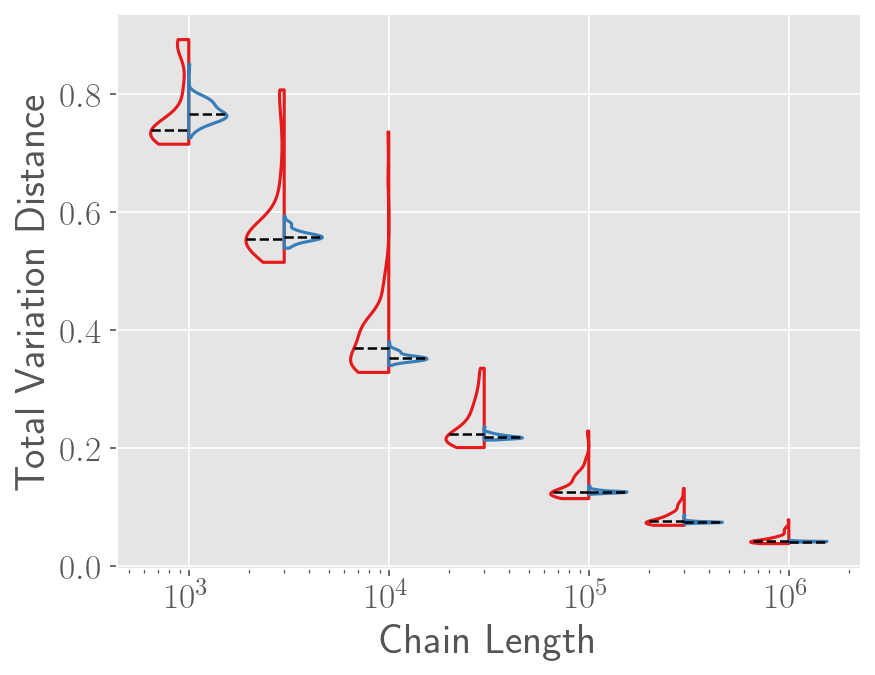}
  \caption{}
  \label{fig:gauss_planes_tvd_cmh_intrepid}
\end{subfigure}%
\begin{subfigure}{.33\textwidth}
  \centering
  \includegraphics[width=\linewidth]{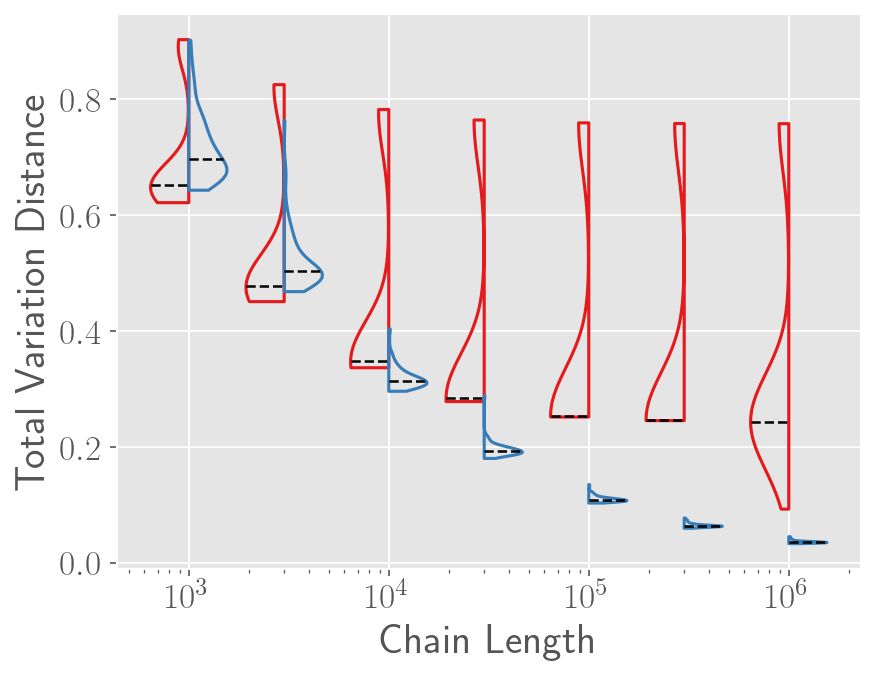}
  \caption{}
  \label{fig:gauss_circles_tvd_cmh_intrepid}
\end{subfigure}
\begin{subfigure}{.33\textwidth}
  \centering
  \includegraphics[width=\linewidth]{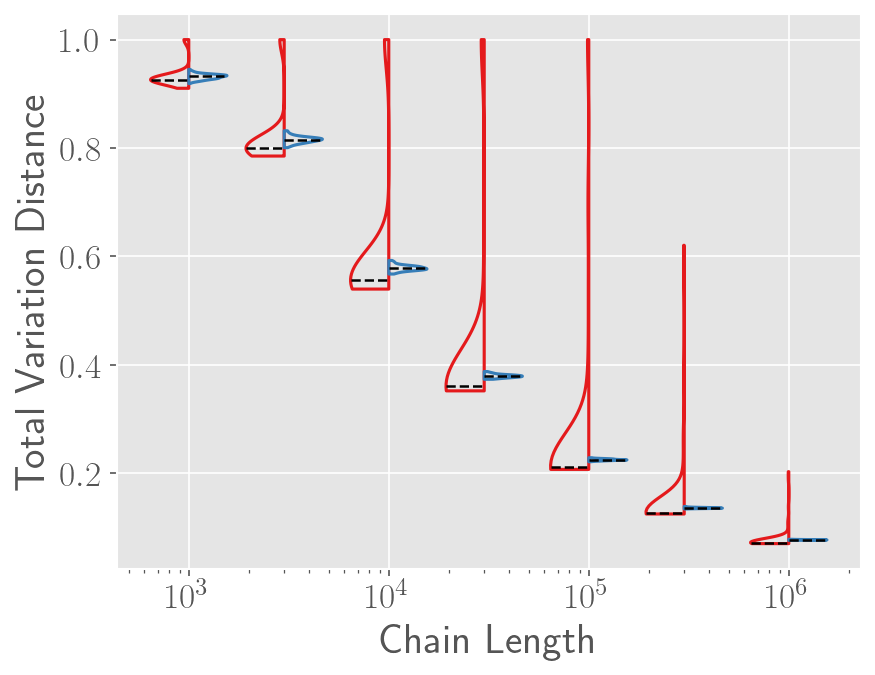}
  \caption{}
  \label{fig:gumbel_ring_tvd_cmh_intrepid}
\end{subfigure}%
\begin{subfigure}{.33\textwidth}
  \centering
  \includegraphics[width=\linewidth]{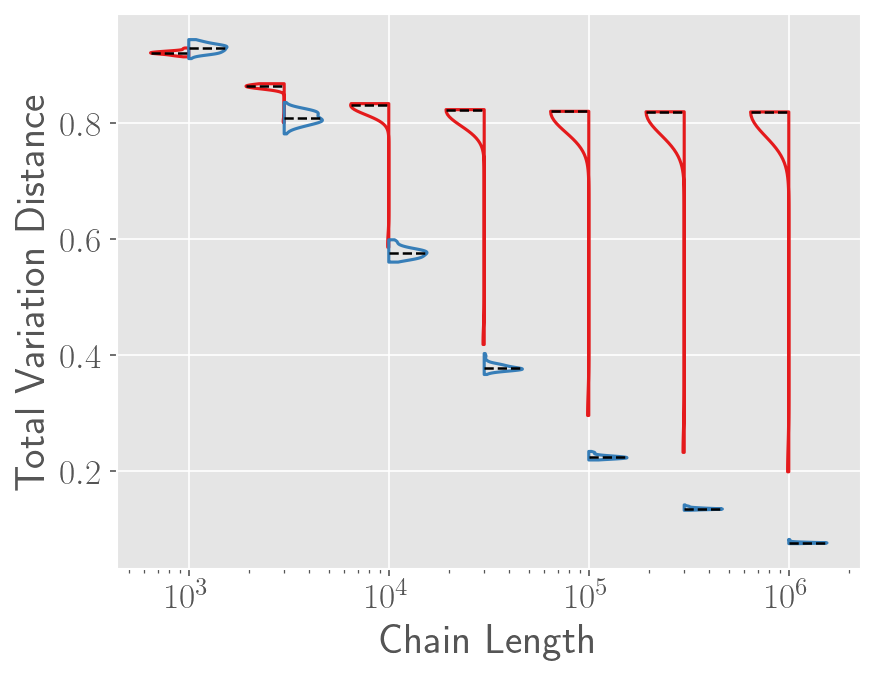}
  \caption{}
  \label{fig:gumbel_planes_tvd_cmh_intrepid}
\end{subfigure}%
\begin{subfigure}{.33\textwidth}
  \centering
  \includegraphics[width=\linewidth]{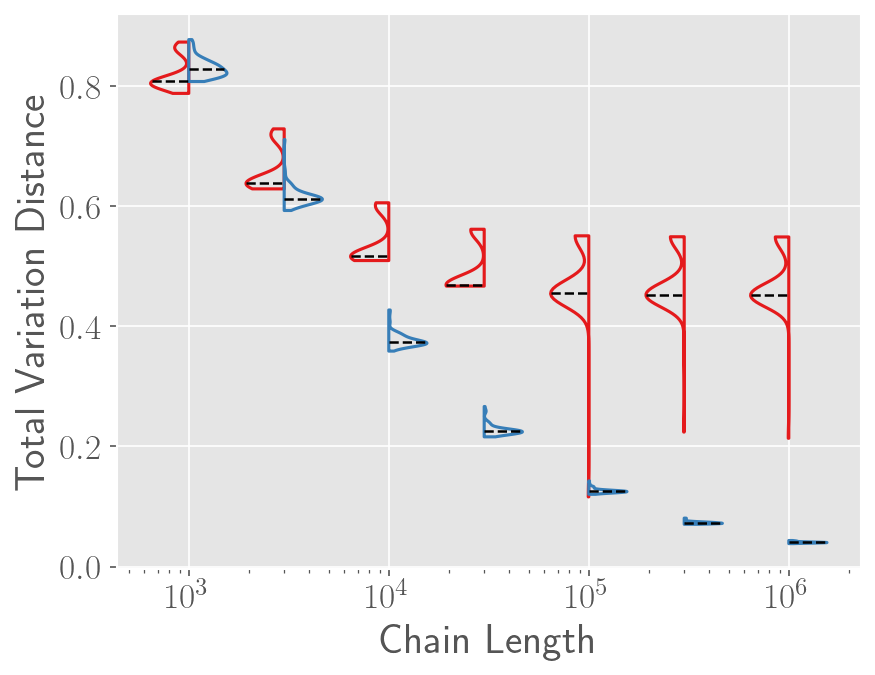}
  \caption{}
  \label{fig:gumbel_circles_tvd_cmh_intrepid}
\end{subfigure}
\begin{subfigure}{.33\textwidth}
  \centering
  \includegraphics[width=\linewidth]{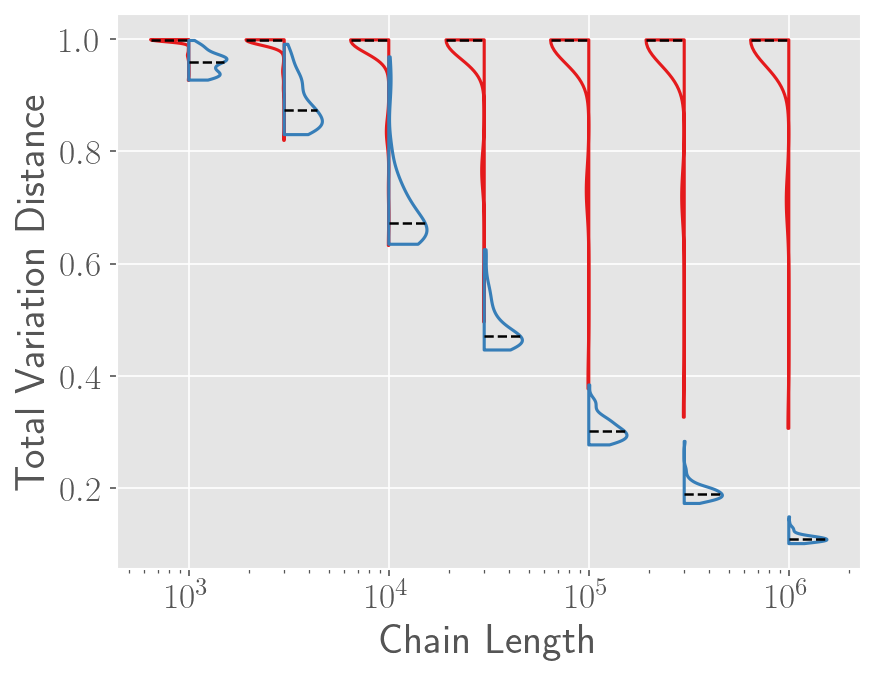}
  \caption{}
  \label{fig:rosenbrock_ring_tvd_cmh_intrepid}
\end{subfigure}%
\begin{subfigure}{.33\textwidth}
  \centering
  \includegraphics[width=\linewidth]{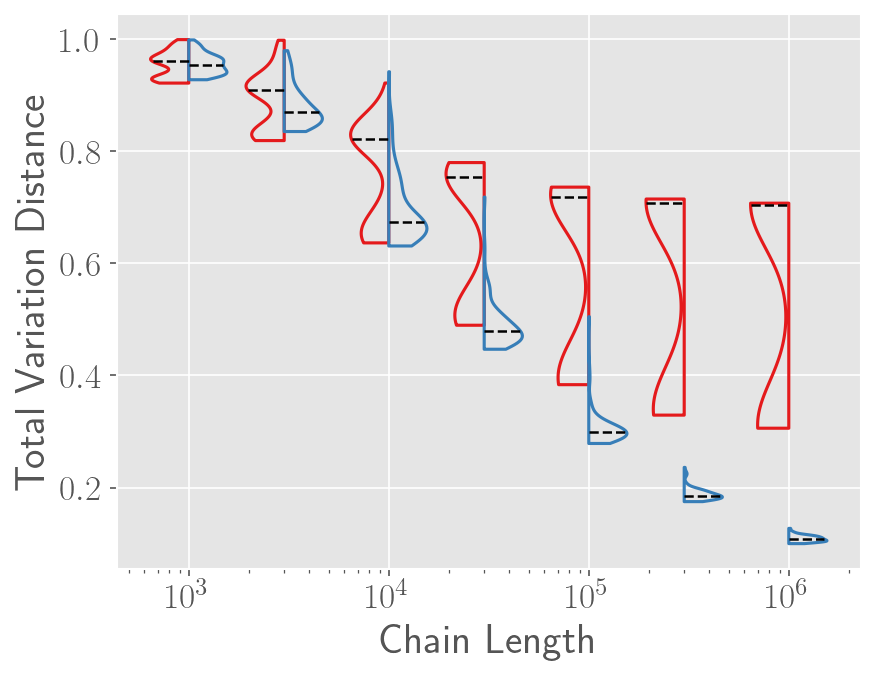}
  \caption{}
  \label{fig:rosenbrock_planes_tvd_cmh_intrepid}
\end{subfigure}%
\begin{subfigure}{.33\textwidth}
  \centering
  \includegraphics[width=\linewidth]{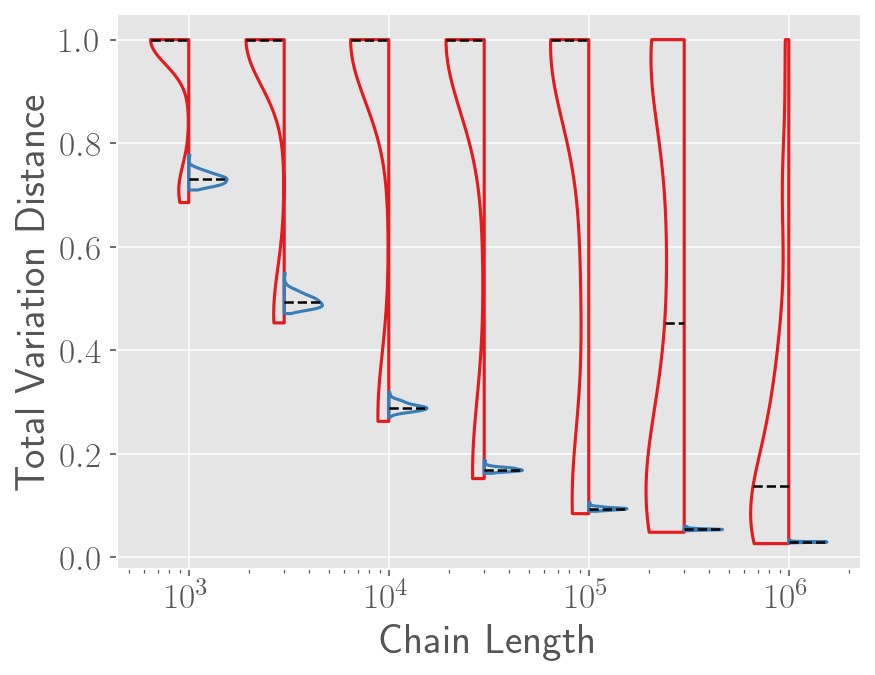}
  \caption{}
  \label{fig:rosenbrock_circles_tvd_cmh_intrepid}
\end{subfigure}
\caption{Violin plots showing convergence of the total variation distance from 100 repeated trials for Intrepid MCMC with $ \beta = 0.1 $ and CMH with increasing chain length for the nine target distributions from Section~\ref{section:distribution_shape_results}: (a) Case 1 (Gauss-Ring), (b) Case 2 (Gauss-Planes), (c) Case 3 (Gauss-Circles), (d) Case 4 (Gumbel-Ring), (e) Case 5 (Gumbel-Planes), (f) Case 6 (Gumbel-Circles), (g) Case 7 (Rosenbrock-Ring), (h) Case 8 (Rosenbrock-Planes), and (i) Case 9 (Rosenbrock-Circles). The legend is common for all subfigures and is provided in (a).}
\label{fig:tvd_chain_length}
\end{figure}

\begin{figure}[!ht]
\centering
\begin{subfigure}{.33\textwidth}
  \centering
  \includegraphics[width=\linewidth]{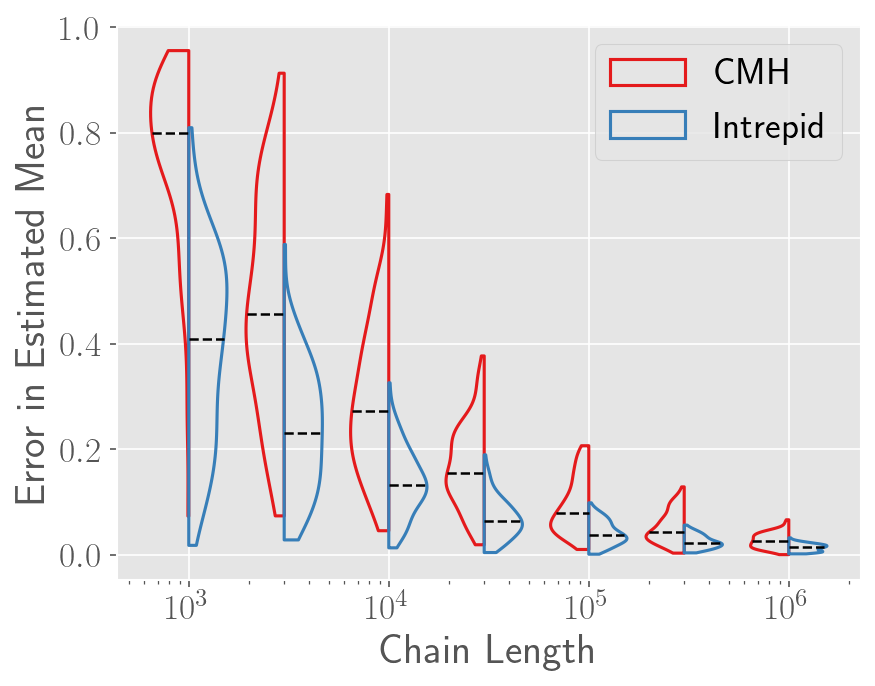}
  \caption{}
  \label{fig:gauss_ring_mean_cmh_intrepid}
\end{subfigure}%
\begin{subfigure}{.33\textwidth}
  \centering
  \includegraphics[width=\linewidth]{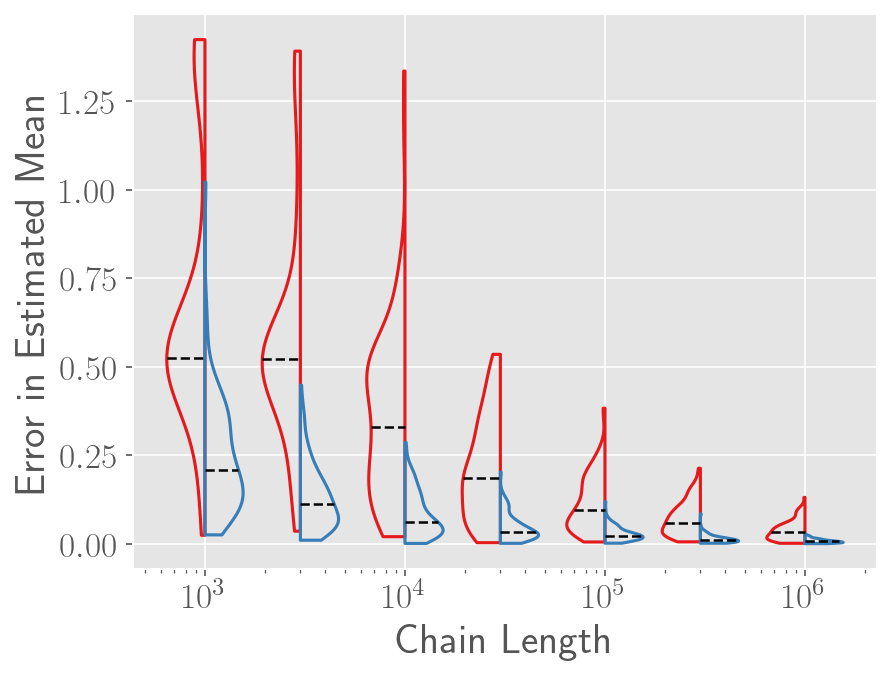}
  \caption{}
  \label{fig:gauss_planes_mean_cmh_intrepid}
\end{subfigure}%
\begin{subfigure}{.33\textwidth}
  \centering
  \includegraphics[width=\linewidth]{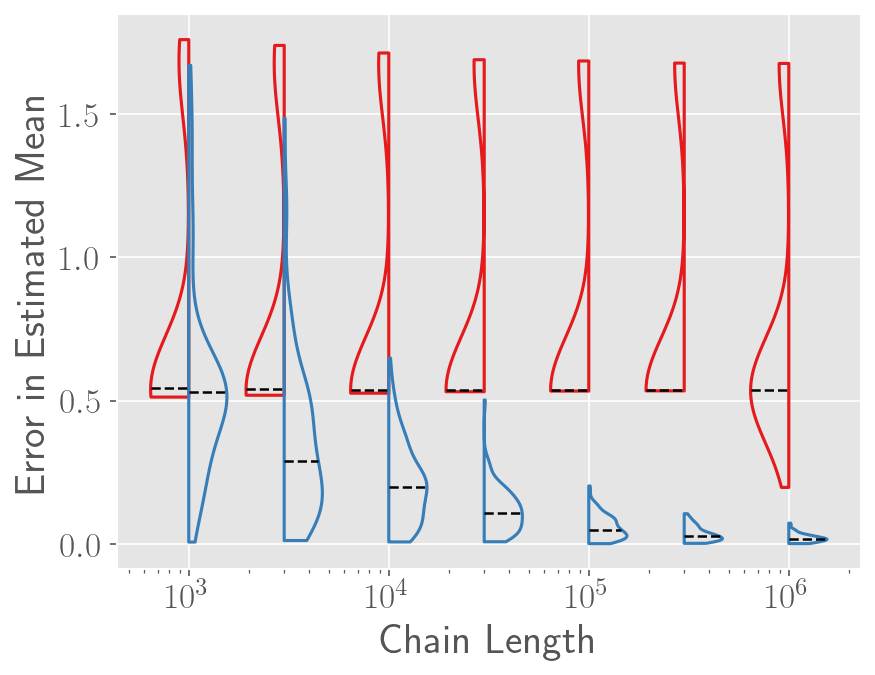}
  \caption{}
  \label{fig:gauss_circles_mean_cmh_intrepid}
\end{subfigure}
\begin{subfigure}{.33\textwidth}
  \centering
  \includegraphics[width=\linewidth]{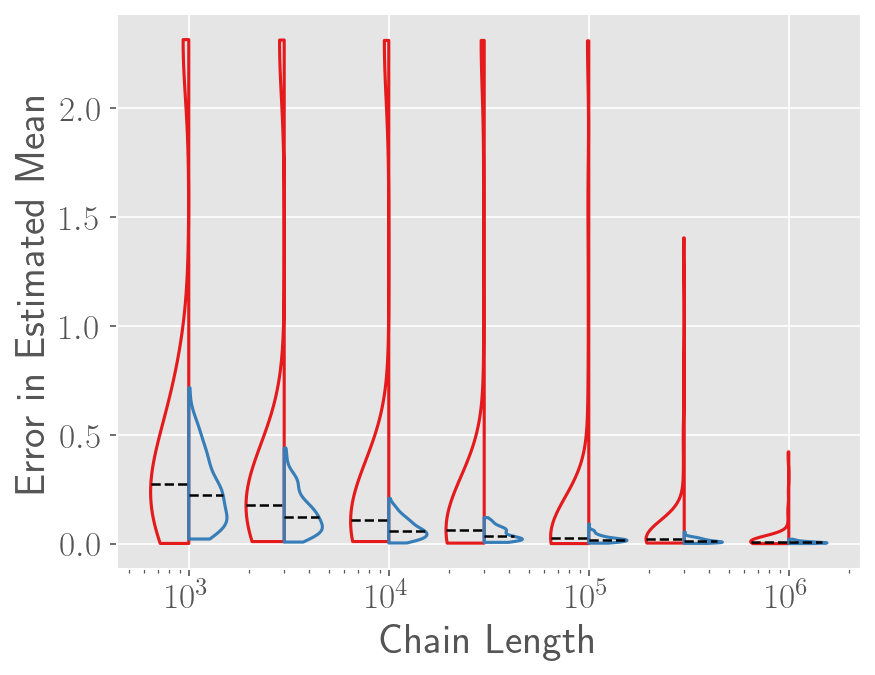}
  \caption{}
  \label{fig:gumbel_ring_mean_cmh_intrepid}
\end{subfigure}%
\begin{subfigure}{.33\textwidth}
  \centering
  \includegraphics[width=\linewidth]{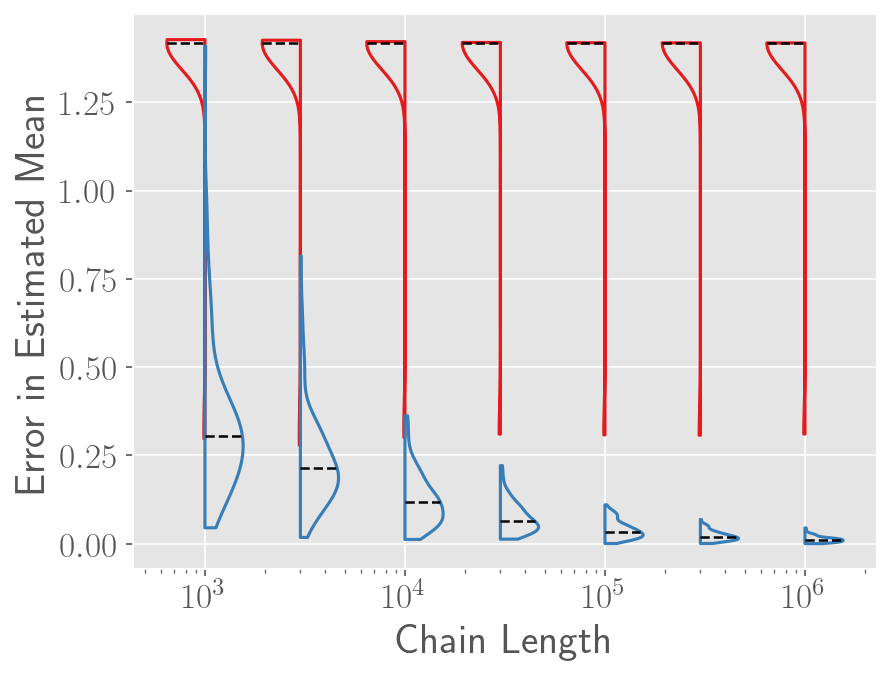}
  \caption{}
  \label{fig:gumbel_planes_mean_cmh_intrepid}
\end{subfigure}%
\begin{subfigure}{.33\textwidth}
  \centering
  \includegraphics[width=\linewidth]{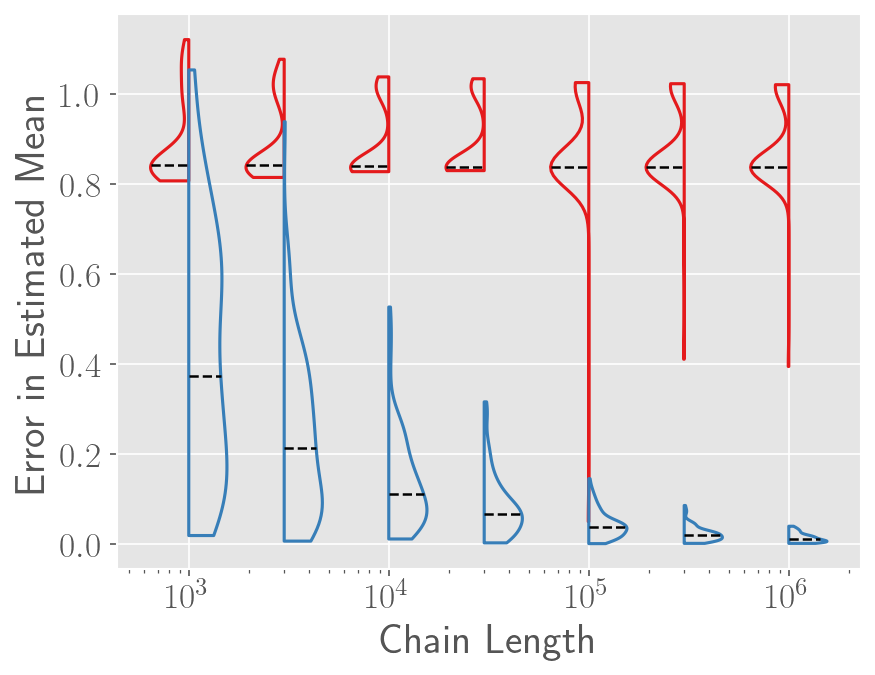}
  \caption{}
  \label{fig:gumbel_circles_mean_cmh_intrepid}
\end{subfigure}
\begin{subfigure}{.33\textwidth}
  \centering
  \includegraphics[width=\linewidth]{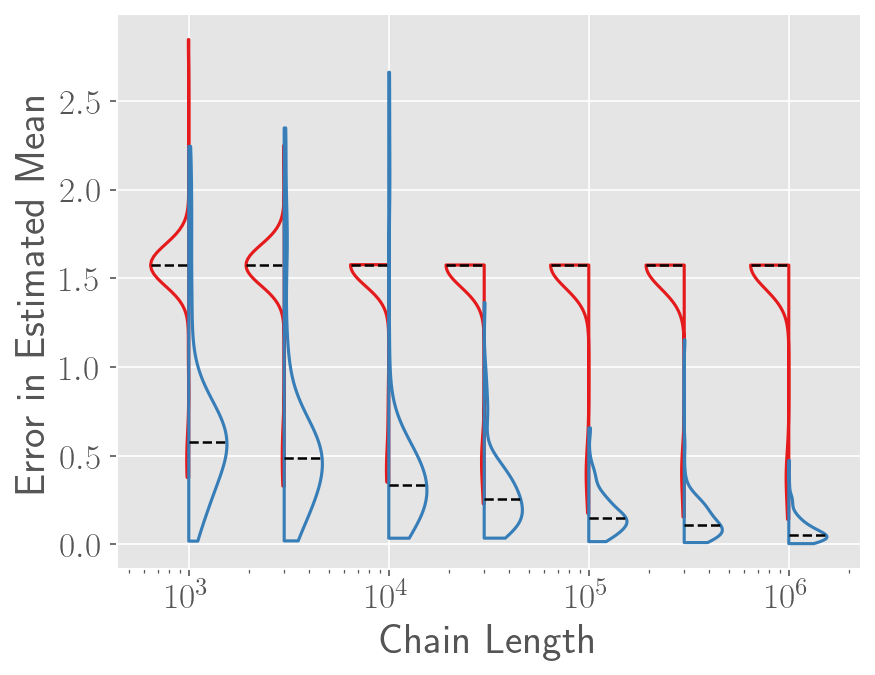}
  \caption{}
  \label{fig:rosenbrock_ring_mean_cmh_intrepid}
\end{subfigure}%
\begin{subfigure}{.33\textwidth}
  \centering
  \includegraphics[width=\linewidth]{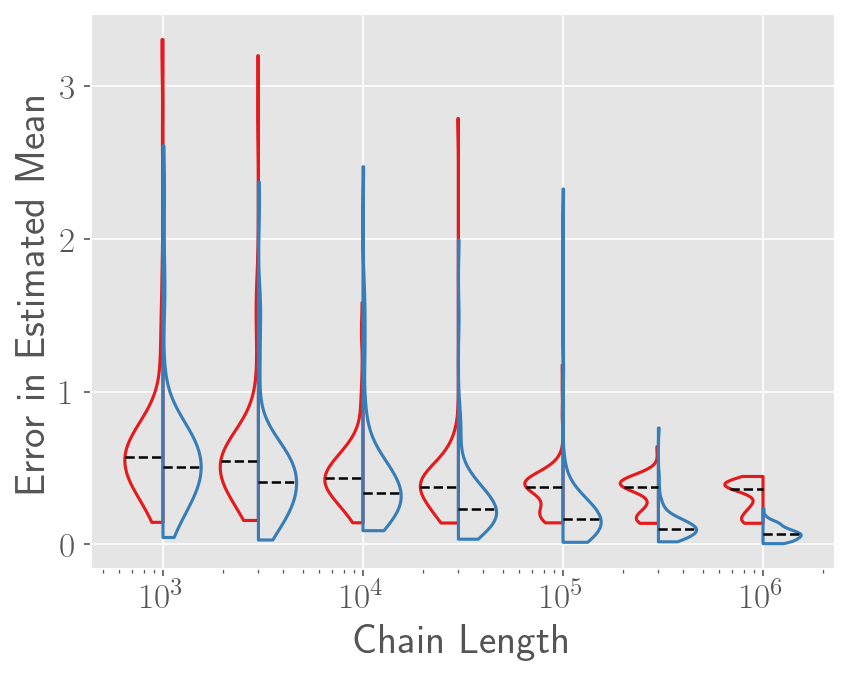}
  \caption{}
  \label{fig:rosenbrock_planes_mean_cmh_intrepid}
\end{subfigure}%
\begin{subfigure}{.33\textwidth}
  \centering
  \includegraphics[width=\linewidth]{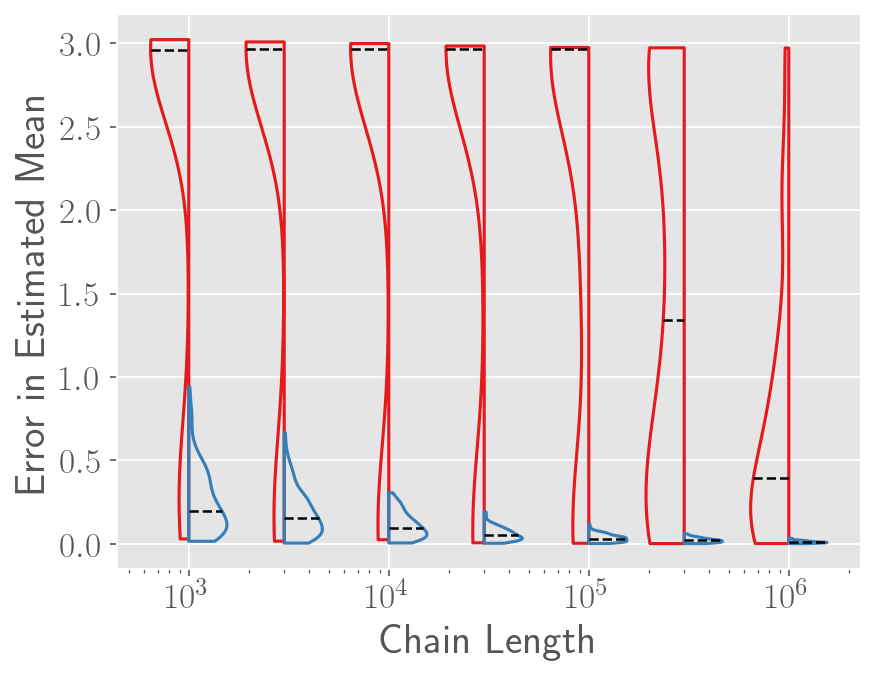}
  \caption{}
  \label{fig:rosenbrock_circles_mean_cmh_intrepid}
\end{subfigure}
\caption{Violin plots showing convergence of the mean from 100 repeated trials for Intrepid MCMC with $ \beta = 0.1 $ and CMH with increasing chain length for the nine target distributions from Section~\ref{section:distribution_shape_results}: (a) Case 1 (Gauss-Ring), (b) Case 2 (Gauss-Planes), (c) Case 3 (Gauss-Circles), (d) Case 4 (Gumbel-Ring), (e) Case 5 (Gumbel-Planes), (f) Case 6 (Gumbel-Circles), (g) Case 7 (Rosenbrock-Ring), (h) Case 8 (Rosenbrock-Planes), and (i) Case 9 (Rosenbrock-Circles). The legend is common for all subfigures and is provided in (a).}
\label{fig:mean_chain_length}
\end{figure}

Figures~\ref{fig:tvd_exploration}--\ref{fig:mean_chain_length} plot statistics from the 100 repeated trials of Intrepid for each target distribution. Figure~\ref{fig:tvd_exploration} shows the TVD for varying exploration ratio, $\beta$, for chain lengths of 10,000 samples and 100,000 samples.
Note again, exploration ratio $\beta=0$ yields CMH. 
It is readily apparent that injecting only a tiny fraction ($ \beta = 0.01 $) of highly exploratory steps can significantly improve convergence to multi-modal target distributions. This is because CMH often gets ``stuck'' in the first mode it finds, whereas Intrepid is designed to explore and identify additional modes. However, if $ \beta $ is too high, the chain wastes too many samples exploring new locations instead of fully populating previously identified modes, and the performance deteriorates. The case of $ \beta = 0.1 $ is seen to consistently perform well. In all cases it outperforms CMH by producing accurate estimates with low TVD with small variation across the many trials. We therefore recommend using $ \beta =0.1 $ and adopt this value for the remaining examples, while leaving optimization of $\beta$ to future work. CMH ($\beta =0$) meanwhile shows either consistently high TVD (always missing modes) or a very high spread in the TVD (finding different modes in different trials
). The latter case results in multi-modal violin plots.   

Figure~\ref{fig:acceptance_rate_exploration} explores the acceptance rate as a function of the exploration ratio $\beta$ for each of the nine distributions. Here, we see that in all cases the acceptance rate diminishes only a small amount by introducing a small fraction of exploration steps ($\beta \le 0.1$). In other words, some exploration can be achieved at the expense of only a modest number of rejections. However, once the exploration ratio grows larger ($\beta \ge 0.3$), exploration begins to come at a significant expense and begins to cause the acceptance rate to drop precipitously. This further reinforces our selection of $\beta=0.1$, which balances high accuracy with only a modest decrease in acceptance rate.

Figures~\ref{fig:tvd_chain_length}--\ref{fig:mean_chain_length} show the convergence of Intrepid MCMC and CMH for increasing chain length.  Here we see several compelling and important features of the Intrepid and CMH convergence. Figure~\ref{fig:tvd_chain_length} shows the TVD for increasing chain length where we clearly see that Intrepid always convergences in distribution to the multi-modal target as the chain length grows larger. Meanwhile, CMH often fails to converge in distribution to the target because it consistently fails to find one or more modes. This results in either a consistently poor TVD (Figure~\ref{fig:rosenbrock_ring_tvd_cmh_intrepid}) or a bimodal violin plot implying that CMH can sometimes find all modes but often cannot (Figures~\ref{fig:gauss_circles_tvd_cmh_intrepid},~\ref{fig:rosenbrock_circles_tvd_cmh_intrepid}). Similarly, Figure~\ref{fig:mean_chain_length} shows convergence in the error of the mean value computed for chains of increasing length. Again, we see that Intrepid consistently converges to very small (near zero) error for long chains while CMH often produces inaccurate mean estimates with large error. Interestingly, we observe that Intrepid MCMC showcases its improved convergence rates differently for different target shapes. In some cases (Gauss-Ring (Case 1), Gauss-Planes (Case 2)), we see that while the medians in TVD plots (Figures~\ref{fig:gauss_ring_tvd_cmh_intrepid},~\ref{fig:gauss_planes_tvd_cmh_intrepid}, respectively) imply that CMH and Intrepid have comparable rates of convergence, the median error in the mean (Figures~\ref{fig:gauss_ring_mean_cmh_intrepid},~\ref{fig:gauss_planes_mean_cmh_intrepid}, respectively) decreases much faster for Intrepid MCMC. On the other hand, sometimes (Rosenbrock-Planes (Case 8)) the median errors in the mean (Figure~\ref{fig:rosenbrock_planes_mean_cmh_intrepid}) remain comparable for both methods while the median TVD (Figure~\ref{fig:rosenbrock_planes_tvd_cmh_intrepid}) decreases more rapidly for Intrepid MCMC. 

\subsection{Varying Dimensions}
\label{section:dimension_results}

To study the performance of Intrepid MCMC in higher dimensions, we consider a general $d-$dimensional version of Case 2 (Gauss-Planes) from Section~\ref{section:distribution_shape_results}. Here, the target distribution takes the form
\begin{equation}
    \label{eqn:high_dimension_targets}
    \pi (\mathbf{x}) = I_1 (\mathbf{x}) \varphi_d (\mathbf{x})
\end{equation}
where $ I_1 (\mathbf{x}) $ is as defined in Table~\ref{tab:indicator_definitions}, and $ \varphi_d (\mathbf{x}) $ is the $ d $-dimensional standard Gaussian density function. We use both CMH and Intrepid MCMC with $ \beta = 0.1 $ to sample from the target for $ d = \left\{ 3, 5, 10, 30, 50 \right\} $, again performing 100 independent trials.
The proposal distributions used for the Markov chains were $ \mathfrak{q}_{(d-1)} \left( \phi | \theta_{s, (d-1)} \right) \equiv \textit{Uniform} \left( - \theta_{s, (d-1)} , 2 \pi - \theta_{s, (d-1)} \right) $, $ \mathfrak{q}_{j} \left( \phi | \theta_{s, j} \right) \equiv \textit{Uniform} \left( - \theta_{s,j} , \pi - \theta_{s,j} \right) $, $ j = \left\{ 1, \dots, (d-2) \right\} $, and $ \mathfrak{q}_r \left( \gamma \right) \equiv \textit{Uniform} \left( 0.5, 2.0 \right) $ for the Intrepid kernel, and $ q_{L_i} \left( z | x_{s, i} \right) \equiv \mathcal{N} \left( x_{s, i}, 1 \right) $, $ i = \left\{ 1, \dots, d \right\} $ for the CMH kernel.

\begin{figure}[h!t]
\centering
\includegraphics[width=0.75\linewidth]{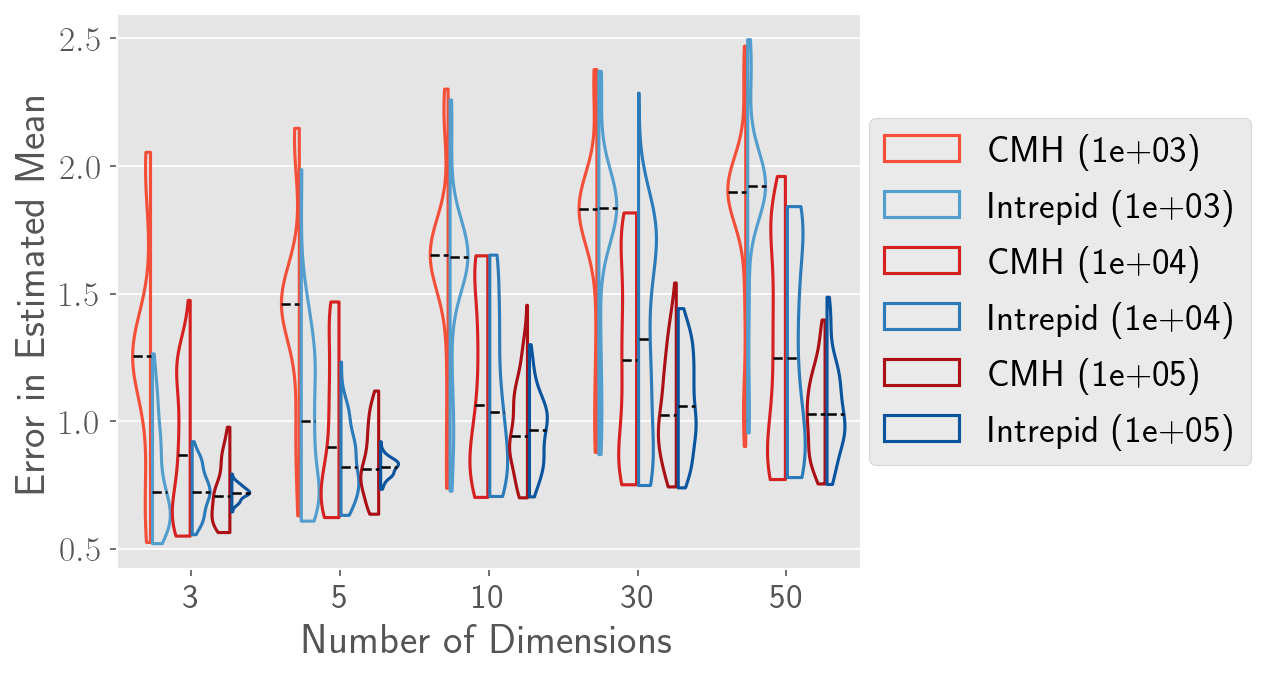}
\caption{Convergence in mean plot for Section~\ref{section:dimension_results}. The plotted quantity is the Euclidean distance between the true mean and the mean estimated from MCMC, normalized by the square root of the trace of the true covariance. 100 independent Markov chains started at randomly chosen locations are used to generate the violin plots. The legend lists the chain length associated with each violin in parentheses.}
\label{fig:multi_dim_mean}
\end{figure}

\begin{figure}[h!t]
\centering
\includegraphics[width=0.75\linewidth]{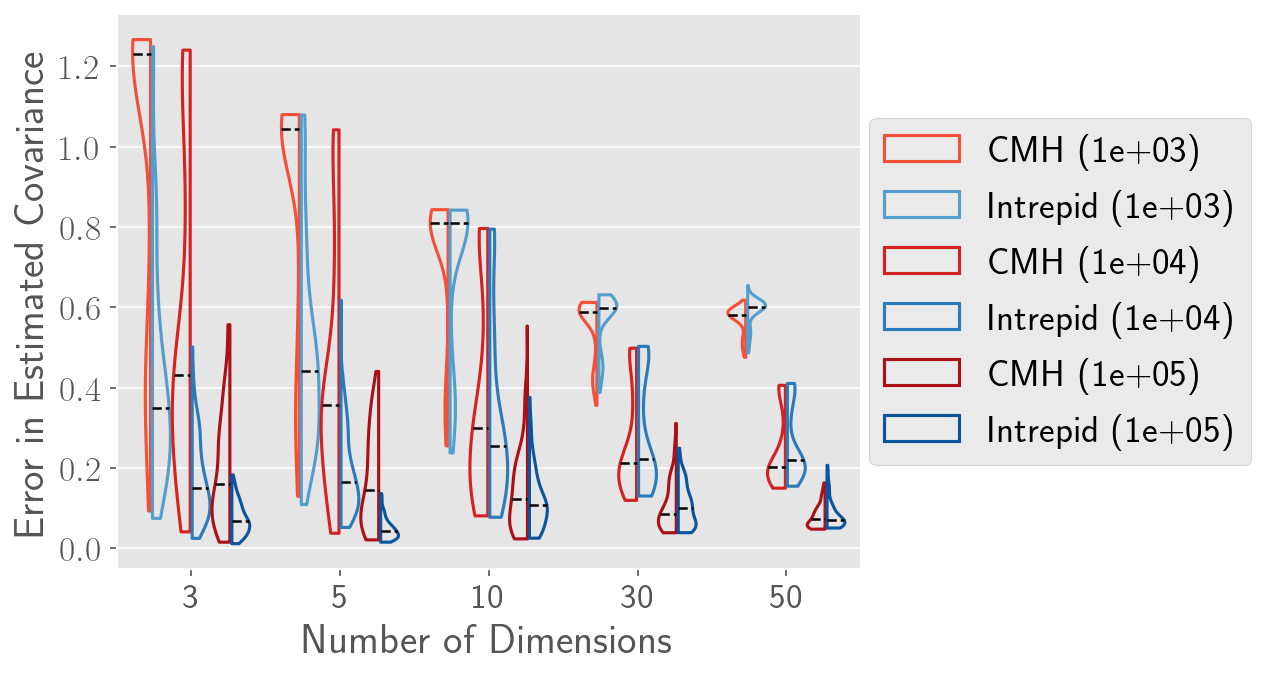}
\caption{Convergence in covariance plot for Section~\ref{section:dimension_results}. The plotted quantity is the Frobenius norm between the true covariance and the covariance estimated from MCMC, normalized by the square root of the trace of the true covariance. 100 independent Markov chains started at randomly chosen locations are used to generate the violin plots. The legend lists the chain length associated with each violin in parentheses.}
\label{fig:multi_dim_cov}
\end{figure}

Convergence is assessed by considering the error in the mean and the covariance estimated using the two MCMC methods, with the true statistics calculated using 50 million IID samples generated from each target by rejection sampling. Figures~\ref{fig:multi_dim_mean} and~\ref{fig:multi_dim_cov} plot the convergence statistics for CMH and Intrepid MCMC ($ \beta = 0.1 $) for different dimensions and three different chain lengths. Error in the mean is expressed through the $ l^2 $-norm of the difference between the estimated and true means 
normalized by the square root of the trace of the true target covariance. Error in the covariance is assessed through the Frobenius norm between the covariance matrices, again normalized by the square root of the trace of the true covariance matrix. 
For smaller dimensions, Intrepid MCMC converges faster than CMH, having lower overall error, and/or smaller spread in estimated error. As dimensionality increases, the performance of Intrepid MCMC degrades to become similar to that of CMH. However, it is noteworthy that even for a large number of dimensions, Intrepid MCMC does not perform worse than CMH. As before, when the violins show two distinct modes, each mode corresponds to an MCMC chain favoring one of the two modes of the target distribution. We further note that additional considerations may be explored to improve performance in moderate-to-high dimension, such as taking component-wise exploratory steps, but these are beyond the scope of this work and are left for future study.  

\subsection{Intrepid Markov Chain Mixing Behavior}
\label{section:intrepid_chain_behavior}

To study the mixing behavior of the Intrepid Markov chain, we revisit Case 7 (Rosenbrock-Ring) from Section~\ref{section:distribution_shape_results}, which is perhaps the most difficult distribution to sample as evidenced by the convergence plots in Figures~\ref{fig:tvd_exploration}--\ref{fig:mean_chain_length} showing that CMH cannot sample from this distribution.
Using the proposal parameters listed in Section~\ref{section:distribution_shape_results} and $ \beta = 0.1 $, we run multiple Intrepid Markov chains, all starting at the same point $ \mathbf{x}_0 = \begin{bmatrix} 0.0 & -4.2 \end{bmatrix}^T $. This starting point is chosen as a potential ``worst-case'' scenario because the sample lies within a very low probability mode of the target distribution (see lower contour line in Figure~\ref{fig:rosenbrock_ring_target} that is very hard to escape due to its separation from the important modes.

\begin{figure}[h!t]
\centering
\begin{subfigure}{.47\textwidth}
  \centering
  \includegraphics[width=\linewidth]{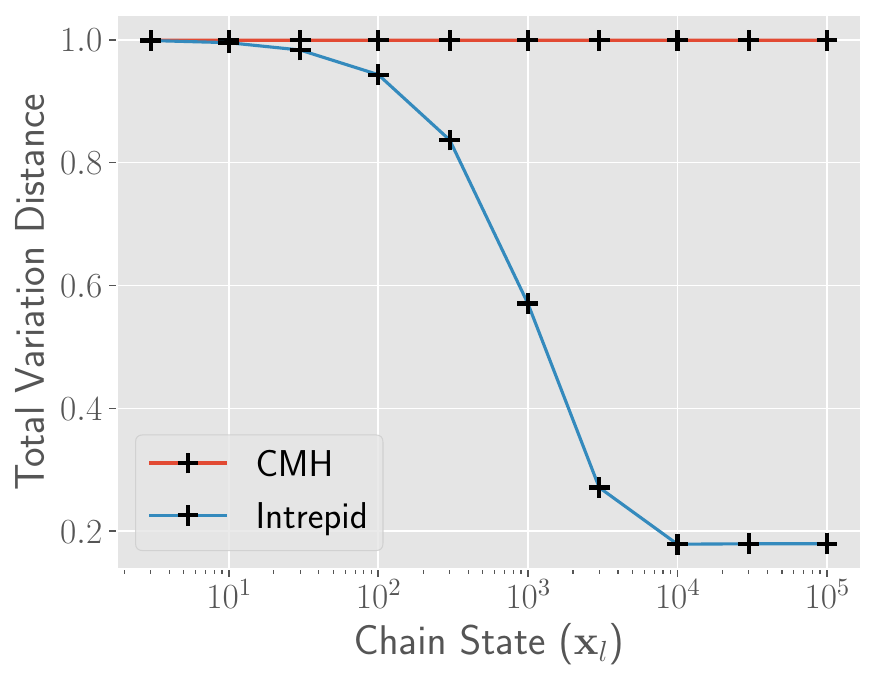}
  \caption{}
  \label{fig:ensemble_tvd}
\end{subfigure}%
\begin{subfigure}{.5\textwidth}
  \centering
  \includegraphics[width=\linewidth]{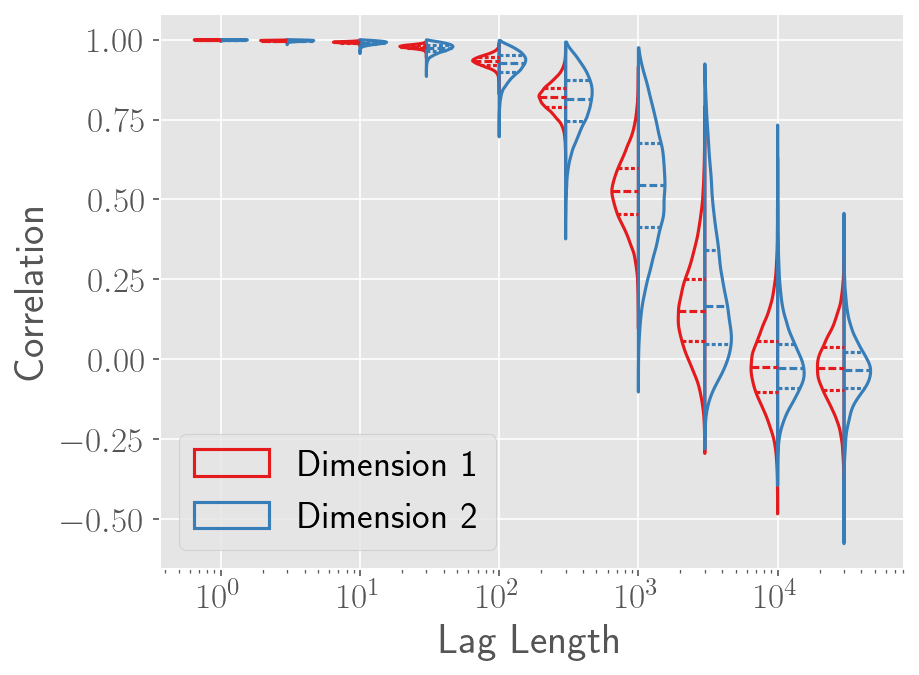}
  \caption{}
  \label{fig:ensemble_lag_k_correlation}
\end{subfigure}
\caption{Results for Section~\ref{section:intrepid_chain_behavior}. (a) Plot of total variation distance of the ensemble distribution of the $ l $-th state of the Markov chain ($ \mathbf{x}_l $) compared to the target distribution, computed using 148,000 independent chains started from the same location for both Intrepid MCMC and CMH. (b) Plot of the lag-$ k $ correlation of the Intrepid Markov chain along both dimensions, with statistics compiled from 29,600 independent Markov chains.}
\label{fig:chain_behavior_results}
\end{figure}

First, we consider the ensemble distribution of state $ \mathbf{x}_l $ of the chain. Figure~\ref{fig:ensemble_tvd} plots the TVD between the target distribution and the ensemble distribution of $ \mathbf{x}_l $ calculated using 148,000 independent Markov chains for increasing values of $l$. Here, we see that CMH never converges because it is unable to escape the low probability mode. On the other hand, the ensemble distribution of the Intrepid Markov chain begins to converge after a few thousand states have been generated. The plot also indicates that our chosen burn-in length of 10,000 samples (used in Sections~\ref{section:distribution_shape_results},~\ref{section:dimension_results}, and~\ref{section:bayesian_example}) is likely sufficient.

Next, we study the correlation of the Intrepid Markov chains.Figure~\ref{fig:ensemble_lag_k_correlation} plots the lag-$ k $ correlation statistics for 29,600 independent Markov chains. Due to the extremely narrow nature of the modes of the target distribution, the correlation is quite high in the beginning. However, for higher lag lengths, the correlation rapidly goes to zero, indicating that the chain has dispersed over all the modes of the target distribution.

\subsection{Bayesian Inference for a 2DOF Moment Resisting Frame under Free Vibration}
\label{section:bayesian_example}

Lastly, we analyze the undamped two-degree-of-freedom shear building model from~\cite{AuBeck2002bayesian, StraubPapaioannou2015bus}, using Bayesian inference to identify the inter-story stiffnesses. The mass of the stories are $ m_1 = 16.531 \cdot 10^3 $ \si{\kilogram} and $ m_2 = 16.131 \cdot 10^3 $ \si{\kilogram}, while the stiffness of each story is stochastic and parameterized as $ k_1 = k_0 x_1 $ and $ k_2 = k_0 x_2 $, where $ k_0 = 29.7 \cdot 10^6 $ \si{\newton \per \meter}, and $ \mathbf{x} = \begin{bmatrix} x_1 & x_2 \end{bmatrix}^T $ is the random vector of inputs. The parent distribution $ p (\mathbf{x}) $ is the prior distribution on $ \mathbf{x} $, which is given as the product of two independent Lognormal distributions for $ x_1 $ and $ x_2 $, with modes at $ 1.3 $ and $ 0.8 $, respectively, and standard deviations of unity. The transformation function $ T (\mathbf{x}) $ is the Likelihood, formulated using the modal measure-of-fit function $ J (\mathbf{x}) $ (as described in~\cite{AuBeck2002bayesian}).
\begin{gather}
    T (\mathbf{x}) = \mathcal{L} (\mathbf{x}) \propto \exp \left[ - \frac{J (\mathbf{x})}{2 \sigma_{\varepsilon}} \right] \label{eqn:bayesian_example_transformation_function} \\
    J (\mathbf{x}) = \sum_{j=1}^2 \left[ \frac{\hat{f}_j^2 (\mathbf{x})}{\Tilde{f}_j^2} - 1 \right]^2 \label{eqn:modal_measure_of_fit}
\end{gather}
where $ \hat{f}_j (\mathbf{x}) $ is the $ j $-th eigenfrequency of the building predicted for the parameter values $ \mathbf{x} $, $ \Tilde{f}_j $ is the corresponding measured frequency ($ \Tilde{f}_1 = 3.13 $ \si{\hertz}, $ \Tilde{f}_2 = 9.83 $ \si{\hertz}), and $ \sigma_{\varepsilon} = \nicefrac{1}{16} $ is the standard deviation of the predicted error.

The resulting target distribution $ \pi (\mathbf{x}) = T(\mathbf{x}) p(\mathbf{x}) $ is the posterior distribution on $ \mathbf{x} $ and is bimodal with modes that are effectively disconnected from each other. We compare the TVD, normalized error in estimated mean, and acceptance rate for Intrepid MCMC with $ \beta = 0.1 $ against CMH, all calculated using the same framework as described in Section~\ref{section:distribution_shape_results}. The number and seeding of chains, burn-in length, and angular and radial proposals used are also the same as in Section~\ref{section:distribution_shape_results}. However, two different proposal distributions are used for the CMH kernel; $ q_{L_i} \left( z | x_{s, i} \right) \equiv \mathcal{N} \left( x_{s, i}, \sigma_b \right) $, $ i = 1, 2 $, for $ \sigma_b = 1 $ and $ \sigma_b = 0.25 $. 

\begin{figure}[h!t]
\centering
\begin{subfigure}{.31\textwidth}
  \centering
  \includegraphics[width=\linewidth]{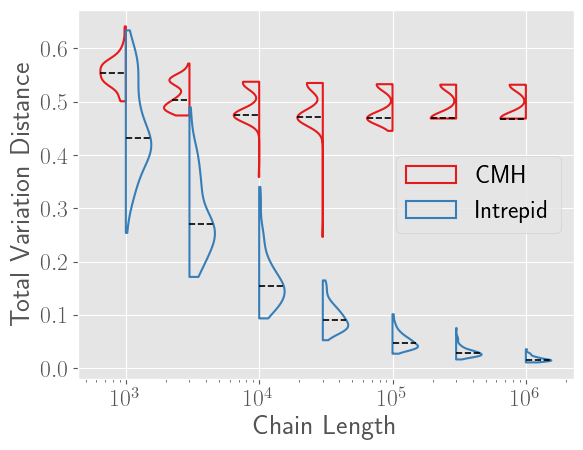}
  \caption{}
  \label{fig:bayesian_large_tvd}
\end{subfigure}%
\begin{subfigure}{.31\textwidth}
  \centering
  \includegraphics[width=\linewidth]{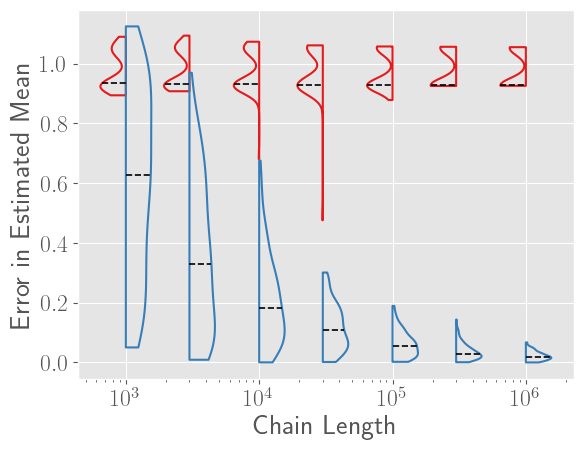}
  \caption{}
  \label{fig:bayesian_large_mean}
\end{subfigure}%
\begin{subfigure}{.31\textwidth}
  \centering
  \includegraphics[width=\linewidth]{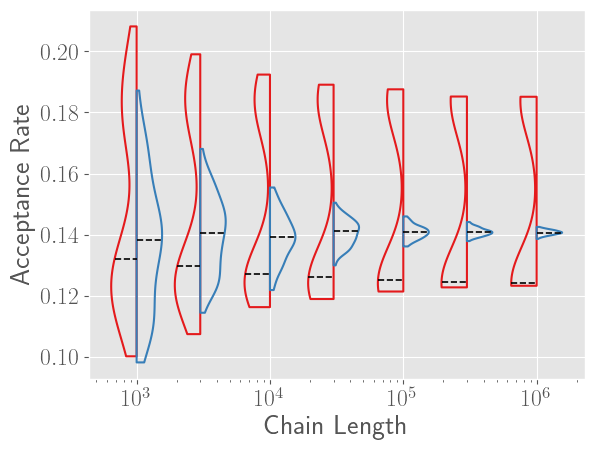}
  \caption{}
  \label{fig:bayesian_large_acceptance_rate}
\end{subfigure}
\begin{subfigure}{.31\textwidth}
  \centering
  \includegraphics[width=\linewidth]{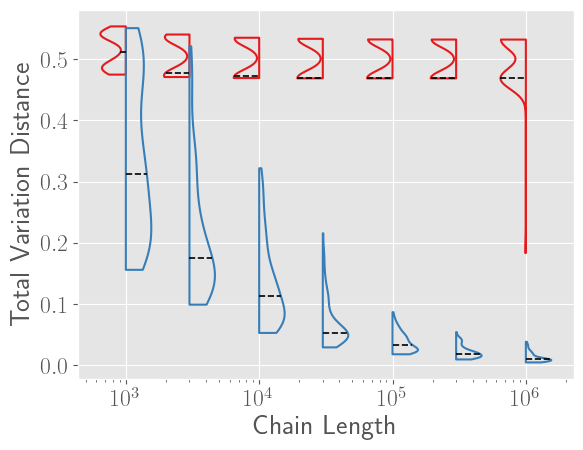}
  \caption{}
  \label{fig:bayesian_small_tvd}
\end{subfigure}%
\begin{subfigure}{.31\textwidth}
  \centering
  \includegraphics[width=\linewidth]{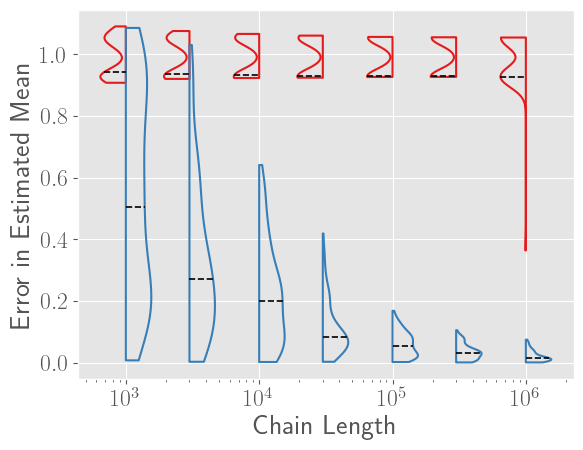}
  \caption{}
  \label{fig:bayesian_small_mean}
\end{subfigure}%
\begin{subfigure}{.31\textwidth}
  \centering
  \includegraphics[width=\linewidth]{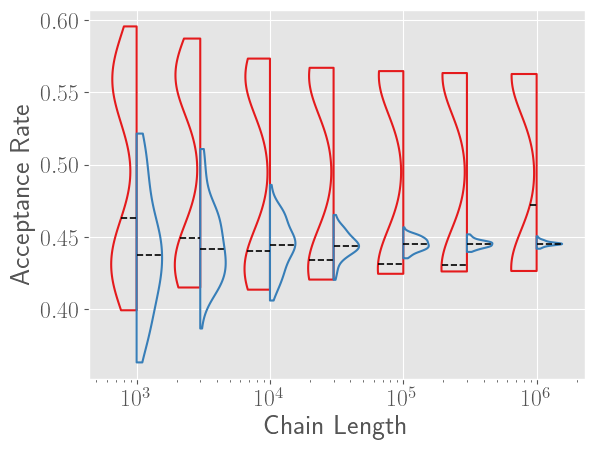}
  \caption{}
  \label{fig:bayesian_small_acceptance_rate}
\end{subfigure}
\caption{Results for the Bayesian inference problem (Section~\ref{section:bayesian_example}). (a) Plot of Total Variation Distance vs. chain length for CMH and Intrepid MCMC with $ \beta = 0.1 $, for CMH proposal with $ \sigma_b = 1 $, (b) plot of error in estimated mean vs. chain length for CMH and Intrepid MCMC with $ \beta = 0.1 $, for CMH proposal with $ \sigma_b = 1 $, (c) plot of acceptance rate vs. chain length for CMH and Intrepid MCMC with $ \beta = 0.1 $, for CMH proposal with $ \sigma_b = 1 $, (d) plot of Total Variation Distance vs. chain length for CMH and Intrepid MCMC with $ \beta = 0.1 $, for CMH proposal with $ \sigma_b = 0.25 $, (e) plot of error in estimated mean vs. chain length for CMH and Intrepid MCMC with $ \beta = 0.1 $, for CMH proposal with $ \sigma_b = 0.25 $, (f) plot of acceptance rate vs. chain length for CMH and Intrepid MCMC with $ \beta = 0.1 $, for CMH proposal with $ \sigma_b = 0.25 $. In all cases, 100 independent Markov chains seeded at randomly selected points were used to compute the convergence statistics. The legend is common for all subfigures and is provided in (a).}
\label{fig:bayesian_results}
\end{figure}

\begin{figure}[h!t]
\centering
\begin{subfigure}{.45\textwidth}
  \centering
  \includegraphics[width=\linewidth]{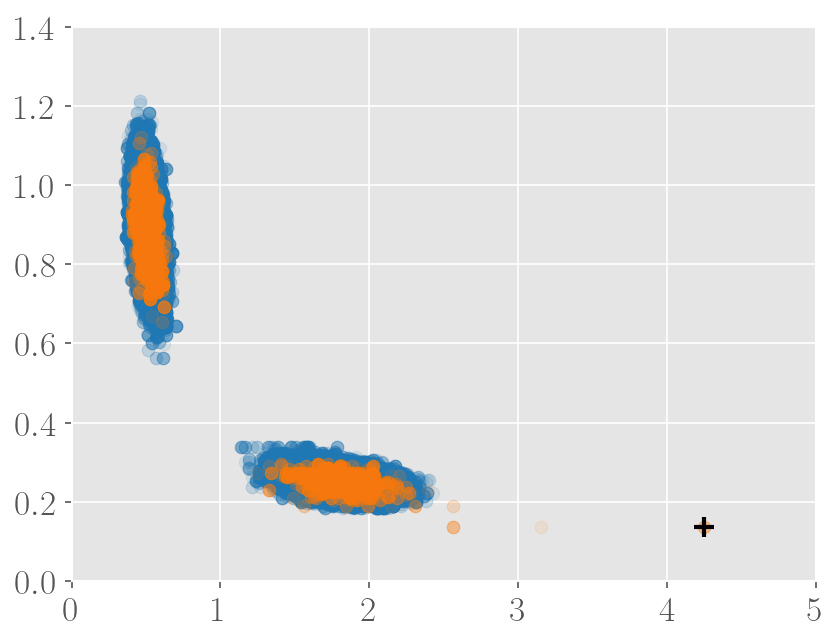}
  \caption{}
  \label{fig:intrepid_scatterplot}
\end{subfigure}%
\begin{subfigure}{.45\textwidth}
  \centering
  \includegraphics[width=\linewidth]{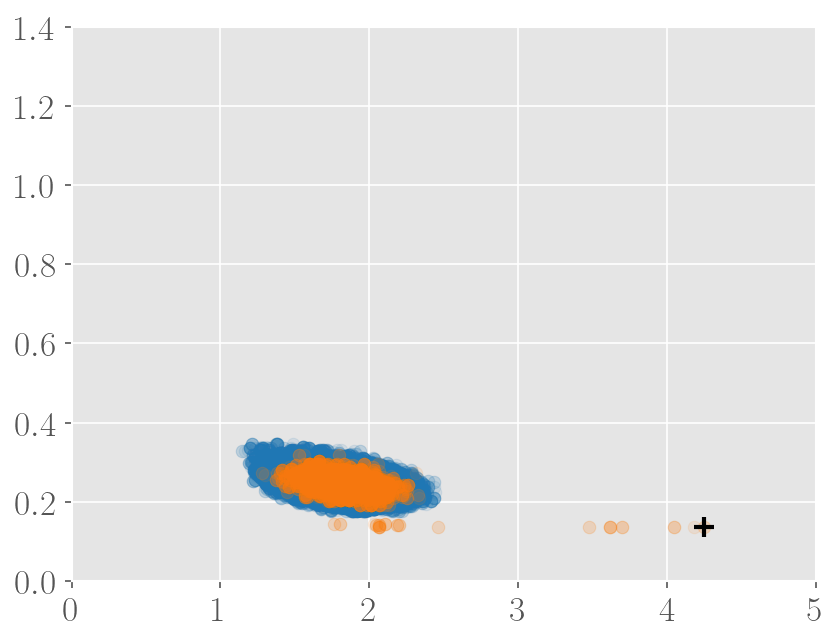}
  \caption{}
  \label{fig:cmh_scatterplot}
\end{subfigure}%
\caption{Representative scatterplot of samples generated by (a) CMH and (b) Intrepid MCMC ($ \beta = 0.1 $). Both cases use the CMH proposal with $ \sigma_b = 1 $. The 10,000 burn-in samples are plotted in orange, all subsequent samples are plotted in blue, and the initial state of the Markov chain is highlighted by the black cross.}
\label{fig:bayesian_large_proposal_scatterplot}
\end{figure}

The results plotted in Figure~\ref{fig:bayesian_results} clearly highlight that Intrepid MCMC does a significantly better job of sampling from the target distribution for both CMH proposals. The bimodality of the convergence measures for CMH -- arising from the bimodality of the target distribution as explained in Section~\ref{section:distribution_shape_results} -- is stark, strongly indicating that CMH gets stuck in one mode or the other nearly every time. By contrast, convergence statistics for Intrepid MCMC are almost completely unimodal in all cases, indicating that every independent trial explores both modes. This observation is further supported by the representative scatterplots in Figure~\ref{fig:bayesian_large_proposal_scatterplot}
, which compare the samples generated by one Intrepid MCMC chain to those generated by one CMH chain (with $ \sigma_b = 1 $); the plot clearly shows how CMH gets stuck in a single mode and is unable to locate the second mode, while Intrepid MCMC is able to locate both modes even in the burn-in phase.

\section{Observations and Conclusions}

Efficient and robust sampling techniques are essential for analyzing engineering systems under uncertainty. MCMC is a class of useful sampling methods for sampling from arbitrary distributions that provides theoretical guarantees on convergence to unnormalized target distributions with arbitrary shapes. Within MCMC, the Metropolis-Hastings family of algorithms is hugely popular due to their simplicity and ease of implementation. However, despite theoretical guarantees, usual constructions of MH struggle to sample from multimodal distributions. In this paper, we introduce Intrepid MCMC as a special case of MH explicitly designed to mitigate this issue.

The proposed Intrepid MCMC maintains most of the simplicity of vanilla MH\footnote{In this manuscript, CMH is used as the representative algorithm for vanilla MH methods.}. But, instead of using a single proposal density tuned to balance exploiting local information and exploring the parameter space, Intrepid MCMC uses two separate proposals -- one for locating new modes in previously unexplored regions (the globally explorative proposal) and another for locally sampling known modes (the locally exploitative proposal). At each iteration, one of the proposals is chosen at random with probability $\beta$ to generate a candidate point. In this way, Intrepid MCMC switches between populating the local region of the parameter space and sweeping across the full space. Through a series of examples, we demonstrate how introducing even a small fraction of these highly exploratory steps to an otherwise vanilla MH scheme results in Intrepid MCMC quickly identifying and sampling all modes of the target distribution. We show how this results in accelerated and more robust convergence to complex (multi-modal) target distributions. 

We further demonstrate how Intrepid steps improve the robustness of MH sampling to the initial state.  While standard MH schemes have the propensity to get stuck in the local mode where the chain initiates, Intrepid chains are much less sensitive to the starting point (see also Appendix~\ref{appendix:shape_results_scatterplots}). As a consequence, standard practices for MH sampling often involve initiating
multiple independent chains started from a dispersed set of initial states. 
Intrepid MCMC, on the other hand, provides reliable sampling without the need to generate a large number of chains. 
For this reason, we intentionally do not compare against multi-chain MCMC methods. Such investigation is an open topic for future research.

Despite the clear improvements in the convergence behavior, future research into Intrepid MCMC can potentially enhance its performance. For example, we observe that the Intrepid MCMC convergence behavior deteriorates with increasing dimensionality, becoming comparable to CMH for approximately 10 or more dimensions. Further work is needed to reduce the dimension dependence of Intrepid MCMC. This may be potentially addressed by considering component-wise intrepid steps or through other carefully design exploration steps. Furthermore, due to the Markovian property (the chain has no memory), the Intrepid Markov chain jumps has no inbuilt mechanism for returning to the previously explored modes besides luck. Possible solutions to this issue include storing some history to inform the exploitative proposal or splitting chains when a candidate from the explorative proposal is accepted. These and many other ideas -- including integration of Intrepid MCMC with existing enhancements like delayed rejection, adaptive tuning, and multi-chain methods -- can potentially improve Intrepid MCMC even further.

In conclusion, the key takeaway from this work is that the convergence behavior of random-walk Metropolis methods can be enhanced for complex distributions by enabling the chain to occasionally take jumps to new regions of the parameter space.
We present one possible way to construct such a proposal. But the Intrepid proposal is likely not the only proposal that can lead to similar improvements. 
To our knowledge, no MCMC methods exist that propose blind exploratory jumps 
in an effort to locate undiscovered regions of the target distribution. Perhaps this is due to apprehensions about wasting computational effort or large rejection rates. Rather, MCMC methods aimed at sampling multimodal distributions rely on expensive gradient information, prior optimization runs, or other sophisticated constructions (as summarized in section~\ref{section:MH_description}), which are much more difficult to implement than random-walk Metropolis and may be computationally expensive in their own right. However, using the proposed algorithm, we show that strategies exist that can make random-walk Metropolis viable for sampling from highly complex distribution shapes, as long as one is willing to be \textit{intrepid}~\footnote{\textit{Intrepid} : Adjective : Characterized by resolute fearlessness, fortitude, and endurance. (From the Merriam-Webster dictionary.)}.

\section*{Acknowledgments}
This research was supported through the INL Laboratory Directed Research and Development (LDRD) Program under DOE Idaho Operations Office Contract DE-AC07-05ID14517. 
Portions of this work were carried out at the Advanced Research Computing at Hopkins (ARCH) core facility (https://www.arch.jhu.edu/), which is supported by the National Science Foundation (NSF) Grant Number OAC1920103. 

The authors are grateful to Ayush Basu and Shayamal Singh for illuminating discussions around Theorem~\ref{thm:Sigma_converges_one_point}, and to Connor Krill for help with data visualization.

\appendix

\section{Derivations associated with the Intrepid Kernel Generating Function}
\label{appendix:intrepid_proposal_and_acceptance_derivation}

For the points $ \mathbf{x}_c, \mathbf{x}_s \in \Omega $ such that $ \mathbf{x}_s \notin d \mathbf{x}_c $, where $ d \mathbf{x}_c $ is an infinitesimal volume element around the point $ \mathbf{x}_c $, we can simplify the globally explorative kernel $ \mathcal{K}_I \left( \mathbf{x}_s, d \mathbf{x}_c \right) $ (Eq.~\eqref{eqn:intrepid_kernel_expansion}) as
\begin{align}
    \mathcal{K}_I (\mathbf{x}_s, d\mathbf{x}_c) &= \kappa_I (\mathbf{x}_s, \mathbf{x}_c) d\mathbf{x}_c + r_I(\mathbf{x}_s) \delta_{x_s} (d\mathbf{x}_c) \\
    &= \kappa_I (\mathbf{x}_s, \mathbf{x}_c) d\mathbf{x}_c \label{eqn:simplified_intrepid_kernel_when_s_not_close_to_c} \\
    &= q_I (\mathbf{x}_c | \mathbf{x}_s) \alpha_I (\mathbf{x}_s, \mathbf{x}_c) d\mathbf{x}_c \label{eqn:simplified_expanded_intrepid_kernel_when_s_not_close_to_c}
\end{align}

\textbf{\textit{We first consider the case where the Radial Transformation Function exists.}} Based on the procedure defined in Section~\ref{section:Intrepid_proposal_theory}, the probability of proposing a move to the set $ d \mathbf{x}_c $ is $ q_I (\mathbf{x}_c | \mathbf{x}_s) d\mathbf{x}_c $ and depends on the distributions $ \mathfrak{q}_r (\gamma) $ and $ \mathfrak{q}_{j} (\phi | \theta_{s, j}) $ $ \forall j \in \left\{ 1, \dots, (d-1) \right\} $ as follows:
\begin{equation}
\label{eqn:intrepid_proposal_relationship}
     q_I (\mathbf{x}_c | \mathbf{x}_s) d\mathbf{x}_c = \left[ \mathfrak{q}_r (\gamma) d \gamma \right] \left[ \prod_{j=1}^{d-1} \mathfrak{q}_{j} (\phi_j | \theta_{s, j}) d \phi_j \right]
\end{equation}

Since the proposal distributions are in hyperspherical coordinates, the infinitesimal volume element $ d \mathbf{x}_c $ is also a hyperspherical volume element, i.e.
\begin{equation}
    \label{eqn:volume_element_dy}
    \begin{aligned}
        d\mathbf{x}_c &= r_c^{d-1} \sin^{d-2} \left( \theta_{c, 1} \right) \sin^{d-3} \left( \theta_{c, 2} \right) \dots \sin \left( \theta_{c, (d-2)} \right) dr_c d \theta_{c, 1} d \theta_{c, 2} \dots d \theta_{c, (d-1)} \\
        &= \left[ r_c^{d-1} d r_c \right] \left[ \prod_{j=1}^{d-1} \sin^{\left( d-j-1 \right)} \left( \theta_{c, j} \right) d \theta_{c, j} \right]
    \end{aligned}
\end{equation}

From Eq.~\eqref{eqn:candidate_direction_component_angle}, we see that $ d \theta_{c, j} = d \phi_j $ (since $ d \theta_{s, j} $ is constant). Similarly, from Eq.~\eqref{eqn:intrepid_candidate_radius} (given that the Radial Transformation Function exists), $ r_c = R_{0, c} \left( \gamma R_{s, 0} \left( r_s \right) \right) \Rightarrow d r_c = R_{0, c}' \left( \gamma R_{s, 0} \left( r_s \right) \right) R_{s, 0} \left( r_s \right) d \gamma $, where $ R_{0, c}' \left( \cdot \right) $ is the derivative of $ R_{0, c} \left( \cdot \right) $ with $ r $ given the direction $ \boldsymbol{\theta}_0 $. Using the uniqueness property of the Radial Transformation Function stated in Definition~\ref{definition:radial_transformation_function} (Appendix~\ref{appendix:PRE_and_RTF}), we can say $ R_{c, 0} (r_c) = \gamma R_{s, 0} (r_s) $. Substituting these quantities into Eq.~\eqref{eqn:intrepid_proposal_relationship}, we get
\begin{multline}
     q_I (\mathbf{x}_c | \mathbf{x}_s) \left[ \left\{ R_{0, c} \left( \gamma R_{s, 0} \left( r_s \right) \right) \right\}^{d-1} R_{0, c}' \left( \gamma R_{s, 0} \left( r_s \right) \right) R_{s, 0} \left( r_s \right) d \gamma \right] \\ \left[ \prod_{j=1}^{d-1} \sin^{\left( d-j-1 \right)} \left( \theta_{c, j} \right) d \phi_j \right] = \left[ \mathfrak{q}_r (\gamma) d \gamma \right] \left[ \prod_{j=1}^{d-1} \mathfrak{q}_{j} (\phi_j | \theta_{s, j}) d \phi_j \right]
\end{multline}
\begin{equation}
    \Rightarrow q_I (\mathbf{x}_c | \mathbf{x}_s) = \frac{\mathfrak{q}_r (\gamma) \left[ \prod_{j=1}^{d-1} \mathfrak{q}_{j} (\phi_j | \theta_{s, j}) \right]}{\left[ \left\{ R_{0, c} \left( \gamma R_{s, 0} \left( r_s \right) \right) \right\}^{d-1} R_{0, c}' \left( \gamma R_{s, 0} \left( r_s \right) \right) R_{s, 0} \left( r_s \right) \right] \left[ \prod_{j=1}^{d-1} \sin^{\left( d-j-1 \right)} \left( \theta_{c, j} \right) \right]}
\end{equation}
\begin{equation}
    \label{eqn:intrepid_proposal_forward}
    \Rightarrow q_I (\mathbf{x}_c | \mathbf{x}_s) = \frac{\mathfrak{q}_r (\gamma) \left[ \prod_{j=1}^{d-1} \mathfrak{q}_{j} (\phi_j | \theta_{s, j}) \right]}{\left[ \left\{ r_c \right\}^{d-1} R_{0, c}' \left( R_{c, 0} \left( r_c \right) \right) R_{s, 0} \left( r_s \right) \right] \left[ \prod_{j=1}^{d-1} \sin^{\left( d-j-1 \right)} \left( \theta_{c, j} \right) \right]}
\end{equation}

The reverse step (i.e., $ \mathbf{x}_c \to \mathbf{x}_s $) is achieved when the value of the radial perturbation is $ \nicefrac{1}{\gamma} $ and the value of the perturbation angle for the $ j $-th component is $ - \phi_j $. Making the necessary changes to Eq.~\eqref{eqn:intrepid_proposal_forward}, we get
\begin{equation}
    \label{eqn:intrepid_proposal_backward_unsimplified}
    q_I (\mathbf{x}_s | \mathbf{x}_c) = \frac{\mathfrak{q}_r (\nicefrac{1}{\gamma}) \left[ \prod_{j=1}^{d-1} \mathfrak{q}_{j} (- \phi_j | \theta_{c, j}) \right]}{\left[ \left\{ r_s \right\}^{d-1} R_{0, s}' \left( R_{s, 0} \left( r_s \right) \right) R_{c, 0} \left( r_c \right) \right] \left[ \prod_{j=1}^{d-1} \sin^{\left( d-j-1 \right)} \left( \theta_{s, j} \right) \right]}
\end{equation}
\begin{equation}
    \label{eqn:intrepid_proposal_backward_simplified}
    \Rightarrow q_I (\mathbf{x}_s | \mathbf{x}_c) = \frac{\mathfrak{q}_r (\nicefrac{1}{\gamma}) \left[ \prod_{j=1}^{d-1} \mathfrak{q}_{j} (- \phi_j | \theta_{c, j}) \right]}{\left[ \left\{ r_s \right\}^{d-1} R_{0, s}' \left( R_{s, 0} \left( r_s \right) \right) \gamma R_{s, 0} \left( r_s \right) \right] \left[ \prod_{j=1}^{d-1} \sin^{\left( d-j-1 \right)} \left( \theta_{s, j} \right) \right]}
\end{equation}

Finally, we can write $ \rho_I (\mathbf{x}_s, \mathbf{x}_c) $ as defined in Eq.~\eqref{eqn:Intrepid_acceptance_rate}
\begin{align}
\label{eqn:rho_with_transformation}
    \rho_I (\mathbf{x}_s, \mathbf{x}_c) &= \frac{\pi(\mathbf{x}_c) q_I (\mathbf{x}_s | \mathbf{x}_c)}{\pi(\mathbf{x}_s) q_I (\mathbf{x}_c | \mathbf{x}_s)} \\
    &= \frac{1}{\gamma} \frac{\pi(\mathbf{x}_c)}{\pi(\mathbf{x}_s)} \frac{\mathfrak{q}_r (\nicefrac{1}{\gamma})}{\mathfrak{q}_r (\gamma)} \left[ \frac{r_c}{r_s} \right]^{(d-1)} \left[ \frac{R_{0, c}' \left( R_{c, 0} \left( r_c \right) \right)}{R_{0, s}' \left( R_{s, 0} \left( r_s \right) \right)} \right] \left[ \prod_{j=1}^{d-1} \frac{\mathfrak{q}_{j} (- \phi_j | \theta_{c, j})}{\mathfrak{q}_{j} (\phi_j | \theta_{s, j})} \right] \left[ \prod_{j=1}^{d-1} \frac{\sin^{\left( d-j-1 \right)} \left( \theta_{c, j} \right)}{\sin^{\left( d-j-1 \right)} \left( \theta_{s, j} \right)} \right] \label{eqn:intrepid_acceptance_ratio_fraction}
\end{align}

\textbf{\textit{If the Radial Transformation Function does not exist,}} we have $ r_c = \gamma r_s $ (Eq.~\eqref{eqn:intrepid_candidate_radius}) $ \Rightarrow d r_c = r_s d \gamma $. Also, the radial proposal distribution depends on $ \boldsymbol{\theta}_c $. Thus, we get
\begin{equation}
     q_I (\mathbf{x}_c | \mathbf{x}_s) \left[ \left\{ \gamma r_s \right\}^{d-1} r_s d \gamma \right] \left[ \prod_{j=1}^{d-1} \sin^{\left( d-j-1 \right)} \left( \theta_{c, j} \right) d \phi_j \right] = \left[ \mathfrak{q}_r (\gamma | \boldsymbol{\theta}_c) d \gamma \right] \left[ \prod_{j=1}^{d-1} \mathfrak{q}_{j} (\phi_j | \theta_{s, j}) d \phi_j \right]
\end{equation}
\begin{equation}
    \label{eqn:fwd_step_no_radial_equivalence}
     \Rightarrow q_I (\mathbf{x}_c | \mathbf{x}_s) = \frac{ \mathfrak{q}_r (\gamma | \boldsymbol{\theta}_c) \left[ \prod_{j=1}^{d-1} \mathfrak{q}_{j} (\phi_j | \theta_{s, j}) \right]}{ \gamma ^{(d-1)} r_s^d \left[ \prod_{j=1}^{d-1} \sin^{\left( d-j-1 \right)} \left( \theta_{c, j} \right) \right]}
\end{equation}

Similar to Eq.~\eqref{eqn:intrepid_proposal_backward_unsimplified}, the reverse step in this case can be given by
\begin{align}
     \Rightarrow q_I (\mathbf{x}_s | \mathbf{x}_c) &= \frac{ \mathfrak{q}_r \left( \left( \nicefrac{1}{\gamma} \right) | \boldsymbol{\theta}_s \right) \left[ \prod_{j=1}^{d-1} \mathfrak{q}_{j} (-\phi_j | \theta_{c, j}) \right]}{ \left( \nicefrac{1}{\gamma} \right) ^{(d-1)} r_c^d \left[ \prod_{j=1}^{d-1} \sin^{\left( d-j-1 \right)} \left( \theta_{s, j} \right) \right]} \\
     &= \frac{ \mathfrak{q}_r \left( \left( \nicefrac{1}{\gamma} \right) | \boldsymbol{\theta}_s \right) \left[ \prod_{j=1}^{d-1} \mathfrak{q}_{j} (-\phi_j | \theta_{c, j}) \right]}{ \gamma r_s^d \left[ \prod_{j=1}^{d-1} \sin^{\left( d-j-1 \right)} \left( \theta_{s, j} \right) \right]} \label{eqn:back_step_no_radial_equivalence}
\end{align}

Therefore, $ \rho_I (\mathbf{x}_s, \mathbf{x}_c) $ as defined in Eq.~\eqref{eqn:Intrepid_acceptance_rate} when no Radial Transformation Function exists, is
\begin{equation}
    \label{eqn:rho_with_no_transformation}
    \rho_I (\mathbf{x}_s, \mathbf{x}_c) = \gamma^{(d-2)} \frac{\pi(\mathbf{x}_c)}{\pi(\mathbf{x}_s)} \frac{\mathfrak{q}_r \left( \left( \nicefrac{1}{\gamma} \right) | \boldsymbol{\theta}_s \right)}{\mathfrak{q}_r (\gamma | \boldsymbol{\theta}_c)} \left[ \prod_{j=1}^{d-1} \frac{\mathfrak{q}_{j} (- \phi_j | \theta_{c, j})}{\mathfrak{q}_{j} (\phi_j | \theta_{s, j})} \right] \left[ \prod_{j=1}^{d-1} \frac{\sin^{\left( d-j-1 \right)} \left( \theta_{c, j} \right)}{\sin^{\left( d-j-1 \right)} \left( \theta_{s, j} \right)} \right]
\end{equation}

\section{Probabilistic Radial Equivalence and the Radial Transformation Function}
\label{appendix:PRE_and_RTF}

To define probabilistic radial equivalence for a distribution $ p (\mathbf{x}) $, we first choose an anchor point $ \mathbf{x}_a \in \Omega $ \footnote{For the entirety of this appendix, we will assume that $ \Omega $ is simply connected.} and define a hyperspherical coordinate system so that we may consider the behavior of $ p (\mathbf{x}) $ along the radial coordinate for different directions. Consider $ \Psi_{\theta} (r) $, the unnormalized radial conditional of $ p (\mathbf{x}) $ for the direction $ \boldsymbol{\theta} $, as defined in Eq.~\eqref{eqn:radial_conditional}.
Two directions $ \boldsymbol{\theta}_1 $ and $ \boldsymbol{\theta}_2 $ can be said to be ``\textit{radially transformable}'' with respect to the continuous density function $ p (\mathbf{x}) $ if and only if there exists a unique bijective mapping from every continuous interval of $ \Psi_{\boldsymbol{\theta}_1} \left( r \right) $ into a corresponding continuous interval of $ \Psi_{\boldsymbol{\theta}_2} \left( r \right) $, such that points mapped onto each other have the same value of $ p (\mathbf{x}) $.

\begin{figure}[!htbp]
\centering
\begin{subfigure}{.31\textwidth}
  \centering
  \includegraphics[width=\linewidth]{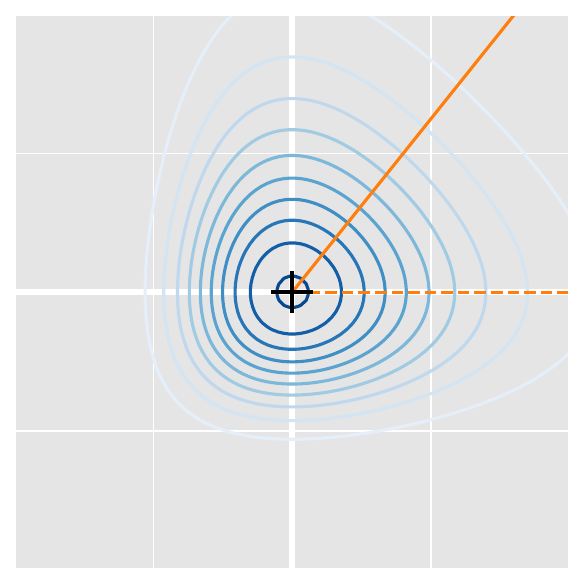}
  \caption{}
  \label{fig:convex_parent}
\end{subfigure}%
\begin{subfigure}{.31\textwidth}
  \centering
  \includegraphics[width=\linewidth]{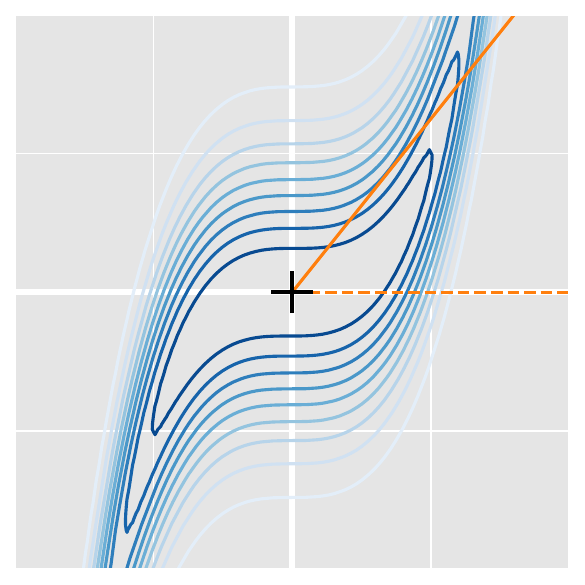}
  \caption{}
  \label{fig:non-convex_parent}
\end{subfigure}%
\begin{subfigure}{.31\textwidth}
  \centering
  \includegraphics[width=\linewidth]{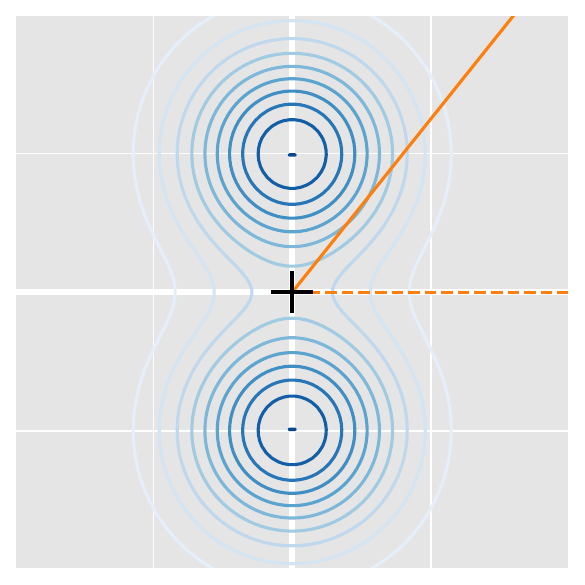}
  \caption{}
  \label{fig:bimodal_parent}
\end{subfigure}
\caption{Visualization of (a) a unimodal distribution with convex contours, (b) a unimodal distribution with non-convex contours, and (c) a bimodal distribution. Two different directional rays are plotted (one solid line and one dashed line). In case (a), both directional ray intersects all contours only once, suggesting it has probabilistic radial equivalence. For cases (b) and (c), the dashed ray intersects each contour only once or not at all, but the solid ray intersects multiple contours more than once, which implies that they do not have probabilistic radial equivalence.}
\label{fig:parent_visualization_for_radial_transformation}
\end{figure}

Intuitively, this mapping stretches or contracts different continuous intervals of $ \Psi_{\boldsymbol{\theta}_1} \left( r \right) $ until $ \Psi_{\boldsymbol{\theta}_2} \left( r \right) $ is reconstructed. Figure~\ref{fig:parent_visualization_for_radial_transformation} plots three different kinds of distributions (unimodal with convex contours, unimodal with non-convex contours, and bimodal) to highlight how the shape of the contours of a distribution can affect its behavior along different directions. For example, if a distribution $ p (\mathbf{x}) $ has directional rays along two different directions that do not have exactly the same number of intersection points with an arbitrarily chosen contour, then the two directions are not radially transformable. This is because the mapping cannot be one-to-one and also preserve the value of $ p (\mathbf{x}) $. This is the case in Figure~\ref{fig:parent_visualization_for_radial_transformation}(b) and (c), where the dashed ray intersects each contour once or not at all, but the solid ray intersects several contours more than once.

If every pair of directions is radially transformable for $ p (\mathbf{x}) $, then we say that $ p (\mathbf{x}) $ has \textit{probabilistic radial equivalence} and the mapping between $ \Psi_{\boldsymbol{\theta}_1} \left( r \right) $ and $ \Psi_{\boldsymbol{\theta}_2} \left( r \right) $ for any two directions $ \boldsymbol{\theta}_1 $ and $ \boldsymbol{\theta}_2 $ is called the \textit{Radial Transformation Function} (RTF). More rigorously, we define the following:

\begin{definition}[Radial Transformation Function]
\label{definition:radial_transformation_function}
The Radial Transformation Function of $ p (\mathbf{x}) $ is a mapping $ R_{1, 2}: \mathcal{X}_1 \to \mathcal{X}_2 $, where set $ \mathcal{X}_1 = \left\{ \mathbf{x} \in \Omega: \mathbf{x} = \mathbf{x}_a + r_1 \boldsymbol{\theta}_1 | \mathbf{x}_a, \boldsymbol{\theta}_1 \right\} $, set $ \mathcal{X}_2 = \left\{ \mathbf{x} \in \Omega: \mathbf{x} = \mathbf{x}_a + r_2 \boldsymbol{\theta}_2 | \mathbf{x}_a, \boldsymbol{\theta}_2 \right\} $, $ r_1, r_2 \in \left[ 0, \infty \right) $, and $ \boldsymbol{\theta}_1, \boldsymbol{\theta}_2 \in \left[ 0, 2 \pi \right) \times \left[ 0, \pi \right]^{(d-2)} $. Given $ r_1 $, $ \boldsymbol{\theta}_1 $, and $ \boldsymbol{\theta}_2 $, we get $ r_2 = R_{1, 2} (r_1) $ such that the following conditions are satisfied:
\begin{enumerate}
    \item $ p (\mathbf{x}_a + r_1 \boldsymbol{\theta}_1 ) = p (\mathbf{x}_a + r_2 \boldsymbol{\theta}_2 ) $
    \item If we also have $ r'_1 $ and $ r'_2 = R_{1, 2} (r'_1) $ such that $ r'_1 < r_1 $, then $ r'_2 < r_2 $ (i.e., the RTF is ``\textit{order-preserving}'').
\end{enumerate}
\end{definition}

To construct the general form of the RTF and derive a condition for its existence, we first define the following notation. For a direction $ \boldsymbol{\theta} $, define the set of radial coordinates $ \mathcal{R}_{\boldsymbol{\theta}} = \left\{ \rho_{\boldsymbol{\theta}}^{(i)}, i \in \left\{0, 1, \dots m_{\theta} \right\} \right\} $ where $ \rho_{\boldsymbol{\theta}}^{(0)} = 0 $ and $ \rho_{\boldsymbol{\theta}}^{(i)} < \rho_{\boldsymbol{\theta}}^{(i+1)} $, such that each $ \rho_{\boldsymbol{\theta}}^{(i)} $ falls in one of the following cases:
\begin{enumerate}[(a)]
    \item $ \Psi_{\boldsymbol{\theta}} \left( r \right) $ has a local minimum at $ r = \rho_{\boldsymbol{\theta}}^{(i)} $,
    \item $ \Psi_{\boldsymbol{\theta}} \left( r \right) $ has a local maximum at $ r = \rho_{\boldsymbol{\theta}}^{(i)} $,
    \item $ \rho_{\boldsymbol{\theta}}^{(i)} $ is the start of an interval of uniformity (i.e. $ \Psi_{\boldsymbol{\theta}} \left( r \right) $ is constant $ \forall r \in \left[ \rho_{\boldsymbol{\theta}}^{(i)}, \rho_{\boldsymbol{\theta}}^{(i+1)} \right] $), or
    \item $ \rho_{\boldsymbol{\theta}}^{(i)} $ is the end of an interval of uniformity (i.e. $ \Psi_{\boldsymbol{\theta}} \left( r \right) $ is constant $ \forall r \in \left[ \rho_{\boldsymbol{\theta}}^{(i-1)}, \rho_{\boldsymbol{\theta}}^{(i)} \right] $).
\end{enumerate} 
$ \mathcal{R}_{\boldsymbol{\theta}} $ is constructed such that there is no $ r \in \left[ 0, \infty \right) $ such that $ r \notin \mathcal{R}_{\boldsymbol{\theta}} $ satisfies one of the above cases for $ \Psi_{\boldsymbol{\theta}} \left( r \right) $. In other words, $ \mathcal{R}_{\boldsymbol{\theta}} $ contains all points along $ \Psi_{\boldsymbol{\theta}} \left( r \right) $ that satsify one of the conditions above. Next, define the intervals $ \mathcal{D}_{\boldsymbol{\theta}}^{(i)} = \left[ \rho_{\boldsymbol{\theta}}^{(i)}, \rho_{\boldsymbol{\theta}}^{(i+1)} \right) $ for $ i \in \left\{ 0, \dots m_{\theta}-1 \right\} $, and $ \mathcal{D}_{\boldsymbol{\theta}}^{(m_{\theta})} = \left[ \rho_{\boldsymbol{\theta}}^{(m_{\theta})}, \infty \right) $. Thus, $ \mathcal{D}_{\boldsymbol{\theta}}^{(i)} $ $ \forall i \in \left\{ 0, \dots, m_{\theta} \right\} $ form a partition of $ \left[ 0, \infty \right) $ (which is the domain of $ r $). Notice that within each interval, $ \Psi_{\boldsymbol{\theta}} \left( r \right) $ is either monotonically increasing, monotonically decreasing, or constant. We can now present a necessary and sufficient condition for the RTF to exist between any two directions (Theorem~\ref{thm:RTF_existence_direction_pair}).

\begin{theorem}[Existence of RTF for a Pair of Directions]
    \label{thm:RTF_existence_direction_pair}
    For a continuous density function $ p (\mathbf{x}) $ and directions $ \boldsymbol{\theta}_1 $ and $ \boldsymbol{\theta}_2 $, the RTF exists iff
    \begin{enumerate}
        \item $ m_1 = m_2 = m $ (the number of elements in $ \mathcal{R}_{\boldsymbol{\theta}_1} $ is the same as in $ \mathcal{R}_{\boldsymbol{\theta}_2} $)
        \item $ \Psi_{\boldsymbol{\theta}_1} \left( \rho_{\boldsymbol{\theta}_1}^{(i)} \right) = \Psi_{\boldsymbol{\theta}_2} \left( \rho_{\boldsymbol{\theta}_2}^{(i)} \right) $ $ \forall i \in \left\{ 0, \dots, m \right\} $
        \item $ \rho_{\boldsymbol{\theta}_1}^{(i)} $ and $ \rho_{\boldsymbol{\theta}_2}^{(i)} $ are of the same type $ \forall i \in \left\{ 0, \dots, m \right\} $ (i.e., local maximum, local minimum, or start or end of an interval of uniformity, as described above)
    \end{enumerate}
\end{theorem}

\begin{proof}[Proof of Theorem~\ref{thm:RTF_existence_direction_pair}]
    Since $ m_1 = m_2 $, the radial coordinate in both directions have been partitioned into the same number of intervals. Comparing $ \mathcal{D}_{\boldsymbol{\theta}_1}^{(i)} $ to $ \mathcal{D}_{\boldsymbol{\theta}_2}^{(i)} $, it is easy to see from the conditions 
    of the theorem that $ \Psi_{\boldsymbol{\theta}_1} \left( r \right) $ has the same behavior in the interval $ \mathcal{D}_{\boldsymbol{\theta}_1}^{(i)} $ that $ \Psi_{\boldsymbol{\theta}_2} \left( r \right) $ has in $ \mathcal{D}_{\boldsymbol{\theta}_2}^{(i)} $ (monotonically increasing, monotonically decreasing, or constant). Additionally, since $ p (\mathbf{x}) $ is continuous, $ \Psi_{\boldsymbol{\theta}_1} \left( r \right) $ and $ \Psi_{\boldsymbol{\theta}_2} \left( r \right) $ are both also continuous. This implies that $ \forall r_1 \in \mathcal{D}_{\boldsymbol{\theta}_1}^{(i)} $, $ \exists r_2 \in \mathcal{D}_{\boldsymbol{\theta}_2}^{(i)} $ such that $ \Psi_{\boldsymbol{\theta}_1} \left( r_1 \right) = \Psi_{\boldsymbol{\theta}_2} \left( r_2 \right) $.

    If $ \Psi_{\boldsymbol{\theta}_1} \left( r \right) $ and $ \Psi_{\boldsymbol{\theta}_2} \left( r \right) $ are both either monotonically increasing or monotonically decreasing in $ \mathcal{D}_{\boldsymbol{\theta}_1}^{(i)} $ and $ \mathcal{D}_{\boldsymbol{\theta}_2}^{(i)} $, define
    \begin{equation}
    \label{eqn:RTF_monotonic_interval}
        R_{1, 2} (r_1) = \Psi_{\boldsymbol{\theta}_2}^{-1} \left( \Psi_{\boldsymbol{\theta}_1} \left( r_1 \right) \right)
    \end{equation}
    where we only consider $ r_1 \in \mathcal{D}_{\boldsymbol{\theta}_1}^{(i)} $ and $ r_2 = R_{1, 2} (r_1) \in \mathcal{D}_{\boldsymbol{\theta}_2}^{(i)} $, thus the inverse exists because $ \Psi_{\boldsymbol{\theta}_2} \left( r \right) $ is monotonic and continuous in this interval.

    If instead $ \Psi_{\boldsymbol{\theta}_1} \left( r \right) $ and $ \Psi_{\boldsymbol{\theta}_2} \left( r \right) $ are both constant in $ \mathcal{D}_{\boldsymbol{\theta}_1}^{(i)} $ and $ \mathcal{D}_{\boldsymbol{\theta}_2}^{(i)} $, define
    \begin{equation}
    \label{eqn:RTF_uniform_interval}
        R_{1, 2} (r_1) = \rho_{\boldsymbol{\theta}_2}^{(i)} + \left( r_1 - \rho_{\boldsymbol{\theta}_1}^{(i)} \right) \left( \frac{\rho_{\boldsymbol{\theta}_2}^{(i+1)} - \rho_{\boldsymbol{\theta}_2}^{(i)}}{\rho_{\boldsymbol{\theta}_1}^{(i+1)} - \rho_{\boldsymbol{\theta}_1}^{(i)}} \right)
    \end{equation}
    where again we only consider $ r_1 \in \mathcal{D}_{\boldsymbol{\theta}_1}^{(i)} $ and $ r_2 = R_{1, 2} (r_1) \in \mathcal{D}_{\boldsymbol{\theta}_2}^{(i)} $.

    Both Eqs.~\eqref{eqn:RTF_monotonic_interval} and~\eqref{eqn:RTF_uniform_interval} result in unique transformations within their intervals, and the intervals themselves are disjoint and order-preserving. Thus, defining $ R_{1, 2} (r) $ in this way satisfies all the conditions required of the RTF as per Definition~\ref{definition:radial_transformation_function}. Hence, under the conditions of the theorem, the RTF exists and is unique, and is defined by Eqs.~\eqref{eqn:RTF_monotonic_interval} and~\eqref{eqn:RTF_uniform_interval}.
    
\end{proof}

Unfortunately, Theorem~\ref{thm:RTF_existence_direction_pair} is not always useful to judge the existence of the RTF in practice, as it is not possible to check every pair of directions for a distribution. Additionally, the RTF depends explicitly on the choice of the anchor point $ \mathbf{x}_a $, but Theorem~\ref{thm:RTF_existence_direction_pair} provides no guidance for its selection. We provide below another sufficient criterion for the existence of the RTF, along with a natural choice for the anchor point. Some necessary definitions are included first.

Let $ L_\zeta = \left\{ \mathbf{x} \in \Omega : p (\mathbf{x}) = \zeta \right\} $ be a level set of $ p (\mathbf{x}) $, and $ l_{\zeta, k} $ be a connected component of $ L_\zeta $. By definition, the family of sets $ l_{\zeta, k} $ $ \forall k $ forms a partition of $ L_\zeta $. Then we can say that $ p (\mathbf{x}) $ is a ``\textit{non-plateauing convex contoured density function}'' based on the following definition.

\begin{definition}[Non-plateauing Convex Contoured Density Function]
\label{definition:non_plateauing_convex_contoured_density}
    A density function $ p (\mathbf{x}) $ is a non-plateauing convex contoured density function if
    \begin{enumerate}
        \item $ p (\mathbf{x}) $ is continuous.
        \item $ \forall \mathbf{x} \in \Omega $, $ \exists \mathbf{y} \in \Omega $ where $ \lVert y - x \rVert_2 < \delta $ (for an arbitrarily small $ \delta > 0 $) such that $ p (\mathbf{x}) \neq p (\mathbf{y}) $. (Hence, $ p (\mathbf{x}) $ has no plateaus.)
        \item $ l_{\zeta, k} $ follows one of two conditions $ \forall \zeta, k $ (ensuring it has convex contours)
        \begin{enumerate}
            \item $ l_{\zeta, k} $ is a singleton set, i.e., it contains exactly one point, or
            \item $ l_{\zeta, k} $ is the boundary of a \textit{convex body with non-empty interior}, i.e., the boundary of the convex hull of $ l_{\zeta, k} $ is $ l_{\zeta, k} $ itself \footnote{Note that since $ \Omega \subseteq \mathbb{R}^d $ and a convex body is compact, such a body in $ \Omega $ must be bounded. Then, as a consequence of the \textit{Separation Theorem}, any convex body in $ \Omega $ is the convex hull of its boundary (see Lemma 1.4.1 in~\cite{schneider2013convex}).}.
        \end{enumerate}
    \end{enumerate}
\end{definition}

For non-plateauing convex contoured density functions, we further define the following notation:
\begin{itemize}
    \item $ \Lambda = \left\{ \left( \zeta, k \right) : l_{\zeta, k} \text{ contains exactly one point} \right\} $
    \item $ \Sigma = \left\{ \left( \zeta, k \right) : l_{\zeta, k} \text{ is the boundary of a convex body with non-empty interior} \right\} $
\end{itemize}
Thus, $ \left[ \bigcup_{\left( \zeta, k \right) \in \Lambda} l_{\zeta, k} \right] \cup \left[ \bigcup_{\left( \zeta, k \right) \in \Sigma} l_{\zeta, k} \right] = \bigcup_\zeta L_\zeta = \Omega $ and $ \Lambda \cap \Sigma = \emptyset $. Finally, notate $ S_{\zeta, k} = Conv(l_{\zeta, k}) $. As a consequence, $ l_{\zeta, k} = Bd(S_{\zeta, k}) $ and $ Int(S_{\zeta, k}) = \left( S_{\zeta, k} \setminus l_{\zeta, k} \right) $. (See Table~\ref{tab:set_notation} for the notation used henceforth.)

\begin{table}[!ht]
    \centering
    \begin{tabular}{c|c}
        Notation & Description \\
        \hline \hline
        $ Int(A) $ & The interior of set $ A $ \\
        $ Bd(A) $ & The boundary of set $ A $ \\
        $ Cl(A) $ & The closure of set $ A $ \\
        $ Conv(A) $ & The convex hull of set $ A $ \\
        $ diam(A) $ & The diameter of set $ A $ (using Euclidean distance) \\
        \hline
    \end{tabular}
    \caption{Some general notation for Appendix~\ref{appendix:PRE_and_RTF}.}
    \label{tab:set_notation}
\end{table}

\begin{theorem}
    \label{thm:Lambda_has_one_point}
    For a non-plateauing convex contoured density function, if $ \bigcap_{\left( \zeta, k \right) \in \Lambda} l_{\zeta, k} \neq \emptyset $, then $ \Lambda $ has only one element. 
\end{theorem}
\begin{proof}[Proof of Theorem~\ref{thm:Lambda_has_one_point}]
    By definition of connected components, $ l_{\zeta_1, k_1} \cap l_{\zeta_2, k_2} = \emptyset $. Thus, by contradiction, $ \Lambda $ cannot have more than one element. 
\end{proof}

When $ \Lambda $ has only one element, we will name it $ \left( \zeta_0, 0 \right) $, such that $ l_{\zeta_0, 0} $ is the only $ l_{\zeta, k} $ that is a singleton.

\begin{theorem}
    \label{thm:Sigma_converges_one_point}
    For a non-plateauing convex contoured density function $ p (\mathbf{x}) $, if $ \left[ \bigcap_{\left( \zeta, k \right) \in \Lambda} l_{\zeta, k} \right] \cap \left[ \bigcap_{\left( \zeta, k \right) \in \Sigma} S_{\zeta, k} \right] \neq \emptyset $, then 
    \begin{enumerate}
        \item $ \bigcap_{\left( \zeta, k \right) \in \Lambda} l_{\zeta, k} = l_{\zeta_0, 0} = \left\{ \mathbf{a} \right\} $, where $ \mathbf{a} \in \Omega $
        \item $ \bigcap_{\left( \zeta, k \right) \in \Sigma} S_{\zeta, k} = \left\{ \mathbf{a} \right\} $
        \item $ \left[ \bigcap_{\left( \zeta, k \right) \in \Lambda} l_{\zeta, k} \right] \cap \left[ \bigcap_{\left( \zeta, k \right) \in \Sigma} S_{\zeta, k} \right] = \left\{ \mathbf{a} \right\} $
    \end{enumerate}
\end{theorem}
\begin{proof}[Proof of Theorem~\ref{thm:Sigma_converges_one_point}]
    If $ \left[ \bigcap_{\left( \zeta, k \right) \in \Lambda} l_{\zeta, k} \right] \cap \left[ \bigcap_{\left( \zeta, k \right) \in \Sigma} S_{\zeta, k} \right] \neq \emptyset $, then
    \begin{gather}
        \bigcap_{\left( \zeta, k \right) \in \Lambda} l_{\zeta, k} \neq \emptyset \label{eqn:singletons_nonempty_intersection} \\
        \bigcap_{\left( \zeta, k \right) \in \Sigma} S_{\zeta, k} \neq \emptyset \label{eqn:convexes_nonempty_intersection}
    \end{gather}

Point 1 follows directly from Eq.~\eqref{eqn:singletons_nonempty_intersection} and Theorem~\ref{thm:Lambda_has_one_point}, and we simply name the single point as $ \mathbf{a} $.

For point 2, realize that as a consequence of Eq.~\eqref{eqn:convexes_nonempty_intersection} and Lemma~\ref{lemma:S_ordered_non_plateau}, we can put all the $ S_{\zeta, k} $ in a nested sequence, i.e.
\begin{equation}
    \label{eqn:sequence_of_S}
    S_{\zeta_1, k_1} \supset S_{\zeta_2, k_2} \supset \dots \supset S_{\zeta_j, k_j} \supset \dots
\end{equation}
Since $ \Omega \subseteq \mathbb{R}^d \Rightarrow S_{\zeta_j, k_j} \subseteq \mathbb{R}^d $ $ \forall j $. Each $ S_{\zeta_j, k_j} $ is non-empty and closed by definition. $ \lim_{j \to \infty} diam \left( S_{\zeta_j, k_j} \right) = 0 $ from Corollary~\ref{corollary:diam_S_to_0}. Then, using \textit{Cantor's Intersection Theorem for complete metric spaces}, since $ \mathbb{R}^d $ is a complete metric space, we know
\begin{equation}
\label{eqn:S_intersect_one_point_j_index}
    \bigcap_{\forall j} S_{\zeta_j, k_j} = \left\{ \mathbf{u} \right\}
\end{equation}
where $ \mathbf{u} \in \Omega $.
\begin{equation}
    \label{eqn:S_intersect_one_point}
    \Rightarrow \bigcap_{\forall \left( \zeta, k \right) \in \Sigma } S_{\zeta, k} = \left\{ \mathbf{u} \right\}
\end{equation}

Finally, $ \left[ \bigcap_{\left( \zeta, k \right) \in \Lambda} l_{\zeta, k} \right] \cap \left[ \bigcap_{\left( \zeta, k \right) \in \Sigma} S_{\zeta, k} \right] \neq \emptyset \Rightarrow \mathbf{u} = \mathbf{a} $, proving point 2 and point 3.
\end{proof}

\begin{theorem}
\label{them:RTF_definition_contour_based}
    Under the conditions of Theorem~\ref{thm:Sigma_converges_one_point}, if $ \mathbf{a} $ is chosen as the anchor point, then the RTF exists and is defined in the following manner:

    If $ \left( r_1, \boldsymbol{\theta}_1 \right) $ is given such that $ \left( \mathbf{x}_a + r_1 \boldsymbol{\theta}_1 \right) \in l_{\zeta, k} $, then, for any $ \boldsymbol{\theta}_2 $, $ R_{1, 2} (r_1) = r_2 $ such that $ \left( \mathbf{x}_a + r_2 \boldsymbol{\theta}_2 \right) \in l_{\zeta, k} $.
\end{theorem}
\begin{proof}[Proof of Theorem~\ref{them:RTF_definition_contour_based}]
    By definition, every point on $ l_{\zeta, k} $ has the same value of $ p (\mathbf{x}) $. From the definition of convexity, since $ \mathbf{a} \in S_{\zeta, k} $ $ \forall \left( \zeta, k \right) \in \Sigma $, any ray $ \mathbf{x}_a + r \boldsymbol{\theta} $ (where $ r \in \left[ 0, \infty \right) $, $ \boldsymbol{\theta} \in \left[ 0, 2 \pi \right) \times \left[ 0, \pi \right]^{(d-2)} $) intersects $ l_{\zeta, k} $ exactly once, which ensures the uniqueness of the RTF as defined in the theorem statement. And Lemma~\ref{lemma:S_ordered_non_plateau} ensures the order-preserving nature of the RTF. Thus, the RTF formed in this way satisfies all the conditions necessary for Definition~\ref{definition:radial_transformation_function}.
\end{proof}

The above condition for the existence of the RTF can be extended (Theorem~\ref{thm:convex_contour_RTF_exists}) even for distributions that have regions of uniformity as long as certain notions of convexity are maintained (Definition~\ref{definition:convex_contoured_density}). For any general density function $ p (\mathbf{x}) $, let us partition $ \Omega $ as follows.
\begin{gather}
    \Omega_U = Cl \left( \left\{ \mathbf{x} \in \Omega : \exists \delta > 0 \text{ such that } p(\mathbf{y}) = p (\mathbf{x}) \; \forall \mathbf{y} \in \Omega \text{ where } \lVert \mathbf{y} - \mathbf{x} \rVert_2 < \delta \right\} \right) \label{eqn:plateau_domain} \\
    \Omega_N = \Omega \setminus \Omega_U \label{eqn:non_plateau_domain}
\end{gather}

We can now define a more general class of density functions as
\begin{definition}[Convex Contoured Density Function]
\label{definition:convex_contoured_density}
    A density function $ p (\mathbf{x}) $ is a convex contoured density function if
    \begin{enumerate}
        \item $ p (\mathbf{x}) $ is continuous.
        \item $ l_{\zeta, k} $ follows one of two conditions $ \forall \zeta, k $ such that $ l_{\zeta, k} \subseteq \Omega_N $
        \begin{enumerate}
            \item $ l_{\zeta, k} $ is a singleton set, i.e., it contains exactly one point, or
            \item $ l_{\zeta, k} $ is the boundary of a \textit{convex body with non-empty interior}, i.e., the boundary of the convex hull of $ l_{\zeta, k} $ is $ l_{\zeta, k} $ itself.
        \end{enumerate}
        \item $ \forall \zeta, k $ such that $ l_{\zeta, k} \subseteq \Omega_U $, if we define $ \lambda_{\zeta, k, \tau} $ as a connected component of $ Bd \left( l_{\zeta, k} \right) $, then it must be the boundary of a \textit{convex body with non-empty interior} $ \forall \zeta, k, \tau $; i.e., the boundary of the convex hull of $ \lambda_{\zeta, k, \tau} $ is $ \lambda_{\zeta, k, \tau} $ itself.
    \end{enumerate}
\end{definition}

For convex contoured density functions, we define $ \Xi = \left\{ \left( \zeta, k \right) : l_{\zeta, k} \subseteq \Omega_U \right\} $ and $\Xi_{T} = \left\{ \left( \zeta, k, \tau \right) : \lambda_{\zeta, k, \tau} \subseteq \Omega_U \right\} $. Thus, $ \left[ \bigcup_{\left( \zeta, k \right) \in \Lambda} l_{\zeta, k} \right] \cup \left[ \bigcup_{\left( \zeta, k \right) \in \Sigma} l_{\zeta, k} \right] = \Omega_N $, $ \left[ \bigcup_{\left( \zeta, k, \tau \right) \in \Xi_T} \lambda_{\zeta, k} \right] = \left[ \bigcup_{\left( \zeta, k \right) \in \Xi} l_{\zeta, k} \right] = \Omega_U $, and the sets $ \Lambda $, $ \Sigma $, and $ \Xi $ are all mutually exclusive. Finally, notate $ S_{\zeta, k, \tau} = Conv(\lambda_{\zeta, k, \tau}) $. As a consequence, $ \lambda_{\zeta, k, \tau} = Bd(S_{\zeta, k, \tau}) $ and $ Int(S_{\zeta, k, \tau}) = \left( S_{\zeta, k, \tau} \setminus \lambda_{\zeta, k, \tau} \right) $.

\begin{theorem}
    \label{thm:convex_contour_RTF_exists}
    For a convex contoured density function $ p (\mathbf{x}) $, the RTF exists if 
    \begin{equation}
        \left[ \bigcap_{\left( \zeta, k \right) \in \Lambda} l_{\zeta, k} \right] \cap \left[ \bigcap_{\left( \zeta, k \right) \in \Sigma} S_{\zeta, k} \right] \cap \left[ \bigcap_{\left( \zeta, k, \tau \right) \in \Xi_T} S_{\zeta, k, \tau} \right] \neq \emptyset
    \end{equation}
\end{theorem}
\begin{proof}[Proof of Theorem~\ref{thm:convex_contour_RTF_exists}]
    From Lemmas~\ref{lemma:S_ordered_non_plateau},~\ref{lemma:S_ordered_for_plateaus} and~\ref{lemma:S_ordered_all}, we can see that we can order all $ S_{\zeta, k} $ and $ S_{\zeta, k , \tau} $. Therefore, for $ \mathbf{x}_a \in \left[ \bigcap_{\left( \zeta, k \right) \in \Lambda} l_{\zeta, k} \right] \cap \left[ \bigcap_{\left( \zeta, k \right) \in \Sigma} S_{\zeta, k} \right] \cap \left[ \bigcap_{\left( \zeta, k, \tau \right) \in \Xi_T} S_{\zeta, k, \tau} \right] $, we can construct a transformation as per Theorem~\ref{them:RTF_definition_contour_based} for all $ l_{\zeta, k} \subseteq \Omega_N $ and Lemma~\ref{lemma:line_segment_RTF_transformation_exists_plateau} for all $ l_{\zeta, k} \subseteq \Omega_U $. This transformation will satisfy all the conditions necessary for the RTF presented in Definition~\ref{definition:radial_transformation_function}.
\end{proof}

A quick note related to Theorem~\ref{thm:convex_contour_RTF_exists} is that if the innermost $ Int \left( S_{\zeta, k, \tau} \right) \subseteq \Omega_N $ then the conditions of Theorem~\ref{thm:Sigma_converges_one_point} apply within $ Int \left( S_{\zeta, k, \tau} \right) $ and there is a unique choice of $ \mathbf{x}_a $. On the other hand, if the innermost $ Int \left( S_{\zeta, k, \tau} \right) \subseteq \Omega_U $, then $ \Lambda = \emptyset $, and any point within the innermost $ Int \left( S_{\zeta, k, \tau} \right) $ can be chosen as the anchor point.

\subsection{Additional Lemmas and Proofs}
\label{appendix:lemmas_for_RTF_proofs}

\begin{lemma}
    \label{lemma:convex_body_unique_intersection}
    For any convex body with non-empty interior $ C $ in $ \Omega $, and any point $ \mathbf{v} \in Int \left( C \right) $, the ray $ Z = \left\{ \mathbf{x} \in \Omega : \mathbf{x} = \mathbf{v} + r \boldsymbol{\theta} \;, \; r \in \left[0, \infty \right) \right\} $ (where $ \boldsymbol{\theta} $ is a fixed direction vector) has a unique intersection point with $ C $, i.e., $ \exists $ unique $ r_0 $ such that $ \left( \mathbf{v} + r_0 \boldsymbol{\theta} \right) \in Bd \left( C \right) $.
\end{lemma}
\begin{proof}[Proof of Lemma~\ref{lemma:convex_body_unique_intersection}]
    \begin{enumerate} The proof is presented in the following steps.
        \item Let there be no such intersection point. In this case, $ \forall \mathbf{z} = Z $, $ \mathbf{z} \in Int \left( C \right) $. Now, $ \lVert \mathbf{z} - \mathbf{v} \rVert_2 = r $, and $ r \in \left[0, \infty \right) $. But this is not possible as $ C $ is bounded, therefore $ diam \left( C \right) $ must be finite. Therefore, there must be a non-zero number of intersection points.
        \item Let the ray have two intersection points $ \mathbf{v}_1 $ and $ \mathbf{v}_2 $ for $ r_1 $ and $ r_2 $ such that $ r_1 > r_2 $. Then $ \mathbf{v}_2 $ can be written as a linear combination of $ \mathbf{v} $ and $ \mathbf{v}_1 $. However, since $ \mathbf{v} \in Int \left( C \right) $, every linear combination of $ \mathbf{v} $ and $ \mathbf{v}_1 $ must also lie in $ Int \left( C \right) $. Therefore, there can only be one intersection point.
    \end{enumerate}
\end{proof}

\begin{lemma}
    \label{lemma:nesting_of_convex_bodies}
    For any two convex bodies with non-empty interiors $ C_1 $ and $ C_2 $ in $ \Omega $, if $ Bd \left( C_2 \right) \subseteq Int \left( C_1 \right) $, then $ C_2 \subset C_1 $.
\end{lemma}
\begin{proof}[Proof of Lemma~\ref{lemma:nesting_of_convex_bodies}]
    As a consequence of the \textit{Separation Theorem}, any convex body in $ \Omega $ is the convex hull of its boundary (see Lemma 1.4.1 in~\cite{schneider2013convex}). Thus,
    \begin{equation}
        C_2 = Conv \left( Bd \left( C_2 \right) \right)
    \end{equation}
    But since $ Bd \left( C_2 \right) \subseteq Int \left( C_1 \right) $ and $ C_1 $ is convex, this implies that $ Conv \left( Bd \left( C_2 \right) \right) \subseteq Int \left( C_1 \right) $
    \begin{equation}
        \Rightarrow C_2 \subset C_1
    \end{equation}
\end{proof}

\begin{lemma}
    \label{lemma:non_empty_intersection_implies_nesting}
    For any two convex bodies with non-empty interiors $ C_1 $ and $ C_2 $ in $ \Omega $, if $ C_1 \cap C_2 \neq \emptyset $ but $ Bd \left( C_1 \right) \cap Bd \left( C_2 \right) = \emptyset $, then, without loss of generality $ C_2 \subset C_1 $.
\end{lemma}
\begin{proof}[Proof of Lemma~\ref{lemma:non_empty_intersection_implies_nesting}]
    Since $ Bd \left( C_1 \right) \cap Bd \left( C_2 \right) = \emptyset $,
    \begin{equation}
        \Rightarrow Bd \left( C_2 \right) = \left[ Bd \left( C_2 \right) \cap Int \left( C_1 \right) \right] \cup \left[ Bd \left( C_2 \right) \cap C_1^C \right]
    \end{equation}
    But $ \left[ Bd \left( C_2 \right) \cap Int \left( C_1 \right) \right] $ and $ \left[ Bd \left( C_2 \right) \cap C_1^C \right] $ are separated sets since $ Int \left( C_1 \right) $ and $ C_1^C $ are separated sets. Since $ Bd \left( C_2 \right) $ is a connected set, it cannot be written as the union of two separated sets.
    \begin{equation}
        \Rightarrow \text{ either} \quad Bd \left( C_2 \right) \subseteq Int \left( C_1 \right) \quad \text{or} \quad Bd \left( C_2 \right) \subseteq C_1^C
    \end{equation}
    Similarly
    \begin{equation}
        Bd \left( C_1 \right) \subseteq Int \left( C_2 \right) \quad \text{or} \quad Bd \left( C_1 \right) \subseteq C_2^C
    \end{equation}
    If $ C_1 \cap C_2 \neq \emptyset $,
    \begin{equation}
        \Rightarrow \left[ C_1 \cap Int \left( C_2 \right) \right] \cup \left[ C_1 \cap Bd \left( C_2 \right) \right] \neq \emptyset
    \end{equation}
    Now, if $ Bd \left( C_1 \right) \subseteq C_2^C \Rightarrow C_1 \cap C_2^C \neq \emptyset $. Then, since $ C_1 $ is connected, and
    \begin{equation}
        C_1 = \left[ C_1 \cap Int \left( C_2 \right) \right] \cup \left[ C_1 \cap Bd \left( C_2 \right) \right] \cup \left[ C_1 \cap C_2^C \right].
    \end{equation}
    We can see that $ \left[ C_1 \cap Bd \left( C_2 \right) \right] \neq \emptyset \Rightarrow Bd \left( C_2 \right) \nsubseteq C_1^C \Rightarrow Bd \left( C_2 \right) \subseteq Int \left( C_1 \right) $. Similarly we can show that $ Bd \left( C_1 \right) \subseteq Int \left( C_2 \right) $ if $ Bd \left( C_2 \right) \subseteq C_1^C $. Therefore, without loss of generality $ Bd \left( C_2 \right) \subseteq Int \left( C_1 \right) $. And as a consequence of Lemma~\ref{lemma:nesting_of_convex_bodies} we get $ C_2 \subset C_1 $.
\end{proof}

    

\begin{lemma}
    \label{lemma:S_ordered_non_plateau}
    For a convex contoured density function $ p (\mathbf{x}) $, given any two tuples $ \left( \zeta_1, k_1 \right) $ and $ \left( \zeta_2, k_2 \right) $, if $ \left( \zeta_1, k_1 \right) \neq \left( \zeta_2, k_2 \right) $ and $ \left[ \bigcap_{\left( \zeta, k \right) \in \Sigma} S_{\zeta, k} \right] \neq \emptyset $, then, without loss of generality, $ S_{\zeta_2, k_2} \subset S_{\zeta_1, k_1} $.
\end{lemma}
\begin{proof}[Proof of Lemma~\ref{lemma:S_ordered_non_plateau}]
    First, realize that $ l_{\zeta_1, k_1} \cap l_{\zeta_2, k_2} = \emptyset $ by definition, and $ \left[ \bigcap_{\left( \zeta, k \right) \in \Sigma} S_{\zeta, k} \right] \neq \emptyset \Rightarrow S_{\zeta_1, k_1} \cap S_{\zeta_2, k_2} \neq \emptyset $. Thus, as a direct consequence of Lemma~\ref{lemma:non_empty_intersection_implies_nesting}, we get without loss of generality, $ S_{\zeta_2, k_2} \subset S_{\zeta_1, k_1} $.

\end{proof}

\begin{lemma}
    \label{lemma:S_ordered_for_plateaus}
    For a convex contoured density function $ p (\mathbf{x}) $, given any two tuples $ \left( \zeta_1, k_1, \tau_1 \right) $ and $ \left( \zeta_2, k_2, \tau_2 \right) $, if $ \left( \zeta_1, k_1, \tau_1 \right) \neq \left( \zeta_2, k_2, \tau_2 \right) $ and $ \left[ \bigcap_{\left( \zeta, k, \tau \right) \in \Xi_T} S_{\zeta, k, \tau} \right] \neq \emptyset $, then, without loss of generality, $ S_{\zeta_2, k_2, \tau_2} \subset S_{\zeta_1, k_1, \tau_1} $.
\end{lemma}
\begin{proof}[Proof of Lemma~\ref{lemma:S_ordered_for_plateaus}]
    Similar to Lemma~\ref{lemma:S_ordered_non_plateau}, this is a direct consequence of Lemma~\ref{lemma:non_empty_intersection_implies_nesting}.
\end{proof}

\begin{lemma}
    \label{lemma:S_ordered_all}
    For a convex contoured density function $ p (\mathbf{x}) $, given any two tuples $ \left( \zeta_1, k_1 \right) $ and $ \left( \zeta_2, k_2, \tau_2 \right) $, if $ \left[ \bigcap_{\left( \zeta, k \right) \in \Sigma} S_{\zeta, k} \right] \cap \left[ \bigcap_{\left( \zeta, k, \tau \right) \in \Xi_T} S_{\zeta, k, \tau} \right] \neq \emptyset $, then either $ S_{\zeta_2, k_2, \tau_2} \subset S_{\zeta_1, k_1} $ or $ S_{\zeta_1, k_1} \subset S_{\zeta_2, k_2, \tau_2} $.
\end{lemma}
\begin{proof}[Proof of Lemma~\ref{lemma:S_ordered_all}]
    Once again, this is a direct consequence of Lemma~\ref{lemma:non_empty_intersection_implies_nesting}.
\end{proof}

\begin{lemma}
    \label{lemma:ball_contains_component_level_set}
    For a non-plateauing convex contoured density function $ p (\mathbf{x}) $, if $ \left[ \bigcap_{\left( \zeta, k \right) \in \Lambda} l_{\zeta, k} \right] \cap \left[ \bigcap_{\left( \zeta, k \right) \in \Sigma} S_{\zeta, k} \right] \neq \emptyset $, then $ \exists B_{\delta} $ such that $ \exists \left( \zeta, k \right) \in \Sigma $ for which $ l_{\zeta, k} \subseteq B_{\delta} $ $ \forall \delta > 0 $, where $ B_{\delta} \in \Omega $ notates a closed ball with radius $ \delta $.
\end{lemma}
\begin{proof}[Proof of Lemma~\ref{lemma:ball_contains_component_level_set}]
    Since $ \left[ \bigcap_{\left( \zeta, k \right) \in \Lambda} l_{\zeta, k} \right] \cap \left[ \bigcap_{\left( \zeta, k \right) \in \Sigma} S_{\zeta, k} \right] \neq \emptyset \Rightarrow \left[ \bigcap_{\left( \zeta, k \right) \in \Lambda} l_{\zeta, k} \right] \neq \emptyset $. Therefore, by Theorem~\ref{thm:Lambda_has_one_point}, there is a unique set $ l_{\zeta_0, 0} $ which is the only singleton. Let $ l_{\zeta_0, 0} = \left\{ \mathbf{a} \right\} $. Define
    \begin{equation}
    \label{eqn:delta_ball_around_a}
        B_{\delta,\mathbf{a}} = \left\{ \mathbf{x} \in \Omega : \lVert \mathbf{x} - \mathbf{a} \rVert_2 \leq \delta \right\}
    \end{equation}
    We will show that for this choice of $ B_{\delta} $, our theorem holds; i.e. $ \exists \left( \zeta, k \right) \in \Sigma $ for which $ l_{\zeta, k} \subseteq B_{\delta,\mathbf{a}} $.

    Note that the family of sets $ l_{\zeta, k} $ forms a partition of the level set $ L_{\zeta} $ for a given $ \zeta $, and the family of level sets $ L_{\zeta} $ forms a partition of $ \Omega $. Therefore, the family of sets $ l_{\zeta, k} $ forms a partition of $ \Omega $ (when the tuple $ \left( \zeta, k \right) $ takes all possible values). Hence, $ \forall \mathbf{x} \in B_{\delta,\mathbf{a}} $, $ \mathbf{x} \in l_{\zeta, k} $ for some $ \left( \zeta, k \right) $. Further, since there is only one value of $ \left( \zeta, k \right) \in \Lambda $ (namely $ \left( \zeta, k \right) = \left( \zeta_0, 0 \right) $), $ \Rightarrow \forall \mathbf{x} \in \left( B_{\delta,\mathbf{a}} \setminus \left\{ \mathbf{a} \right\} \right) $, $ \mathbf{x} \in l_{\zeta, k} $ for some $ \left( \zeta, k \right) \in \Sigma $.

    Define $ \mathcal{U} = \left\{ \left( \zeta, k \right) : l_{\zeta, k} \cap \left( B_{\delta,\mathbf{a}} \setminus \left\{ \mathbf{a} \right\} \right) \neq \emptyset \right\} $, and notice that $ \mathcal{U} \subseteq \Sigma $. Since $ \left[ \bigcap_{\left( \zeta, k \right) \in \Lambda} l_{\zeta, k} \right] \cap \left[ \bigcap_{\left( \zeta, k \right) \in \Sigma} S_{\zeta, k} \right] \neq \emptyset $, 
    \begin{equation}
    \label{eqn:a_in_all_S_interior}
        \Rightarrow \mathbf{a} \in Int \left( S_{\zeta, k} \right) \quad \forall \left( \zeta, k \right) \in \mathcal{U}
    \end{equation}
    (This is because $ \mathbf{a} \in l_{\zeta_0, 0} \Rightarrow \mathbf{a} \notin l_{\zeta, k} $ for any other tuple.)

    Further define $ \mathcal{S}_{\mathcal{U}} = \left\{ S_{\zeta, k} : \left( \zeta, k \right) \in \mathcal{U} \right\} $ and $ \mathcal{L}_{\mathcal{U}} = \left\{ l_{\zeta, k} : \left( \zeta, k \right) \in \mathcal{U} \right\} $. Finally, define the shorthand $ \mu \equiv \left( \zeta, k \right) \in \mathcal{U} $.

    Now, let us assume the converse of our goal, i.e. let $ l_{\mu} \cap B_{\delta,\mathbf{a}}^C \neq \emptyset $ $ \forall \mu $. Notice that $ l_{\mu} \cap B_{\delta,\mathbf{a}} \neq \emptyset $ by definition $ \forall \mu $.
    Thus,
    \begin{equation}
    \label{eqn:s_outside_B_exists}
        S_{\mu} \cap B_{\delta,\mathbf{a}}^C \neq \emptyset \Rightarrow S_{\mu} \cap Int \left( B_{\delta,\mathbf{a}} \right)^C \neq \emptyset \quad \forall \mu
    \end{equation}
    and from Eq.~\eqref{eqn:a_in_all_S_interior}
    \begin{equation}
    \label{eqn:s_inside_B_exists}
        Int \left( S_{\mu} \right) \cap Int \left( B_{\delta,\mathbf{a}} \right) \neq \emptyset \quad \forall \mu
    \end{equation}

    Now, $ S_{\mu} $ is closed, implying that $ S_{\mu}^C $ is an open set. And $ Int \left( B_{\delta,\mathbf{a}} \right) $ is open by definition. Thus, $ \left[ S_{\mu}^C \cup Int \left( B_{\delta,\mathbf{a}} \right) \right] = \left[ S_{\mu} \cap Int \left( B_{\delta,\mathbf{a}} \right)^C \right]^C $ is open, implying that $ \left[ S_{\mu} \cap Int \left( B_{\delta,\mathbf{a}} \right)^C \right] $ is a closed set $ \forall \mu $. Consequently, since $ S_{\mu} $ is compact, $ \left[ S_{\mu} \cap Int \left( B_{\delta,\mathbf{a}} \right)^C \right] $ is a compact set as well, $ \forall \mu $ (a closed subset of a closed compact set is compact).

    As a corollary of Lemma~\ref{lemma:S_ordered_non_plateau}, we can order all the elements of $ \mathcal{S}_{\mathcal{U}} $ as
    \begin{equation}
        \label{eqn:order_of_S_mu}
        S_{\mu_1} \supset S_{\mu_2} \supset \dots \supset S_{\mu_j} \supset \dots
    \end{equation}
    where $ \mu_j \in \mathcal{U} $ $ \forall j $. Therefore,
    \begin{equation}
        \label{eqn:order_of_S_mu_interior_B}
        \left( S_{\mu_1} \cap Int \left( B_{\delta,\mathbf{a}} \right)^C \right) \supset \left( S_{\mu_2} \cap Int \left( B_{\delta,\mathbf{a}} \right)^C \right) \supset \dots \supset \left( S_{\mu_j} \cap Int \left( B_{\delta,\mathbf{a}} \right)^C \right) \supset \dots
    \end{equation}

    Therefore the sequence of sets $ \left\{ \left( S_{\mu_j} \cap Int \left( B_{\delta,\mathbf{a}} \right)^C \right) \right\}_{\forall j} $ follows the conditions for \textit{Cantor's Intersection Theorem}, implying
    \begin{equation}
        \label{eqn:nonempy_intersection_S_B}
        \bigcap_{\forall j} \left( S_{\mu_j} \cap Int \left( B_{\delta,\mathbf{a}} \right)^C \right) \neq \emptyset
    \end{equation}

    Thus, $ \exists \mathbf{z} \in \left( S_{\mu} \cap Int \left( B_{\delta,\mathbf{a}} \right)^C \right) $ $ \forall \mu $. Define $ \mathcal{V}_{\mathbf{az}} = \left\{ (1-t) \mathbf{a} + t \mathbf{z} : t \in \left[ 0, 1 \right)] \right\} $. From Eq.~\eqref{eqn:a_in_all_S_interior} and the convexity of $ S_{\mu} $, we know that $ \mathcal{V}_{\mathbf{az}} \subseteq Int \left( S_{\mu} \right) $ $ \forall \mu $. And since $ \mathbf{a} \in Int \left( B_{\delta,\mathbf{a}} \right) $ and $ B_{\delta,\mathbf{a}} $ is compact, $ \Rightarrow \exists \mathbf{v} \in \mathcal{V}_{\mathbf{az}} $ such that $ \mathbf{v} \in Int \left( B_{\delta,\mathbf{a}} \right) $.

    From the partition argument at the beginning, we know that $ \mathbf{v} \in l_{\mu_v} $ for some $ \mu_v \in \mathcal{U} $, which implies that $ \mathbf{v} \notin Int \left( S_{\mu_v} \right) $. But this is a contradiction as $ \mathbf{v} \in \mathcal{V}_{\mathbf{az}} $ and $ \mathcal{V}_{\mathbf{az}} \subseteq Int \left( S_{\mu} \right) $ $ \forall \mu $ (including $ \mu_v $).

    Therefore, $ \exists \mu $ such that $ l_{\mu} \cap B_{\delta,\mathbf{a}} = \emptyset \Rightarrow l_{\mu} \subseteq B_{\delta,\mathbf{a}} $. And since $ \mathcal{U} \subseteq \Sigma $, we get that $ \exists \left( \zeta, k \right) \in \Sigma $ for which $ l_{\zeta, k} \subseteq B_{\delta,\mathbf{a}} $.
\end{proof}

\begin{lemma}
    \label{lemma:S_in_B}
    For a non-plateauing convex contoured density function $ p (\mathbf{x}) $, if $ \left( \zeta, k \right) \in \Sigma $ and $ l_{\zeta, k} \subseteq Int \left( B_{\delta} \right) $, then $ S_{\zeta, k} \subset B_{\delta} $, where $ B_{\delta} $ is a ball of radius $ \delta $.
\end{lemma}
\begin{proof}[Proof of Lemma~\ref{lemma:S_in_B}]
    Define $ C = \left\{ \mathbf{z} = t \mathbf{x}_1 + (1-t) \mathbf{x}_2 | \mathbf{x}_1 , \mathbf{x}_2 \in l_{\zeta, k}, t \in \left[ 0, 1 \right] \right\} $. By definition, $ C \subseteq Int \left( B_{\delta} \right) $ (convex combination of points in the interior of a convex set lie in the interior of the convex set).
    
    Also by definition, $ C $ is the convex hull of $ l_{\zeta, k} $, i.e., $ C = S_{\zeta, k} $.
    \begin{align*}
        \Rightarrow S_{\zeta, k} &\subseteq Int \left( B_{\delta} \right) \\
        \Rightarrow S_{\zeta, k} &\subset B_{\delta}
    \end{align*}
\end{proof}

\begin{corollary}
\label{corollary:diam_S_to_0}
    For a non-plateauing convex contoured density function $ p (\mathbf{x}) $, as a consequence of Lemmas~\ref{lemma:ball_contains_component_level_set} and~\ref{lemma:S_in_B}, we have that the greatest lower bound for $ diam \left( S_{\zeta, k} \right) $ is $ 0 $ $ \forall \left( \zeta, k \right) \in \Sigma $.
\end{corollary}

\begin{lemma}
    \label{lemma:line_segment_RTF_transformation_exists_plateau}
    For a convex contoured density function $ p (\mathbf{x}) $, choose any $ \left( \zeta, k \right) \in \Xi $ such that $ Bd \left( l_{\zeta, k} \right) = \lambda_{\zeta, k, \tau_1} \cup \lambda_{\zeta, k, \tau_2} $ and $ S_{\zeta, k, \tau_2} \subset S_{\zeta, k, \tau_1} $. Then, for any point $ \mathbf{v} \in S_{\zeta, k, \tau_2} $ and given two directions $ \boldsymbol{\theta}_1 $ and $ \boldsymbol{\theta}_2 $, define the line segments $ Z_i = \left\{ \mathbf{x} \in \Omega : \mathbf{x} = \mathbf{v} + r_i \boldsymbol{\theta_i} \; , \; r_i > \left[ 0, \infty \right) \right\} $ for $ i = 1, 2 $. There is a one-to-one transformation between $ \mathcal{Z}_1 = \left[ Z_1 \cap Cl \left( S_{\zeta, k, \tau_1} \setminus S_{\zeta, k, \tau_2} \right) \right] $ and $ \mathcal{Z}_2 = \left[ Z_2 \cap Cl \left( S_{\zeta, k, \tau_1} \setminus S_{\zeta, k, \tau_2} \right) \right] $ that follows the requirements of an RTF as per Definition~\ref{definition:radial_transformation_function}.
\end{lemma}
\begin{proof}[Proof of Lemma~\ref{lemma:line_segment_RTF_transformation_exists_plateau}]
    Since $ \mathcal{Z}_i \subseteq l_{\zeta, k} $, $ i = 1, 2 $, and by definition any point in $ l_{\zeta, k} $ has a density value of $ \zeta $, the first condition of the RTF is trivially satisfied for any mapping between $ \mathcal{Z}_1 $ and $ \mathcal{Z}_2 $.

    Next, since $ \mathbf{v} \in S_{\zeta, k, \tau_1} $, $ \mathbf{v} \in S_{\zeta, k, \tau_2} $, and $ S_{\zeta, k, \tau_1} $ and $ S_{\zeta, k, \tau_2} $ are bounded and convex, the rays $ Z_1 $ and $ Z_2 $ intersect the boundary of each set exactly once each (Lemma~\ref{lemma:convex_body_unique_intersection}). Define $ \left[ Z_i \cap \lambda_{\zeta, k, \tau_j} \right] = \left\{ \mathbf{y}_{ij} \right\} $ for $ i, j = 1, 2 $. These points can be written as
    \begin{equation}
        \begin{aligned}
        \label{eqn:plateau_segments_intersections}
            \mathbf{y}_{11} &= \mathbf{v} + b_1 \boldsymbol{\theta}_1 \qquad & \mathbf{y}_{12} &= \mathbf{v} + a_1 \boldsymbol{\theta}_1 \\
            \mathbf{y}_{21} &= \mathbf{v} + b_2 \boldsymbol{\theta}_2 \qquad & \mathbf{y}_{22} &= \mathbf{v} + a_2 \boldsymbol{\theta}_2
    \end{aligned}
    \end{equation}
    where $ a, b > 0 $. Additionally, $ S_{\zeta, k, \tau_2} \subset S_{\zeta, k, \tau_1} $ implies $ \mathbf{y}_{12} , \mathbf{y}_{22} \in Int \left( S_{\zeta, k, \tau_1} \right) $, and therefore, using the convexity of $ S_{\zeta, k, \tau_1} $, we can also write
    \begin{equation}
        \begin{aligned}
        \label{eqn:plateau_segments_intersections_convexity}
            \mathbf{y}_{12} &= t \mathbf{v} + (1-t) \mathbf{y}_{11} \; , \quad 0 < t < 1 \\
            \mathbf{y}_{22} &= t \mathbf{v} + (1-t) \mathbf{y}_{21} \; , \quad 0 < t < 1
        \end{aligned}
    \end{equation}
    Substituting Eq.~\eqref{eqn:plateau_segments_intersections} into Eq.~\eqref{eqn:plateau_segments_intersections_convexity}, we can say that $ a_1 < b_1 $ and $ a_2 < b_2 $.
    
    Now, it is easy to see that $ \left[ Z_i \cap S_{\zeta, k, \tau_1} \right] = \left\{ \mathbf{x} \in \Omega : \mathbf{x} = \mathbf{v} + r_i \boldsymbol{\theta_i} \; , \; r_i > \left[ 0, b_i \right] \right\} $ and $ \left[ Z_i \cap S_{\zeta, k, \tau_2} \right] = \left\{ \mathbf{x} \in \Omega : \mathbf{x} = \mathbf{v} + r_i \boldsymbol{\theta_i} \; , \; r_i > \left[ 0, a_i \right] \right\} $ for $ i = 1, 2 $. Therefore, the segment $ \mathcal{Z}_i $ can be parameterized by $ r_i \in \left[ a_i, b_i \right] $ which is a continuous and bounded interval. Thus, the mapping
    \begin{equation}
        r_2 = a_2 + \frac{b_2 - a_2}{b_1 - a_1} \left( r_1 - a_1 \right)
    \end{equation}
    satisfies the second condition from Definition~\ref{definition:radial_transformation_function} and is one-to-one.
\end{proof}







\section{Additional Scatterplots for Section~\ref{section:distribution_shape_results} Examples}
\label{appendix:shape_results_scatterplots}

\begin{figure}[!ht]
\centering
\begin{subfigure}{.32\textwidth}
  \centering
  \includegraphics[width=\linewidth]{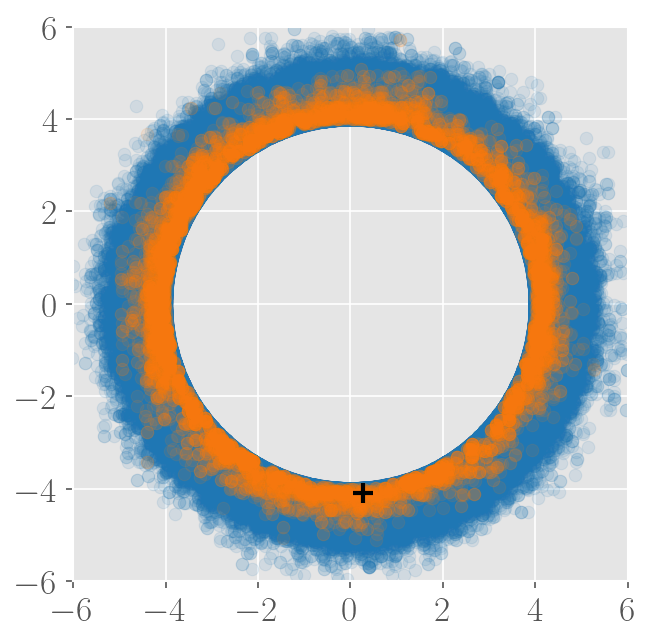}
  \caption{}
  \label{fig:gauss_ring_scatter_intrepid}
\end{subfigure}%
\begin{subfigure}{.32\textwidth}
  \centering
  \includegraphics[width=\linewidth]{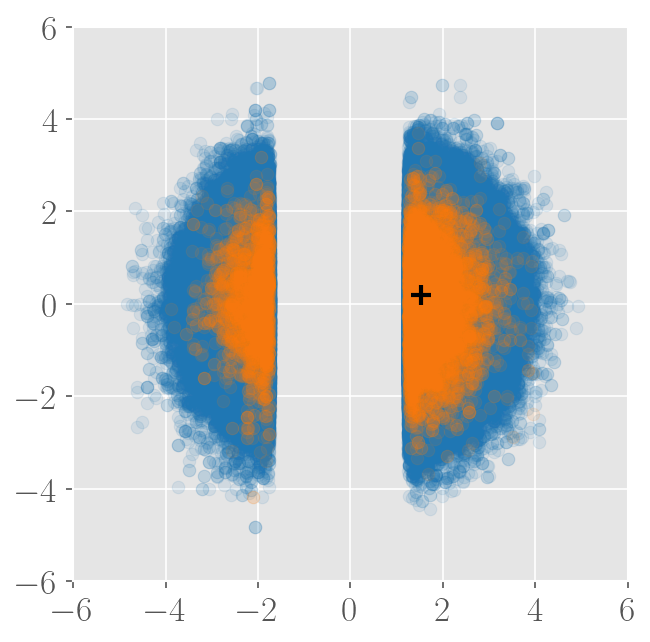}
  \caption{}
  \label{fig:gauss_planes_scatter_intrepid}
\end{subfigure}%
\begin{subfigure}{.32\textwidth}
  \centering
  \includegraphics[width=\linewidth]{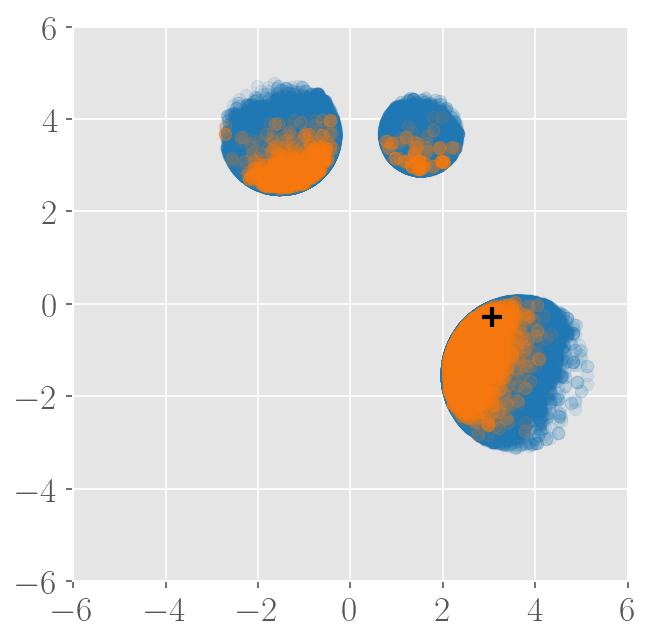}
  \caption{}
  \label{fig:gauss_circles_scatter_intrepid}
\end{subfigure}
\begin{subfigure}{.32\textwidth}
  \centering
  \includegraphics[width=\linewidth]{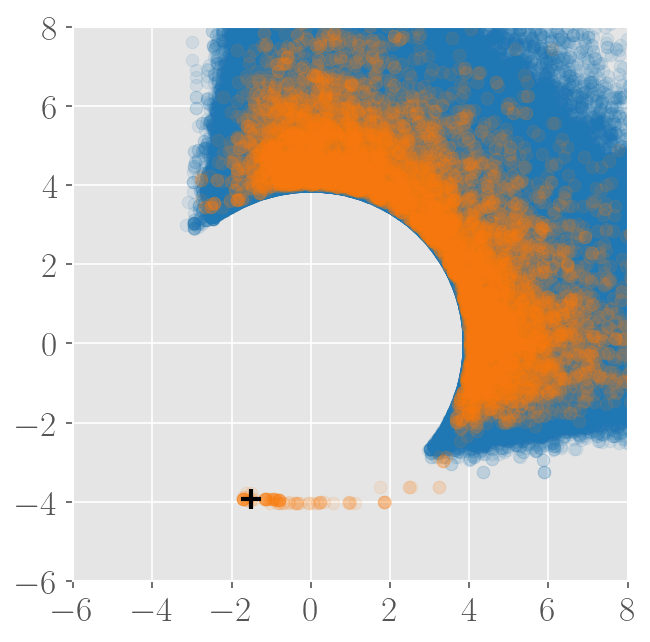}
  \caption{}
  \label{fig:gumbel_ring_scatter_intrepid}
\end{subfigure}%
\begin{subfigure}{.32\textwidth}
  \centering
  \includegraphics[width=\linewidth]{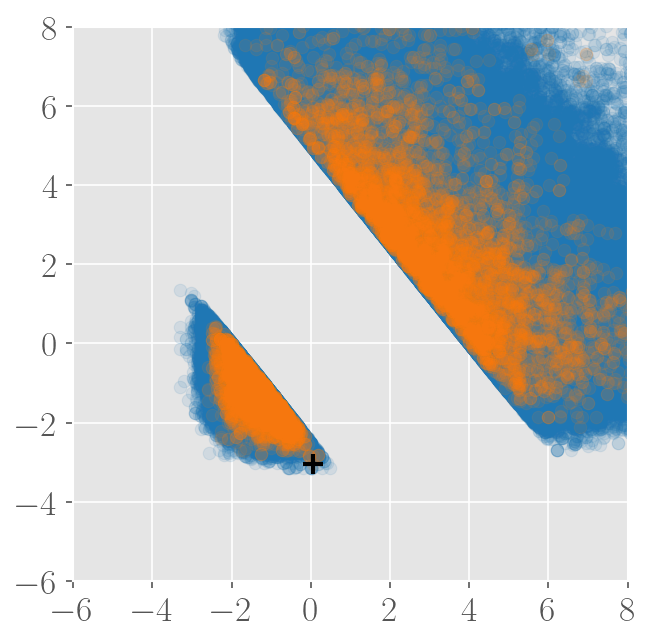}
  \caption{}
  \label{fig:gumbel_planes_scatter_intrepid}
\end{subfigure}%
\begin{subfigure}{.32\textwidth}
  \centering
  \includegraphics[width=\linewidth]{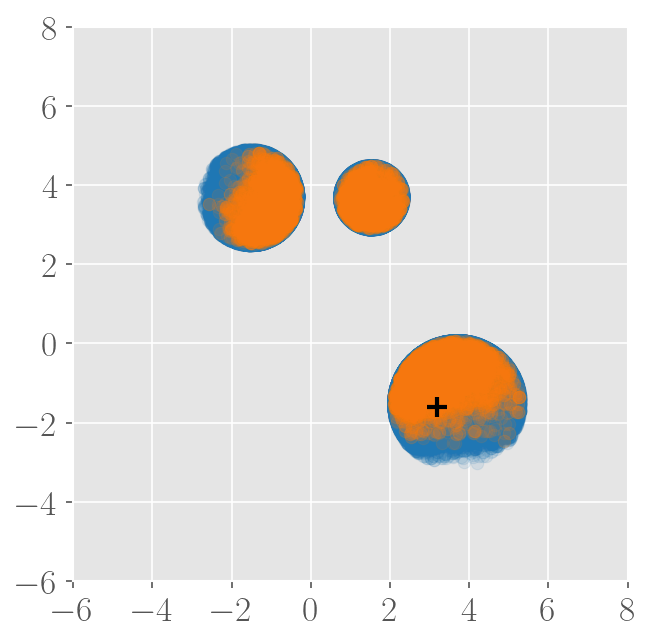}
  \caption{}
  \label{fig:gumbel_circles_scatter_intrepid}
\end{subfigure}
\begin{subfigure}{.32\textwidth}
  \centering
  \includegraphics[width=\linewidth]{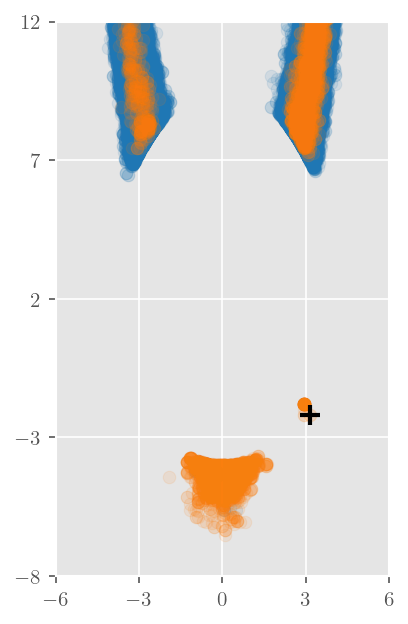}
  \caption{}
  \label{fig:rosenbrock_ring_scatter_intrepid}
\end{subfigure}%
\begin{subfigure}{.32\textwidth}
  \centering
  \includegraphics[width=\linewidth]{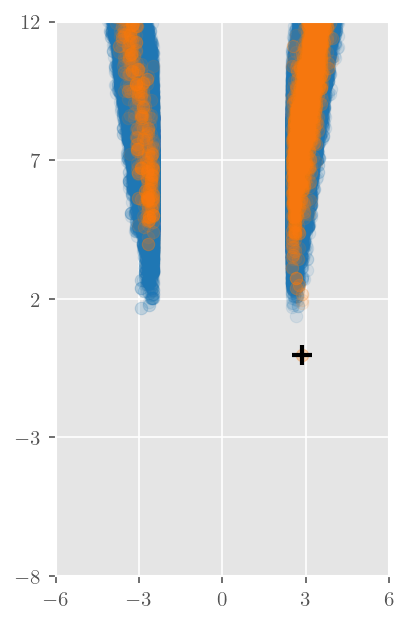}
  \caption{}
  \label{fig:rosenbrock_planes_scatter_intrepid}
\end{subfigure}%
\begin{subfigure}{.32\textwidth}
  \centering
  \includegraphics[width=\linewidth]{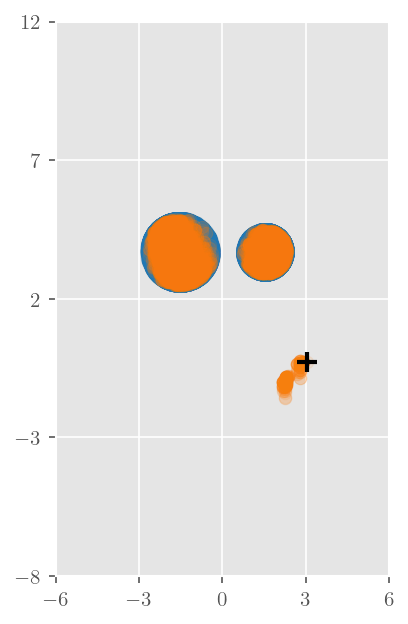}
  \caption{}
  \label{fig:rosenbrock_circles_scatter_intrepid}
\end{subfigure}
\caption{Scatterplot of samples produced by a single Intrepid Markov chain ($ \beta = 0.1 $) for all cases from Section~\ref{section:distribution_shape_results}. (a) Case 1 (Gauss-Ring), (b) Case 2 (Gauss-Planes), (c) Case 3 (Gauss-Circles), (d) Case 4 (Gumbel-Ring), (e) Case 5 (Gumbel-Planes), (f) Case 6 (Gumbel-Circles), (g) Case 7 (Rosenbrock-Ring), (h) Case 8 (Rosenbrock-Planes), and (i) Case 9 (Rosenbrock-Circles). The 10,000 burn-in samples are plotted in orange, all subsequent samples are plotted in blue, and the initial state of the Markov chain is highlighted by the black cross.}
\label{fig:nine_shapes_scatter_intrepid}
\end{figure}

\begin{figure}[!ht]
\centering
\begin{subfigure}{.32\textwidth}
  \centering
  \includegraphics[width=\linewidth]{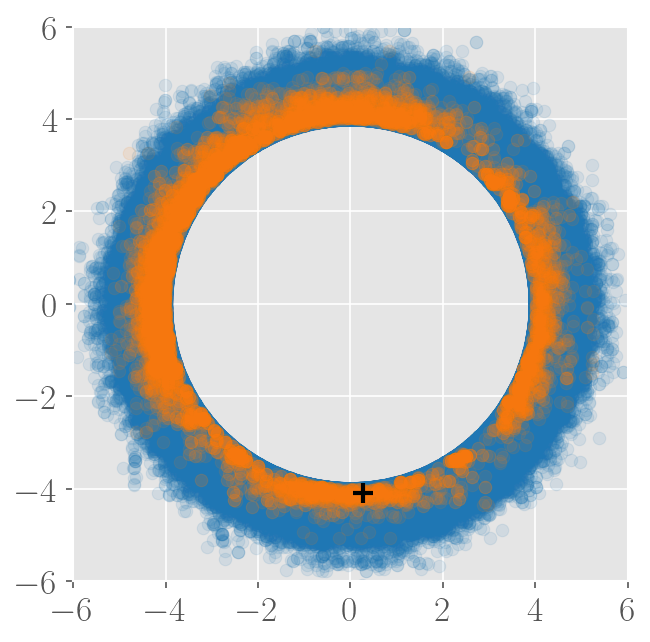}
  \caption{}
  \label{fig:gauss_ring_scatter_cmh}
\end{subfigure}%
\begin{subfigure}{.32\textwidth}
  \centering
  \includegraphics[width=\linewidth]{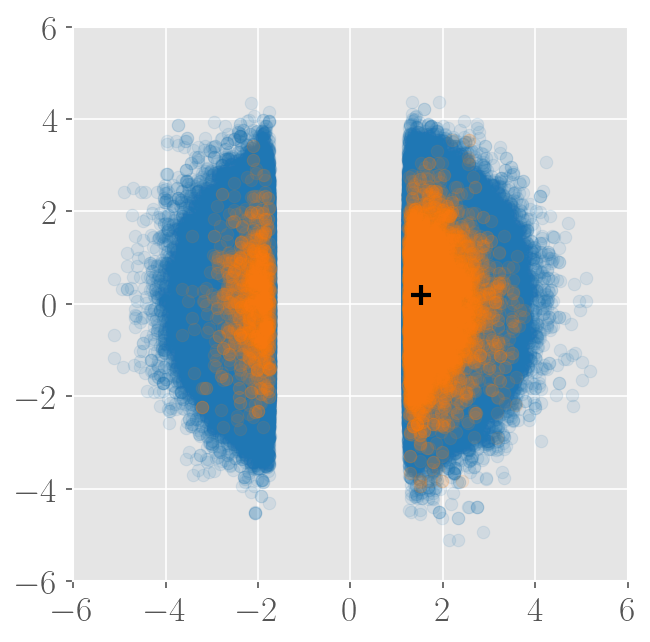}
  \caption{}
  \label{fig:gauss_planes_scatter_cmh}
\end{subfigure}%
\begin{subfigure}{.32\textwidth}
  \centering
  \includegraphics[width=\linewidth]{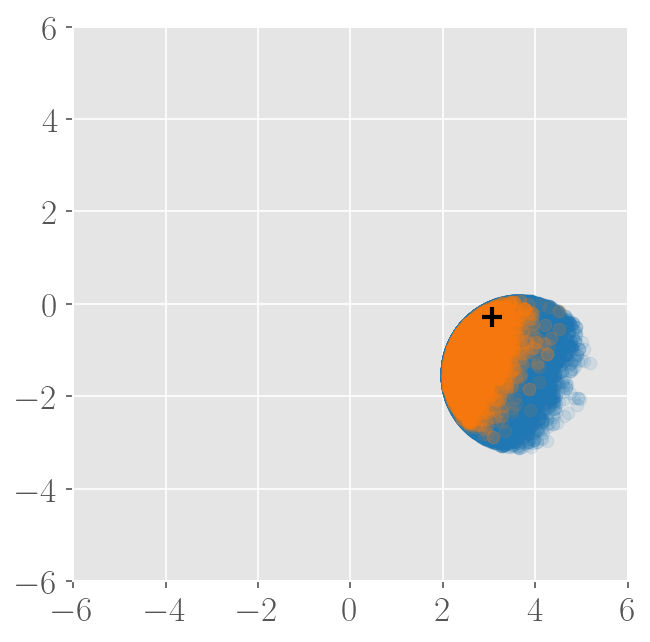}
  \caption{}
  \label{fig:gauss_circles_scatter_cmh}
\end{subfigure}
\begin{subfigure}{.32\textwidth}
  \centering
  \includegraphics[width=\linewidth]{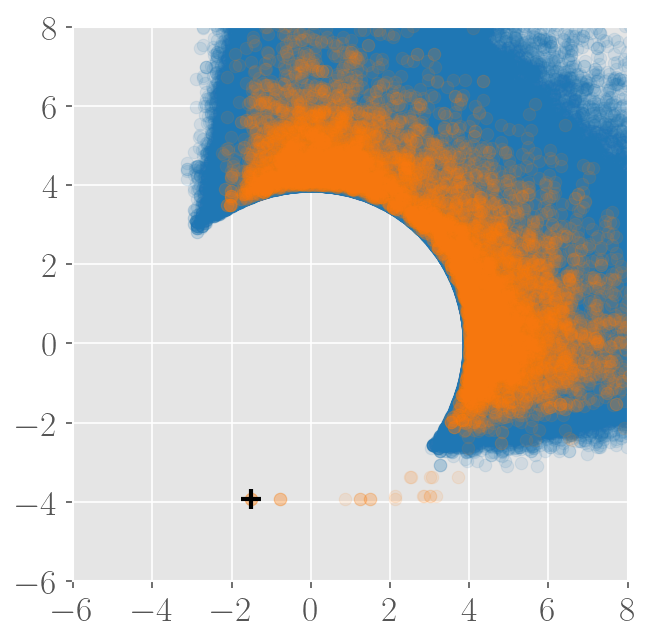}
  \caption{}
  \label{fig:gumbel_ring_scatter_cmh}
\end{subfigure}%
\begin{subfigure}{.32\textwidth}
  \centering
  \includegraphics[width=\linewidth]{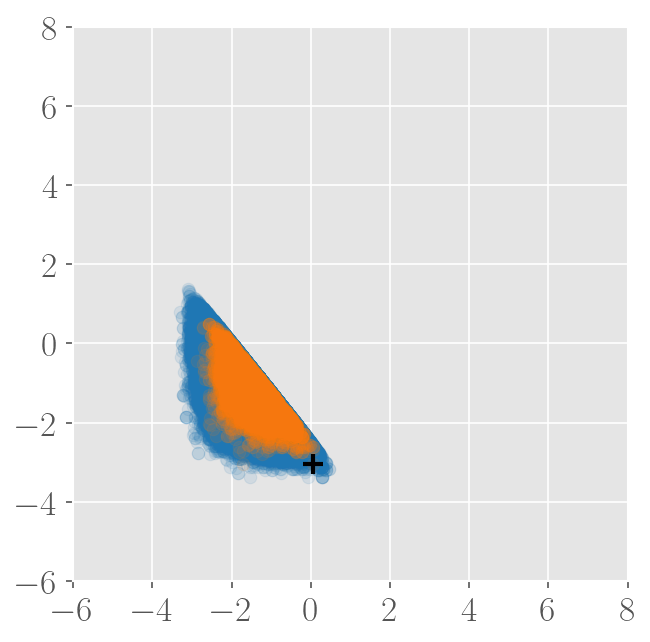}
  \caption{}
  \label{fig:gumbel_planes_scatter_cmh}
\end{subfigure}%
\begin{subfigure}{.32\textwidth}
  \centering
  \includegraphics[width=\linewidth]{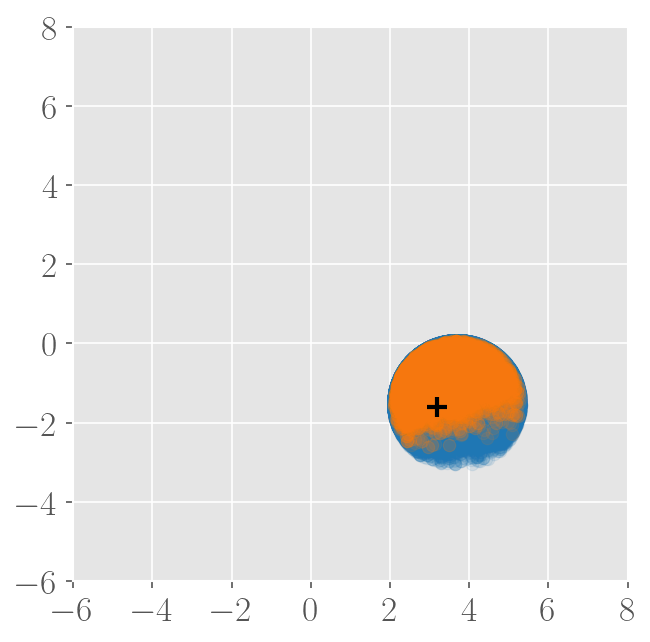}
  \caption{}
  \label{fig:gumbel_circles_scatter_cmh}
\end{subfigure}
\begin{subfigure}{.32\textwidth}
  \centering
  \includegraphics[width=\linewidth]{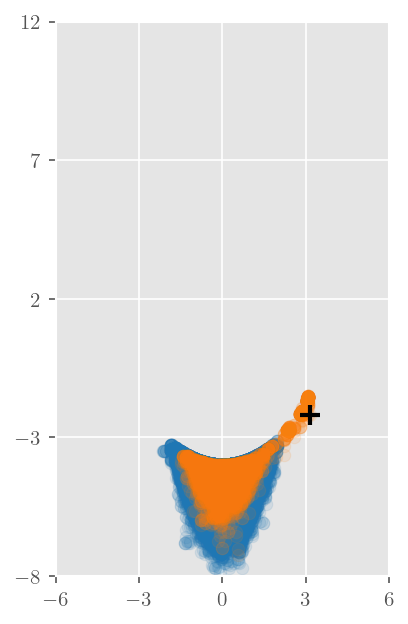}
  \caption{}
  \label{fig:rosenbrock_ring_scatter_cmh}
\end{subfigure}%
\begin{subfigure}{.32\textwidth}
  \centering
  \includegraphics[width=\linewidth]{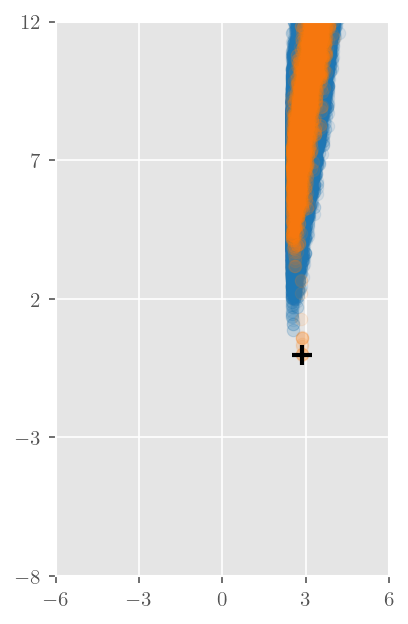}
  \caption{}
  \label{fig:rosenbrock_planes_scatter_cmh}
\end{subfigure}%
\begin{subfigure}{.32\textwidth}
  \centering
  \includegraphics[width=\linewidth]{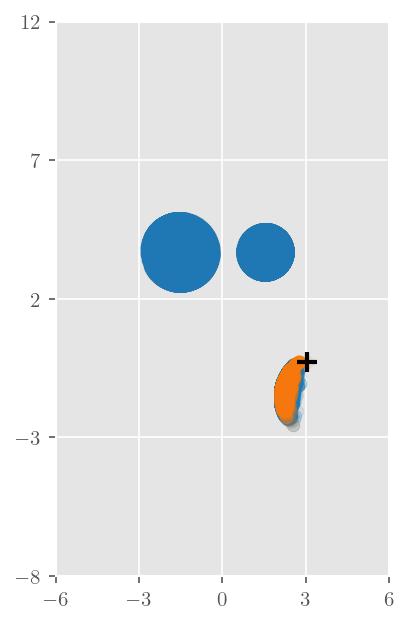}
  \caption{}
  \label{fig:rosenbrock_circles_scatter_cmh}
\end{subfigure}
\caption{Scatterplot of samples produced by a single CMH Markov chain for all cases from Section~\ref{section:distribution_shape_results}. (a) Case 1 (Gauss-Ring), (b) Case 2 (Gauss-Planes), (c) Case 3 (Gauss-Circles), (d) Case 4 (Gumbel-Ring), (e) Case 5 (Gumbel-Planes), (f) Case 6 (Gumbel-Circles), (g) Case 7 (Rosenbrock-Ring), (h) Case 8 (Rosenbrock-Planes), and (i) Case 9 (Rosenbrock-Circles). The 10,000 burn-in samples are plotted in orange, all subsequent samples are plotted in blue, and the initial state of the Markov chain is highlighted by the black cross.}
\label{fig:nine_shapes_scatter_cmh}
\end{figure}

Representative scatterplots of samples produced by a single chain of Intrepid (Figure~\ref{fig:nine_shapes_scatter_intrepid}) and CMH (Figure~\ref{fig:nine_shapes_scatter_cmh}) are presented here to further highlight the behavior of the two methods. The proposal parameters for both methods are as listed in Section~\ref{section:distribution_shape_results}, with Intrepid using $ \beta = 0.1 $. Each chain is propagated to 1 million samples after 10,000 samples are discarded as burn-in.

As stated in the conclusion, the figures in this section highlight the increased robustness to starting location that Intrepid MCMC enjoys over a classical MH method such as CMH. Particularly for the multimodal distributions, CMH struggles to escape the mode closest to the starting location of the chain, and in many cases is unable to sample from the other modes at all. Even in cases where the full length of the CMH chain does populate all regions of the target distribution, inspecting the burn-in samples exposes the tendency of CMH to concentrate samples in some regions. On the other hand, Intrepid MCMC is able to fully populate all regions of the target distribution in each of the nine cases, and does so even within the burn-in length, highlighting a significantly faster convergence. Its behavior is also more uniform, with there being significantly fewer noticeable regions of sample concentration compared to CMH.

\bibliographystyle{unsrt}  


\end{document}